\documentclass[nofootinbib,aps,twocolumn,epsf]{revtex4}

\usepackage{graphicx}
\usepackage{amsmath}
\usepackage{amsfonts}
\usepackage{amssymb}
\def\y{\'{\i}}
\def\ni{\noindent}
\def\bea{\begin{eqnarray}}
\def\eea{\end{eqnarray}}
\def\beq{\begin{equation}}
\def\eeq{\end{equation}}

\begin{document}

\title{Topics on Hydrodynamic Model of Nucleus-Nucleus 
 Collisions} 
\author{Y. Hama$^1$, T. Kodama$^2$ and 
 O. Socolowski Jr.$^1$} 
\affiliation{$^1$ Instituto de F\y sica, Universidade 
             de S\~ao Paulo, C.P. 66318, 05315-970 
             S\~ao Paulo-SP, Brazil \\
             $^2$ Instituto de F\y sica, Universidade 
             Federal do Rio de Janeiro, C.P. 68528, 
             21945-970 Rio de Janeiro-RJ , Brazil} 


\begin{abstract}
A survey is given on the applications of hydrodynamic 
model of nucleus-nucleus collisons, focusing especially 
on i) the resolution of hydrodynamic equations for 
arbitrary configurations, by using the smoothed-particle 
hydrodynamic approach; ii) effects of the event-by-event 
fluctuation of the initial conditions on the 
observables; iii) decoupling criteria; iv) analytical 
solutions; and others. 
\vspace*{1.5cm}
\end{abstract}
\maketitle


\tableofcontents

\vspace*{1.cm} 



\section{Introduction}

Hydrodynamic Model has been proposed by 
Landau~\cite{Landau} in 1953 as an improvement over 
the Fermi statistical model~\cite{Fermi} for the 
multiple particle production phenomena in high-energy 
nuclear collisions. At that time, these phenomena were 
observed in cosmic rays. Although the Fermi model 
offered an ingenious insight into the mechanism of the 
high-energy nuclear collision processes and gave a 
prediction for 
the energy dependence of the multiplicity, which was 
verified by the data, it was known that it had troubles 
in reproducing particle spectra and relative abundance 
of $K$ over $\pi$. This was because, in this model, 
particles were assumed to be emitted directly from the 
hot and dense, thermally equilibrated matter formed in 
high-energy nuclear collisions, which was supposed to be 
at rest, so that the model predicted isotropic momentum 
distribution which did not agree with the observed 
spectra. Furthermore, because of high temperature 
($T>>m_{K}$)  reached in the process, multiplicity ratio 
depended only on the isotopic-spin statistical weight, 
namely $K/\pi =4/3$. This conclusion of the model was 
also not in agreement with data. 
\smallskip

These problems were solved naturally by letting the hot 
and dense matter expand before particle emission takes 
place, reducing thus heavy-particle multiplicities, 
because of the Boltzmann factor, and giving at the same 
time alongated momentum spectra, due to a violent 
longitudinal expansion caused by a large pressure 
gradient in the beam direction. A nice feature of 
this model is that, since the entropy is conserved in 
the ideal case Landau studied, the energy dependence of 
the total particle multiplicity predicted by the Fermi 
model, and verified experimentally, is preserved. 
\smallskip 

When accelerator data on multiparticle production 
began to appear, first in $pp$ collisions at CERN ISR, 
and later in $\bar{p}p$ collisions at $S\bar{p}pS$ 
collider, Carruthers~\cite{Carrut} revived this 
{\it Heretical Model} in 1974, showing that several 
aspects of those phenomena may be well understood 
within Hydrodynamic Model. When laboratory study of 
high-energy heavy-nucleus collisions started, 
Hydrodynamic Model became one of the essential tools 
for these investigations. 
\smallskip

According to Hydrodynamic Model, the description of 
high-energy nuclear collisions goes as follows. At the 
beginning, two Lorentz contracted (in the c.m. frame) 
nuclei collide and it is assumed that, after a complex 
process involving microscopic collisions of nuclear 
constituents, a hot and 
dense matter is formed, which would be in local thermal 
equilibrium. 
The description of this initial thermalization process
is out of the scope of hydrodynamics. In hydrodynamics, 
we simply assume that the local thermal equilibrium is 
attained and these states of matter are specified by some 
appropriate \textit{initial conditions} (IC) in terms of 
distributions of fluid velocity and thermodynamical quantities 
for a given time-like parameter. 
Then, it follows a {\it hydrodynamical 
expansion}, described by the conservation equations of 
energy-momentum, baryon number and other conserved 
numbers, such as strangeness, isotopic spin, etc. 

\bea 
\partial_\nu T^{\mu \nu} = & 0&\,, \label{em_cons} \\
\partial_\mu(n_B u^\mu) = & 0&\,, \label{bn_cons} \\
\partial_\mu(n_S u^\mu) = & 0&\,, \label{ent_cons} \\
\cdots &&\,, \nonumber 
\eea
where 
\beq
T^{\mu\nu}=(\varepsilon + p)u^\mu u^\nu-p g^{\mu\nu} 
\label{Tmn}   
\eeq
is the energy-momentum tensor, $n_B$, $n_S$, $\varepsilon$, $p$ 
are, respectively, the baryon number density, the 
strangeness density, the energy density and the pressure, 
all of them given in the proper frame of reference of 
the fluid element, and $u^\mu$ is the four-velocity of 
the fluid. Moreover, we have to specify some 
{\it equations of state} (EoS), which depend on the 
nature of the hot matter produced. 
\smallskip 

As the expansion proceedes, the fluid becomes cooler 
and cooler and more rarefied, occurring finally 
the {\it decoupling} of the constituent particles, that 
is, they don't interact any more until their detection. 
However, long-lived resonances and other unstable 
particles may decay after this instant of time. The 
observable quantities such as ${dN}/{dy}$, 
${d\sigma}/{dm_T}$, $<v_2>$, $\cdots$ are then computed 
by using these decoupled or free particles. 
\smallskip 

The main object of studies by using Hydrodynamic Model 
is to investigate, through comparison of its 
predictions with data, properties of the matter formed 
during high-energy nuclear collisions, specified by the 
initial conditions, equations of state and freeze-out 
or decoupling conditions. We emphasize that these 
properties are not known a priori. It should also be 
stressed that even the basic assumption of ``local 
equilibrium'' is not granted for a priori. We expect 
that experimental and theoretical studies of some 
appropriate observables may respond these questions. 
Therefore, it is fundamental to find what are these 
``most appropriate observables''. 
\smallskip 

In this survey, we shall discuss some aspects of 
this model, by focusing mostly on those ones, like 
development of hydrodynamic code capable to treat 
problems with highly asymmetrical configurations, 
effects of the initial-condition fluctuaions and 
improvement of the description of decoupling process. 
These are features which have been investigated and 
developped within the S\~ao Paulo - Rio 
de Janeiro Collaboration 
in the last $\sim15$ years. For a review of other 
aspects of recent developments, see for instance 
Ref.~\cite{hirano,huovinen,kolb}. 
\smallskip 

In the following, in the next Section, we discuss the 
initial conditions, by emphasizing the importance of 
the event-by-event fluctuations as shown by some event 
generators. In Section~\ref{EoS}, we describe several 
equations of state, usually employed in these studies. 
Section~\ref{hydroequations} is devoted to the 
resolution of hydrodynamic equations. There, we begin 
describing some analytic solutions, turn to the 
variational formulation and, finally, an application 
of this approach to develop a numerical code, using 
algorithm of smoothed-particle hydrodynamics. Then, 
we consider the decoupling mechanisms in 
Section~\ref{decoupling}, by stressing that, although 
the commonly used Cooper-Frye presciption~\cite{C-F} 
is convenient and can give many good results, more 
realistic treatment of decoupling is needed in order 
to correctly extract information on the hot matter 
formed in the collision process. In 
Section~\ref{application}, we give some results 
obtained with the methodology described here. Finally, 
conclusions are drawn in Section~\ref{conclusion}. 


\section{Initial conditions}\label{IC} 

In usual hydrodynamic approach of high-energy nuclear 
collisions, one customarily assumes some highly 
symmetric and smooth IC, parametrized in a convenient 
way, which would correspond to the mean distributions 
of hydrodynamic variables averaged over several 
events~\cite{huovinen,kolb,morita}. However, our 
systems are not large, so large fluctuations varying 
from event to event are expected, even under the same 
initial conditions of colliding objects, such as the 
incident energy and the impact parameter of the nuclei. 
What are the effects of the {\it event-by-event 
fluctuation of IC}$\,$? Are they sizable? Do they depend 
on EoS? These are some questions which arise regarding 
this subject. 
\smallskip 

As mentioned in the Introduction, IC are determined by 
a complex process involving microscopic collisions of 
nuclear constituents not accounted for by hydrodynamic 
model, so when we want to introduce fluctuations in the 
IC of a hydrodynamic system, we must go beyond the 
hydrodynamic degrees of freedom. Just to see whether 
such event-by-event fluctuations of IC give sizeable 
effects, so merit a more detailed study, in 
\cite{samya}, Paiva {\it et al.} used the Interacting 
Gluon Model~\cite{igm} (IGM) to generating fluctuating 
IC and, using Khalatnikov 1-dimensional 
solution~\cite{Khalatnikov}, showed that the rapidity 
distribution obtained by averaging over results 
starting from fluctuating IC is quite different from 
that obtained starting from the averaged IC. 
\smallskip 

There are some other simulations, which try to 
incorporate, in hydrodynamic computations, fluctuating 
IC given by more elaborate microscopic models: with a 
use of some event generator, {\it e.g.} 
HIJING~\cite{gyulassy}, VNI~\cite{schlei}, 
URASiMA~\cite{nonaka}, NeXuS~\cite{nexus}, or some 
effective theory such as string model~\cite{laszlo}, 
perturbative QCD + saturation of produced 
partons~\cite{eskola} or color glass 
condensate~\cite{nara}. In principle, one could test 
each of these different microscopic models, by 
connecting them to some hydrodynamic code and computing 
several observables to see which are the differences 
among them and which are more suitable for describing 
experimental data, provided the other ingredients of 
the hydrodynamic model are well known, that is not the 
case. Here, we shall instead discuss not the details of 
such models, but more or less model-independent 
consequences of such fluctuations. Anyhow, we have to 
adopt some microscopic model. In the following, we 
shall mainly discuss the recent works of S\~ao 
Paulo-Rio de Janeiro Collaboration, using NeXuS event 
generator, coupled to hydrodynamic code 
SPheRIO\footnote{{\bf S}moothed {\bf P}article 
{\bf h}ydrodynamic {\bf e}volution of 
{\bf R}elativistic heavy {\bf IO}n collisions.}. 
\smallskip

NeXuS is a microscopic model based on the Regge-Gribov 
theory and can give, in the event-by-event basis, 
detailed space distributions of energy-momentum tensor, 
baryon-number, strangeness and charge densities, at a 
given initial time $\tau=\sqrt{t^2-z^2}\sim1\,$fm, for 
any given pair of incident nuclei or hadrons. 
One important point when we use a microscopic model 
to create a set of IC for hydrodynamics is that 
the energy-momentum tensor produced by the 
microscopic model does not necessarily correspond to 
that of local equilibrium. For example, NeXuS 
generates, as its output, the energy-momentum tensor 
$T^{\mu\nu}(x)$ and the current densities of 
conserved quantum numbers, $j^\mu_B(x)$, $j^\mu_S(x)$ 
and $j^\mu_Q(x)$, where $B$, $S$ and $Q$ refer to 
baryon number, strangeness and electric charge, 
respectively. However, the four-velocities 
corresponding to these currents usually do not 
coincide, and more importantly, do not coincide with 
that of the frame where $T^{\mu\nu}$ becomes diagonal. 
Furthermore, the space components of the diagonalized 
$T^{\mu\nu}$ are not necessarily identical (anisotropic 
stress). These facts mean that the matter is not in 
local equilibrium. In order to transform the 
energy-momentum tensor to that of the equilibrated one, 
we adopt the following procedure. First, following 
Landau~\cite{landau2}, we identify the normalized 
time-like eigenvector of $T^{\mu\nu}$ as the 
four-velocity of the fluid and the eigenvalue as the 
energy density, 
\begin{equation}
T^\mu_{\;\;\nu} u^\nu = \varepsilon u^\mu.
\end{equation} 
Using this four-velocity, we calculate the proper baryon 
number density as 
\begin{equation}
n_B = j^\mu_B u_\mu
\end{equation} 
and analogously for the other densities. Once 
$\varepsilon$ and $n$'s are obtained, all the other 
thermodynamical quantities are calculated using the 
equations of state. By this procedure  we force the 
system into a local thermal equilibrium, conserving 
the proper energy density of the system. 

\begin{figure}[!htb]
\begin{center}
\includegraphics*[angle=-90, width=8.5cm]{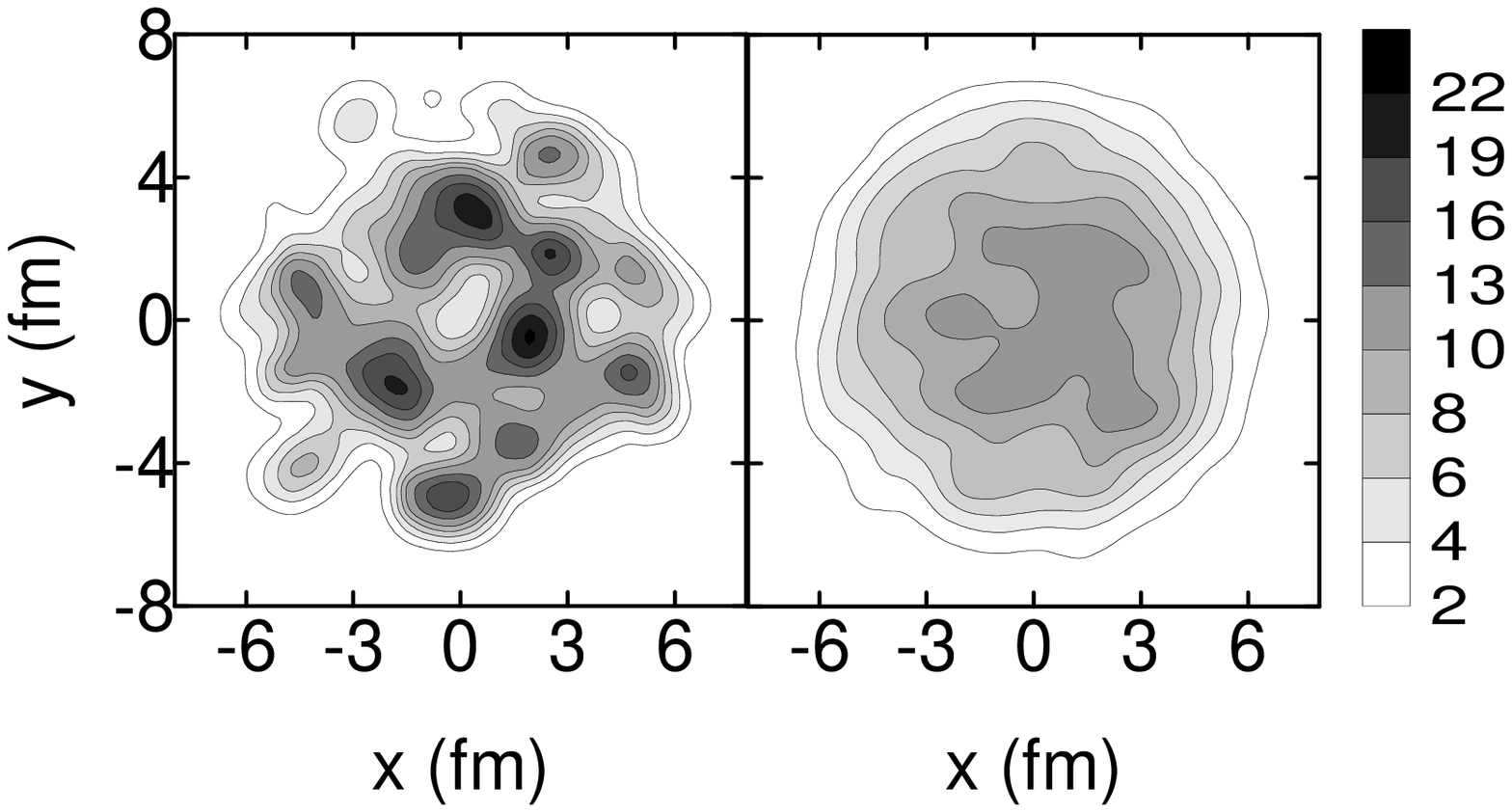}
\end{center}
\caption{Examples of initial conditions for 
central Au+Au collisions given by NeXus at mid-rapidity 
plane. The energy density is plotted in units of 
GeV/fm$^3$. 
Left: one random event. Right: average over 30 random 
events (corresponding to the smooth initial conditions 
in the usual hydro approach).} 
\label{fic}
\end{figure}

\begin{figure}[thb]
\vspace*{-1.5cm} 
\begin{center}
\includegraphics{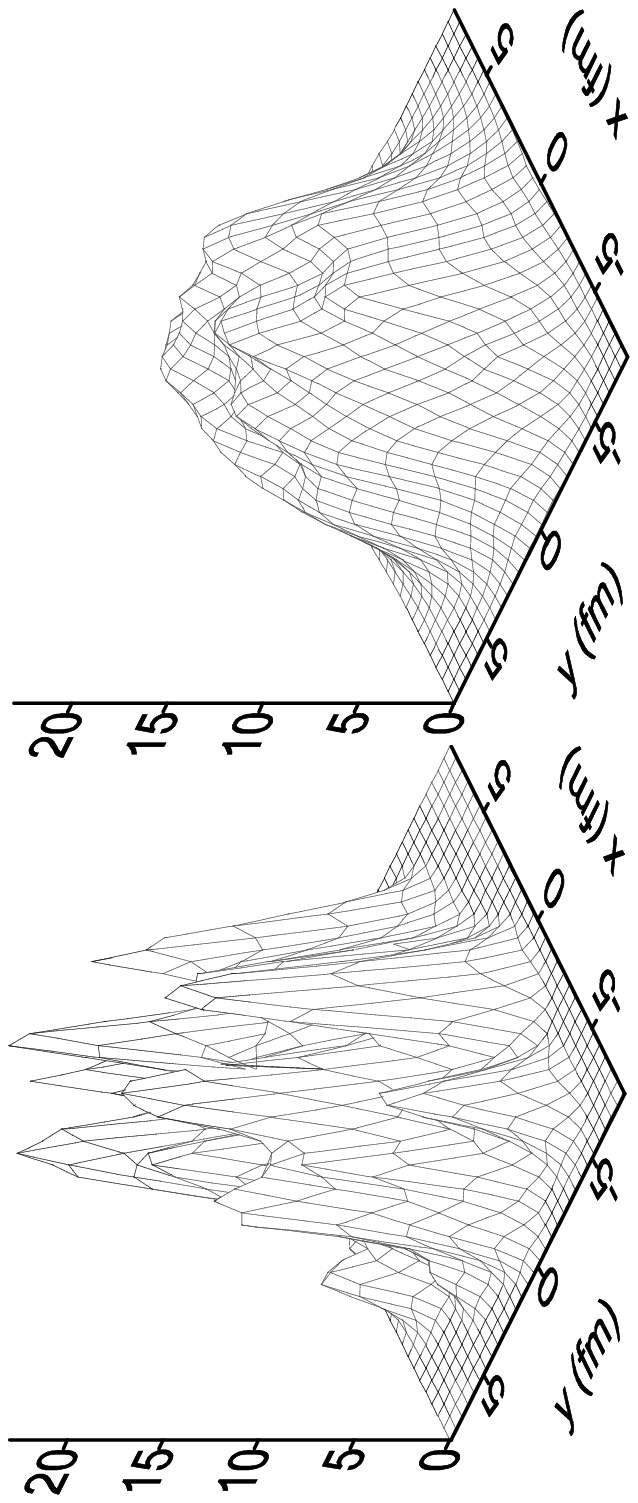}
\end{center}
\vspace{5.6cm} 
\caption{A different representation of the same IC 
shown in FIG. \ref{fic}, at mid-rapidity plane. 
The vertical axis represents the energy density in 
units of GeV/fm$^3$. } 
\label{fic2}
\end{figure}

The IC thus generated are used as inputs for SPheRIO 
code. We show in Figs.~\ref{fic} and \ref{fic2} an 
example of such a fluctuating event, produced by NeXuS 
event generator, for central Au~+~Au collision at 
130A GeV, compared with an average over 30 events. As 
can be seen, the energy-density distribution for a 
single event (left), at the mid-rapidity plane, 
presents several blobs of high-density matter, whereas 
in the averaged IC (right) the distribution is smoothed 
out, even\hfilneg\ 
\smallskip

\ni though the number of events is only 30. Similar 
bumpy event structure was also shown in calculations 
with {HIJING}~\cite{gyulassy}. So, the main question 
here is {\it whether the averages over the observables 
computed starting from such fluctuating IC, like the 
one at the left panel of} Figs.~\ref{fic} {\it and} 
\ref{fic2} {\it are similar or sizeably different from 
the correspondent ones computed from averaged smooth IC 
like that at the right panel there}, or symbolically, 
\bea 
<f>\equiv\frac 1N\sum_{j=1}^N(\mbox{IC}
   \rightarrow f)_j\ 
   &\vbox{\hbox{\lower.1cm\hbox{$\,$?}}
         \hbox{$\simeq$}}&\ 
   \frac 1N\sum_{j=1}^N(\mbox{IC})_j\rightarrow f 
   \nonumber \\
   &\equiv&\ (<IC>\rightarrow f)\,, 
   \label{average} 
\eea 
where $<f>$ is the value of some relevant quantity $f$, 
obtained by averaging over $N$ total number of events 
and (IC$\rightarrow f)_j$ is the value of the same 
quantity in the $j$-th event, with some 
event-dependent IC, whereas $(<IC>\rightarrow f)$ 
represents the same quantity $f$ given by the average 
IC. This is a crucial point in data analyses, because 
the left-hand side is closer to the data point 
experimentalists obtain, whereas the right-hand side 
is the quantity usually computed by theorists for 
that data point, in order to extract properties of 
the matter formed in nuclear collisions. If the bumpy 
structure shown by event generators effectively exists 
in experimental situations, then {\it how do these hot 
spots manifest themselves in the observables?} 
\smallskip

A general conclusion one can draw about this question  
is that the total entropy of the system becomes always 
smaller when one takes such fluctuations into account, 
in comparison to the case without fluctuations, which 
means with average over the event-by-event fluctuating 
IC taken before the expansion. This can be seen by 
observing that, in ideal hydrodynamics, both energy and 
entropy are conserved. Then, considering for simplicity 
an ideal gas so that $S_i=\alpha(E_i)^{3/4}$, with 
$\alpha=$const.$>0$, for each random event, 

\bea 
<E>&=&\frac{1}{N}\sum_i E_i \\
<S>&=&\frac{1}{N}\sum_i S_i \nonumber \\
   &=&\frac{\alpha}{N}\sum_i (E_i)^{3/4} \nonumber \\
   &=&\frac{\alpha}{N}\sum_i<E>^{3/4}
     \left[1+\frac{\Delta E_i}{<E>}\right]^{3/4} 
     \nonumber \\ 
   &\sim&\,\alpha<E>^{3/4}- 
         \frac{3\alpha}{32N}\sum_i<E>^{3/4}
         \left[\frac{\Delta E_i}{<E>}\right]^2 
     \nonumber \\
   &<&\, \alpha <E>^{3/4}. \label{entrop} 
\eea 

\bigskip

\ni Here, $<E>$ and $<S>$ in the left-hand sides mean 
the averaged energy and entropy over the fluctuating 
events, whereas the right-hand side of $<S>$ is the 
entropy corresponding to the averaged initial 
conditions, with the averaged energy $<E>$. The linear 
terms of the expansion in $\Delta E_i/<E>$ are 
cancelled out when the summation is performed. If one 
recalls that particle multiplicity is proportional to 
the entropy for each particle species, one would expect 
that also the multiplicity becomes, in general, smaller 
when one takes such fluctuations into account.  
\smallskip

Other possible manifestations of this inhomogeneity of 
IC that we can expect intuitively are: enhancement of 
high-$p_T$ components due to more violent expansion in 
the surface region, smaller HBT radii, due to 
concentrations of matter in small spots, azimuthal 
asymmetry even in central collisions. However, 
computations are needed to obtain quantitative 
conclusions, whether such discrepancies are meaningful 
or not. We shall discuss this question in 
Sec.~\ref{application}. 


\section{Equations of state} 
 \label{EoS} 

As mentioned already, the basic assumption in 
hydrodynamical models is the local thermal equilibrium. 
Once this condition is satisfied, all the 
thermodynamical relations should be valid in each 
space-time point~\footnote{Recently it is suggested 
that the thermodynamical relations can be satisfied 
without having the thermal equilibrium in the sense of 
Boltzmann distribution~\cite{berges04,kodama04}.}. 
Thus, the energy, pressure 
and temperature are given as functions of baryon number 
and entropy densities, specifying the properties of the 
matter. In this Section, we discuss how to obtain 
simple phenomelogical equations of state (EoS) for the 
hydrodynamical description of relativistic nuclear 
collisions~\cite{kodama}. 

\subsection{Hadronic gas}

The strong interactions among hadrons are very 
complicated and difficult to be incorporated into the 
EoS for practical use. However, for very high energy, 
we may consider that the hadronic gas may be 
approximated as an ideal gas, although the degree of 
approximation can not be evaluated theoretially. The 
recent thermal model for the description of chemical 
abundances~\cite{thermal} show that such an approach 
can reproduce quite well the observed multiplicity 
ratios of produced hadrons. Here we assume that all 
the particles can be treated as quantum ideal gas, 
except for a correction due to the excluded volume. 
We also include a main part of observed resonances in 
Particle Data Tables. The inclusion of resonances can 
be considered as an effective way to consider the 
interactions among hadrons as explained later. 
\smallskip

First, we recall that, in a grand canonical ensemble 
for an ideal gas of quantum particles, the 
thermodynamical potential per volume (the pressure) 
is given by

\beq
p(T,\mu)=\frac{\theta\,g}{(2\pi)^3}
 \int d^3 k\ 
 ln(1+\theta\,e^{\beta\left(\mu-\epsilon(k)\right)}) 
 \label{1}
\eeq 
were $\theta$ $=\pm 1$ ($+$ for fermions, $-$ for 
bosons), $\beta=1/T$ is the inverse of the temperature 
$T$, $\mu$ the chemical potential, $g$ the 
degeneracy factor and $\epsilon(k)=\sqrt{k^2+m^2}$ 
with $m$ the mass of the particle. The number density 
$n$ and the energy density $\varepsilon$ can be 
obtained by the usual thermodynamical relations, 
$n=(\partial p/\partial\mu)_{V,T}\,$, 
$\varepsilon=(\partial p/\partial\beta)_{\lambda}\,$, 
where $\lambda=e^{\beta\mu}$ is the fugacity. The 
entropy density of the gas can be calculated as 
$s=\beta(p+\varepsilon-\mu n)$. 
\smallskip 

For example, in Landau's model~\cite{Landau}, the 
equation of state was taken as that of the massless 
pion gas. For bosons with $m=\mu=0,$ Eq.~(\ref{1}) 
can be integrated analytically to give 
\beq
p(T)=\frac{g\,\pi^2}{90}T^4, 
 \label{p_massless}
\eeq 
and accordingly 
\begin{equation}
s=\frac{g\,\pi^2}{15}T^3,\ \ 
\varepsilon=3p=\frac{g\,\pi^2}{30}T^4. 
\end{equation}
For the pion gas, due to the isospin factor, we can 
take $g=3$.


For more realistic equations of state, we should 
include all the resonance particles in the gas. 
Furthermore, we should also take into account more 
than one type of conserved quantum numbers, such as 
electric charge (equivalently the 3$^{rd}$ component of 
isospin), baryon number and strangeness. In this case, 
the chemical potential must be written as 
$\mu=B\mu_B+S\mu_S+T^{\left(3\right)}\mu_3$ where 
$B,S,T^{\left(3\right)}$ are baryon, strangeness and 
the thrid component of isospin quantum numbers, 
respectively, and $\mu_B, \mu_S$ and $\mu_3$ are the 
corresponding chemical potentials. Thus, for a mixture 
of particles with these conserved quantum numbers, 
Eq.~(\ref{1}) should be generalized to 
\beq 
p_{HG}(T,\mu_B,\mu_S,\mu_3)=\sum_i p_i(T,\mu_i) ,  
 \label{p-HG}
\eeq 
where the sum refers to the particle species (including 
resonances) and 
$\mu_i=B_i\mu_B+S_i\mu_S+T_i^{\left(3\right)}\mu_3$ 
with $B_i,S_i,T_i^{\left(3\right)}$ are the quantum 
numbers of the $i$-th particle species. We verify that 
the baryon number density of the mixture is 
\bea 
n_B&=&\left(\frac{\partial p}{\partial\mu_B}
      \right)_{V,T} \nonumber \\
&=&\sum_i\left(\frac{\partial p_i\left(T,\mu_i\right)} 
   {\partial\mu_B}\right) \nonumber \\ 
&=&\sum_i B_i\ n^{\left(i\right)}, 
\eea 
where $n^{(i)}=(\partial p_i(T,\mu_i)/\partial\mu_i)$ 
is the number density of the $i$-th particle species. 
\smallskip

Except for pions, most of hadrons and resonances can be 
well approximated by the Boltzmann limit. In this case, 
we have 
\beq 
p_i(T,\mu_i)\simeq g_i\frac{T^2 m^2}{2\pi^2} 
 K_2\left(\frac{m_i}{T}\right)e^{\mu_i/T}, 
\end{equation} 
and 
\begin{equation}
n_i=g_i\frac{Tm^2}{2\pi^2}
 K_2\left(\frac{m_i}{T}\right)e^{\mu_i/T}, 
\eeq 
where $K_2$ is a modified Bessel function. From these 
relations, we can see immediately the usual ideal gas 
relation, 
\beq 
p_i=n_i T\,.  
\eeq 

When the widths of the resonances are taken into 
account, Eq.(\ref{p-HG}) must be modified. For an 
interacting gas, the power series expansion of the 
pressure in terms of fugacity, 
\beq 
p(T,\mu)=p^{id}(T,\mu)
        +T\sum_{n=2}^{\infty}b_{n}(T)e^{\beta\mu n},  
\eeq 
is known as the cluster expansion (which is intimately 
related with the virial expansion) and $b_n$ are called 
``virial'' coefficients. Here, $p^{id}$ is the pressure 
of the corresponding ideal gas and, roughly speaking, 
the index $n$ in the sum represents the order of 
multiple particle interactions. For $n=2,$ the 
contribution to the pressure comes from the 2-body 
interactions. Beth and Uhlenbeck~\cite{Beth-Uhlen} 
showed that the second virial coefficient can be 
expressed in terms of the scattering phase-shift of 
constituting particles. This approach was generalized 
to the relativistic Boltzmann gas by Dashen, Ma and 
Bernstein~\cite{dashen} and the result for $b_{2}$ is 
\bea 
b_{2}(T)&=&\frac{T}{2\pi^2}\int_{W_0}^{\infty}dW~W^2 
           K_{2}(\beta W) \nonumber \\ 
        &&\times\frac{1}{\pi}\sum_{\ell}(2\ell+1) 
         \frac{\partial}{\partial W}\delta_{\ell}(W)\,, 
\eea 
where $\delta_{\ell}(W)$ is the phase shift for the 
$\ell$-th partial wave. 
\smallskip

Consider the case of a gas with mass $M$ and suppose 
there exists a resonance in the two particle collision, 
\beq 
M+M\rightarrow M_R\,. 
\eeq 
When the resonance has a width $\Gamma$ and spin $J$, 
only $\ell=J$ dominates the sum and 
\beq 
\delta_{\ell}(W)=\frac{\Gamma}{2}\frac{1}{M_R-W}\,, 
\eeq 
so that we have the Breit-Wigner formula, 
\beq 
\frac{\partial}{\partial W}\delta_{\ell}(W) 
=\frac{\Gamma}{2}\frac{1}{(M_R-W)^2+\Gamma^2/4}\,. 
\eeq 
Therefore, the pressure of the system can be written as 
\[
p=p^{id}+p_R\,, 
\] 
where 
\beq 
p_R=g_R\,\frac{T^2\Gamma}{4\pi^3}
    e^{\beta\mu_R}\int_{W_0}^{\infty}dW
    \frac{W^2 K_2(\beta W)}{(M_R-W)^2+\Gamma^2/4}\ , 
 \label{p_res} 
\eeq 
with 
\begin{eqnarray*}
  g_R&=&2S+1\,,\\
\mu_R&=&2\mu\,.
\end{eqnarray*}
For extremely narrow resonances, 
$\Gamma\rightarrow0\,,$ 
\beq 
p_R\rightarrow g_R
 \frac{T^2 M_R^2}{2\pi^2} K_2(\beta M_R) 
 e^{\beta\mu_R}, 
\eeq 
which is exactly the pressure of the ideal relativistic 
Boltzmann gas made of resonances with mass $M_R\,$. 
Eq.~(\ref{p_res}) suggests that, for more general case, 
the effect of resonance width can be obtained by a 
convolution of the normalized mass spectrum $f(M)$ of 
the resonance, with the pressure $p^{id}(M)$ of the 
ideal gas of mass $M$, as 
\[
p_R=\int dM_R f(M_R)p^{id}(M).  
\]

\subsection{Effect of resonance width as a function of 
 temperature} 

Equation (\ref{p_res}) shows that the effect of the 
resonance width on the pressure of the gas is 
temperature dependent. To see this effect, let us introduce the 
quantity, 
\begin{align} 
F(T,M_R)=&\frac{\Gamma}{M_R^2 K_2(M_R/T)}\hfill
 \nonumber\\
&\times\int_{W_0}^{\infty}dW\frac{W^2 K_2(W/T)}
                      {(M_R-W)^2+\Gamma^2/4}\,,
\end{align}
so that 
\begin{equation}
p_R=p_R^{\Gamma=0}\times F(T,M_R,\Gamma) . 
\end{equation}
Another way to see the effect of the width, we may 
introduce the effective mass of the resonance $M_{eff}$ 
defined\hfilneg\ 

\begin{figure}[!htb] 
\vspace*{-2.2cm}
\begin{center}
\includegraphics*[width=8.cm]{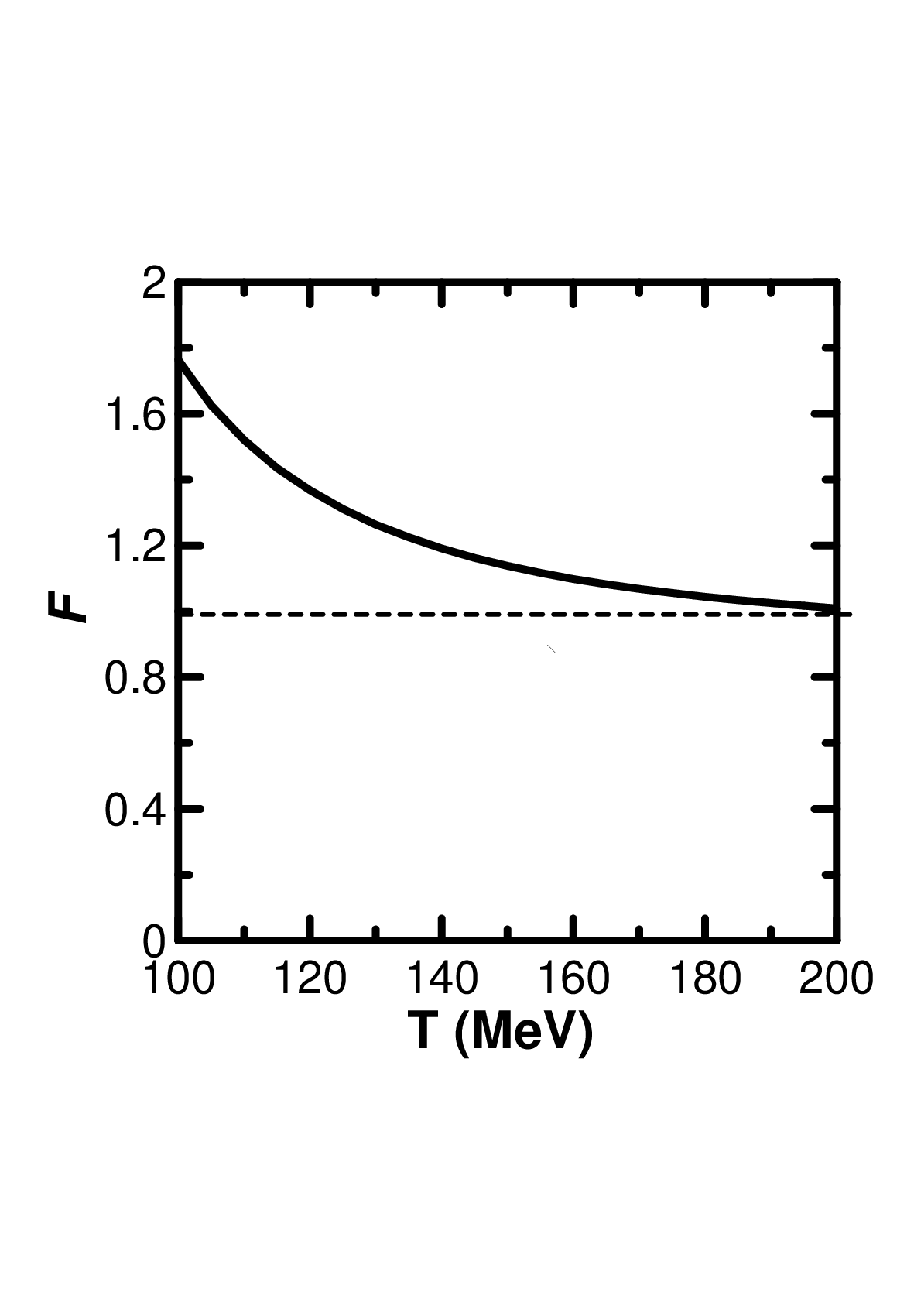} 
\end{center} 
\vspace*{-2.5cm}
\caption{Correction factor $F$ as function of 
temperature $T$ for the resonance $\rho$.} 
\label{FTrho}
\end{figure} 
\medskip

\begin{figure}[!htb]
\begin{center}
\vspace*{-2.cm}
\includegraphics*[width=8.cm]{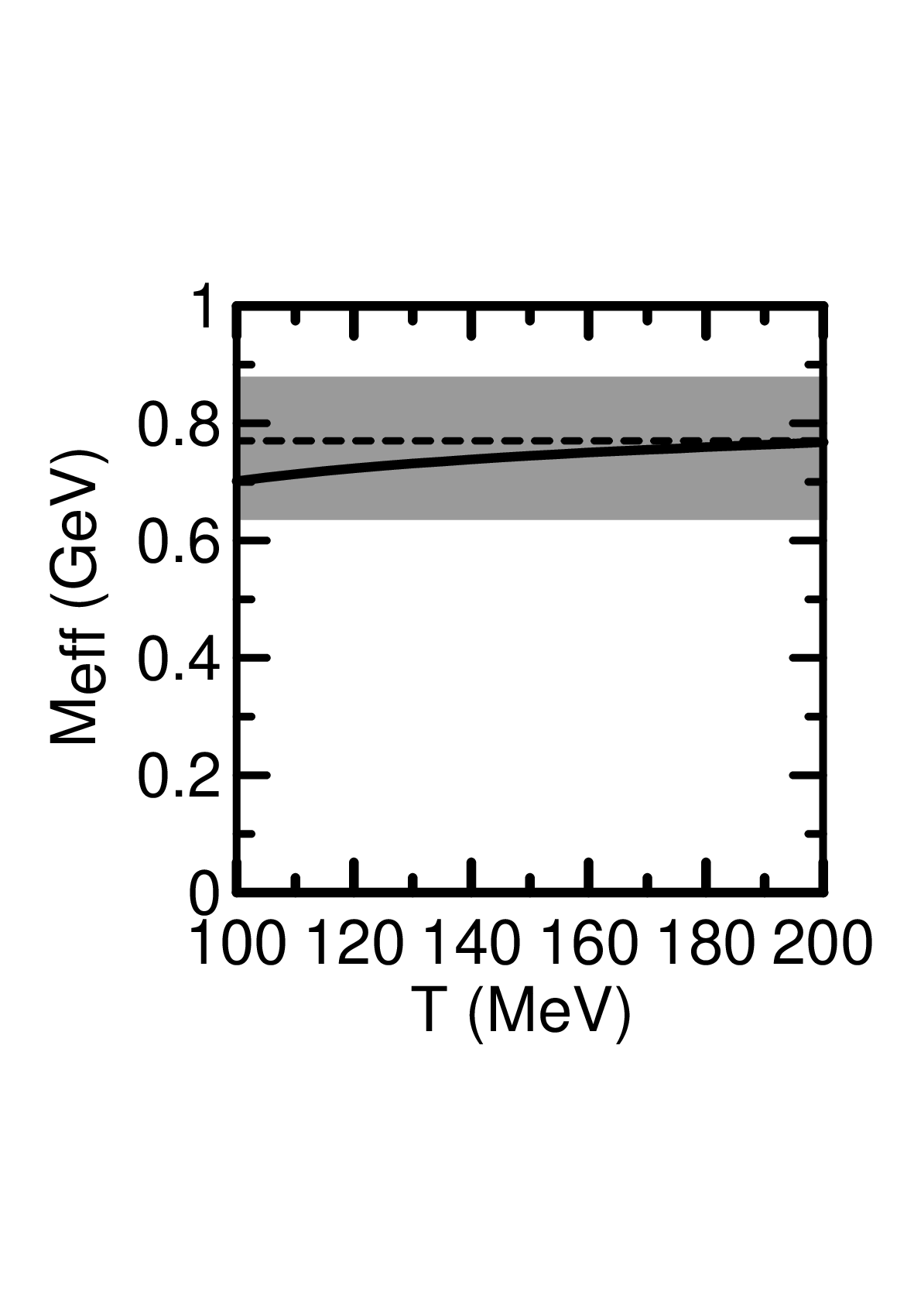} 
\end{center}
\vspace*{-2.5cm}
\caption{Effective mass of the resonance $\rho$ as 
function of temperature $T$. The dark area corresponds 
to the resonance width.} 
\label{MTrho}
\end{figure} 

\begin{figure}[!htb]
\vspace*{-1.8cm}
\begin{center}
\includegraphics*[width=7.6cm]{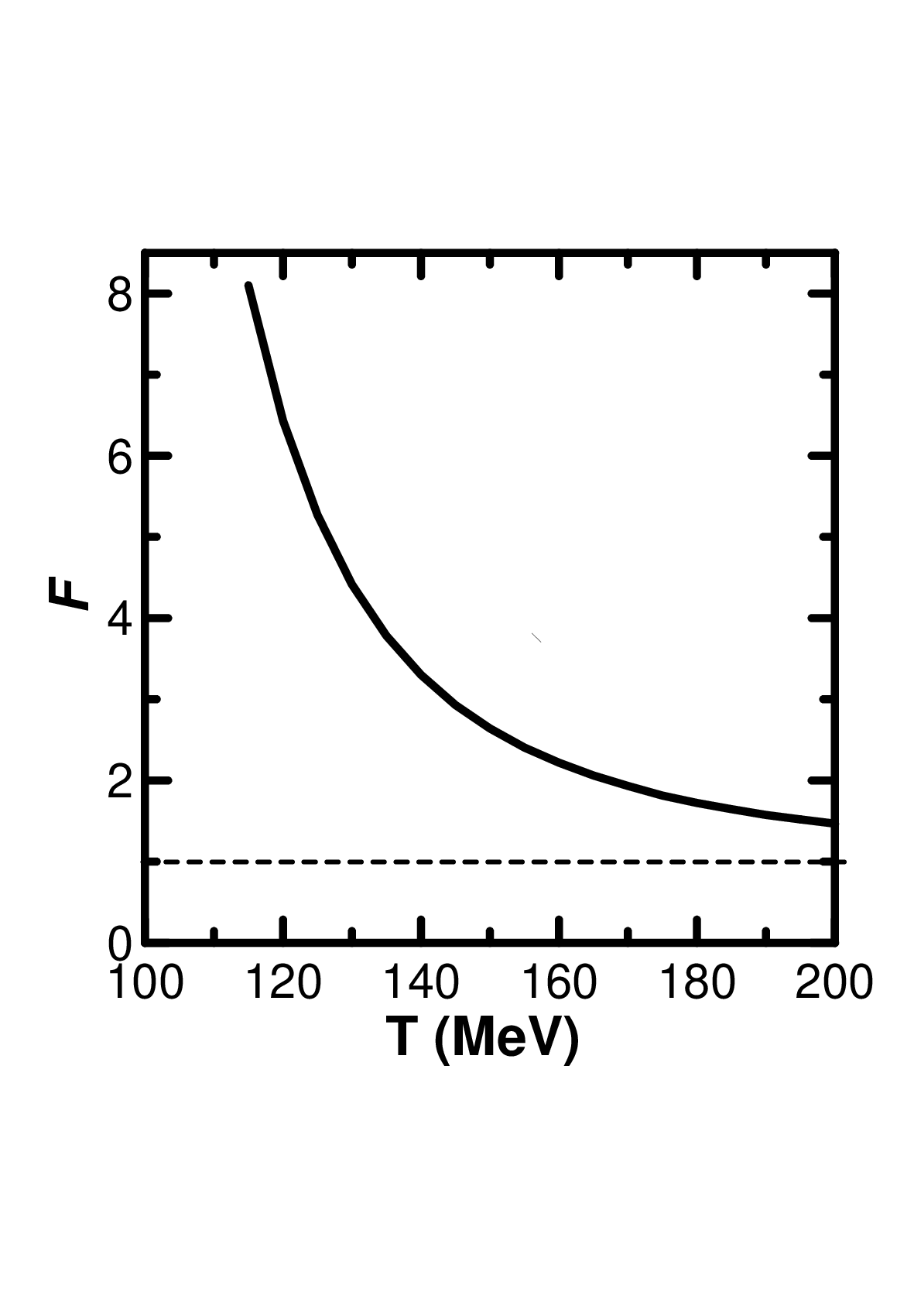} 
\end{center}
\vspace*{-2.3cm}
\caption{Correction factor $F$ as function of 
temperature $T$ for the resonance $f$.} 
\label{FTf}
\end{figure} 

\begin{figure}[!htb]
\vspace*{-2.35cm}
\begin{center}
\includegraphics*[width=7.95cm]{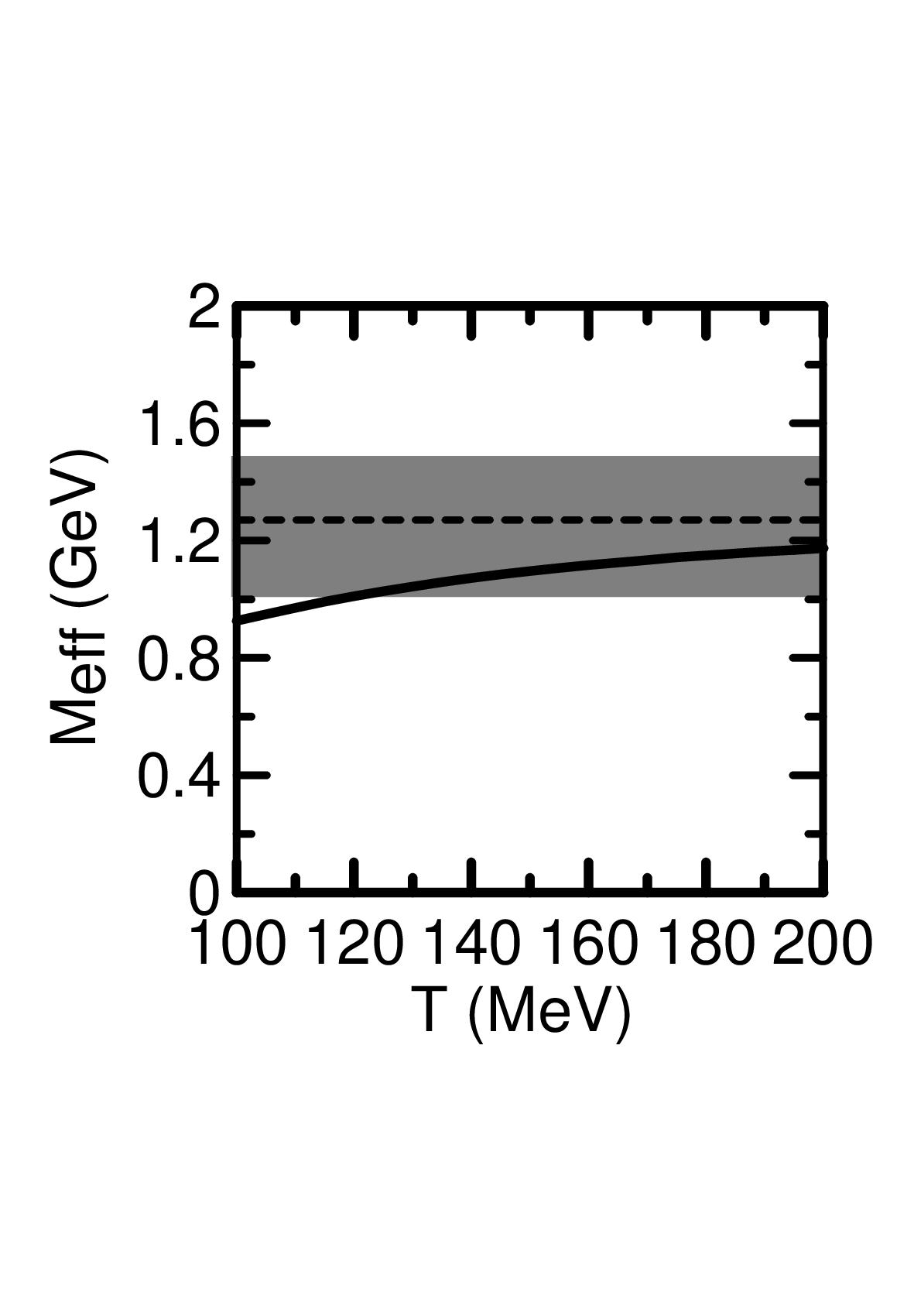} 
\end{center}
\vspace*{-2.5cm}
\caption{Effective mass of the resonance $f$ as 
function of temperature $T$. The dark area corresponds 
to the resonance width.} 
\label{MTf}
\end{figure} 
\ni by 
\[
M_{eff}^2 K_2(\beta M_{eff})\equiv M_R^2 K_2(\beta M_R) 
 F(T,M_R) .
\]
Using this effective mass, we can write the resonance 
pressure as 
\beq 
p_R=g_R\frac{T^2M_{eff}^2}{2\pi^2} 
 K_2(\beta M_{eff})e^{\beta\mu_R},
\eeq 
that is, as if the pressure of ideal particle with mass 
$M_{eff}$. In Figs.~\ref{FTrho},\ref{MTrho},\ref{FTf} 
and \ref{MTf}, we show the temperature 
dependence of $F$ and $M_{eff}$ for two tipical cases, 
one for the resonance $\rho$ (light and narrow width) 
and the resonance $f$ (heavy and large width) 
resonances. As we see, the ideal gas approximation 
deviates substancially for low temperature, especially 
for large width particles. The ideal gas approximation 
is only valid for $T\gg\Gamma$. 

\subsection{Excluded-volume correction}

From the analysis of thermal modesl\cite{thermal}, it 
became clear that the ideal gas description requires 
a modification to adjust the size of the system. The 
volume to fit the particle abundances is found to be 
too small. To avoid this problem, the correction due 
to the excluded volume effect, like a Van der Waals 
hard core correction is introduced~\cite{eos_ev}. 
According to this prescription, Eq.~(\ref{p-HG}) is 
modified by the following coupled equations 
\bea 
p_{HG}(T,\mu_B,\mu_S,\mu_3)
 &=&\sum_{i=1}p_i^{id}(T,\tilde{\mu_i})\,,
    \label{p_HGa} \\ 
 \tilde{\mu_i}&\equiv&\mu_i-v_i\,p_{HG}\,,
    \label{p_HGb}  
\eea 
where as before $\mu_i=B_i\mu_B+S_i\mu_S+T_i^3\mu_3$ 
is the chemical potential and $v_i$ is the excluded 
volume of the $i$-th hadron species. The superscript 
$id$ refers to the ideal gas case. The above equations 
constitute an implicit equation for $p_{HG}$ so that 
these two equations are solved iteratively to obtain 
$p_{HG}$ for a given set of parameters, $T,\mu_B,\mu_S$ 
and $\mu_3$. The number density of the $i$-th hadron is 
given by 
\beq 
n_i^{excl}(T,\mu_i)=\frac{n_i^{id}(T,\tilde{\mu_i})}
          {1+\sum_j v_j\,n_j^{id}(T,\tilde{\mu_j})}\,,
\label{16}
\eeq 
where $n_i^{id}$ is the ideal gas expression of the 
particle density for the $i$-th particle species. 

\subsection{Gas of Quarks and Gluons} 

The simplest way to introduce the phase of quarks and 
gluons in the equations of state is the use of the MIT 
Bag model. The effect of gluon and quark condensate in 
the physical vacuum is expressed as the enerdy density 
of the vacuum (or vacuum pressure). Thus the energy 
density and pressure of an ideal quark-gluon gas 
calculated in the QCD vacuum should be modified 
according to the rule, 
\begin{eqnarray*}
\varepsilon  &\rightarrow &\varepsilon +B, \\
p &\rightarrow &p-B,
\end{eqnarray*} 
where $B$ is the vacuum pressure. Note that this 
vacuum pressure has an analogous property of the 
cosmological constant $\Lambda$ of Einstein. Now, when 
we consider just the $u$ and $d$ quarks and neglect 
their masses, we have 
\begin{eqnarray} 
p_{qgp}&=&\frac{g_q}{6\pi^2}
  \left[\frac{1}{4}\mu_q^4+\frac{\pi^2}{2}\mu_q^2T^2
   +\frac{7\pi^4T^4}{60}\right] \nonumber \\
 &+&\frac{g_G\pi^2}{90}T^4-B\,, 
\end{eqnarray} 
with 
\begin{eqnarray*}
g_q &=&2\times 2\times 3\,, \\
g_G &=&2\times 8\,, 
\end{eqnarray*} 
the statisfical factors of quarks and gluons. For 
quarks, we have $\mu_q=\mu_B/3\,.$ For $\mu_B=0\,,$ 
we have 
\begin{equation}
p_{qgp}^{(ud)}=37\times\frac{\pi^2}{90}T^4-B 
\end{equation} 
or effectively $g_{qgp}=37\,.$ To include the 
strangeness and also charge conservation, we proceed 
in the same way as the hadronic gas and we have 
\begin{align} 
p_{qgp}(T,\mu_B,\mu_S,\mu_3) 
&\!=\frac{g_l}{6\pi^2}\!
  \left[\frac{1}{4}\mu_u^4
  +\frac{\pi^2}{2}\mu_u^2 T^2\! 
  +\!\frac{7\pi^4 T^4}{60}\right] \nonumber\\
&+\frac{g_l}{6\pi^2}\!
  \left[\frac{1}{4}\mu_d^4
  +\frac{\pi^2}{2}\mu_d^2 T^2\!
  +\!\frac{7\pi^4 T^4}{60}\right] \nonumber\\
&+p_s^{id}(T,\mu_s)+p_s^{id}(T,-\mu_s)\nonumber\\
&+\frac{g_G\pi^2}{90}T^4-B\,, 
 \label{p_qgp} 
\end{align} 
where $g_\ell=2\times3$ and 
\begin{align*}
\mu_u & = \frac{1}{3}\mu_B+\frac{1}{2}\mu_3\,, \\
\mu_d & = \frac{1}{3}\mu_B-\frac{1}{2}\mu_3\,, \\
\mu_s & = \frac{1}{3}\mu_B-\mu_S\,.  
\end{align*}

\subsection{Construction of equations of state for the 
practical use}
\label{practicalEOS}

The expressions Eqs.~(\ref{p_HGa},\ref{p_HGb}) or 
Eq.~(\ref{p_qgp}) are, however, not convenient for the 
use in hydrodynamical calculations. This is because 
the variables in such calculations are the conserved 
quantum numbers and the entropy density and not the 
chemical potentials and the temperature. So, we need 
to invert the relations, 
\begin{eqnarray}
n_B &=& n_B(T,\mu_B,\mu_S,\mu_3)\,, \label{nb}\\ 
n_S &=& n_S(T,\mu_B,\mu_S,\mu_3)\,, \label{ns}\\ 
n_3 &=& n_3(T,\mu_B,\mu_S,\mu_3)\,, \label{n3}\\ 
s\ \,&=& s\ \,(T,\mu_B,\mu_S,\mu_3)\,, \label{entro} 
\end{eqnarray}
to get 
\begin{eqnarray}
\mu_B &=& \mu_B(n_B,n_S,n_3,s)\,, \\
\mu_S &=& \mu_S(n_B,n_S,n_3,s)\,, \\
\mu_3 &=& \mu_3(n_B,n_S,n_3,s)\,, \\
  T\  &=&   T\ (n_B,n_S,n_3,s)\,. 
\end{eqnarray}
However, this is a formidable task even numerically. We 
are thus forced to reduce the degrees of freedom for the 
practical application to hydrodynamics. For this purpose, 
we set the isospin and strangeness densities to null 
everywhere. That is, we impose the conditions, 
\begin{eqnarray}
n_S &=& 0\,,\\
n_3 &=& 0\,.
\end{eqnarray}
These conditions together with Eqs.~(\ref{ns}) and 
(\ref{n3}) determine $\mu_S$ and $\mu_3$ as functions 
of $T$ and $\mu_B\,$. Therefore, $n_B$ and $s$ in 
Eqs.~(\ref{nb}) and (\ref{entro}) become now functions 
of two variables $T$ and $\mu_B\,$,  
\begin{eqnarray}
n_B &=& n_B(T,\mu_B)\,,\\
  s &=&   s(T,\mu_B)\,, 
\end{eqnarray} 
which can be inverted numerically and give  
\begin{eqnarray}
T&=&T(n_B,s)\,, \\
\mu_B&=&\mu_B(n_B,s)\,. 
\end{eqnarray}


The above inversion process allows us to write any 
of thermodynamcial quantities as functions of $n_B$ 
and $s$, both in hadronic gas and quark-gluon plasma. 
Then, for a given pair of density parameters $n_B$ 
and $s$, we should determine which phase the physical 
system assumes. When two phases are in equlibrium, we 
must have~\cite{sollf} 
\begin{equation}
p_{HG}(T,\mu_B)=p_{QGP}(T,\mu_B) 
\label{Gibbs} 
\end{equation}
so that it determines the phase boundary line in 
$(\mu_B,T)$ plane and separates this plane into two 
domains. The domain where $p_{HG}>p_{QGP}$ is the 
hadron gas phase and the other $p_{HG}<p_{QGP}$ is the 
QGP phase. These two domains are in contact on the 
phase boundary line Eq.~(\ref{Gibbs}). 

However, the above two domains in $(\mu_B,T)$ plane 
are mapped into two separated domains in the $(n_B,s)$ 
plane and there appeas a new third domain between them. 
That is, the phase boundary line $(\mu_B,T)$ plane 
spreads into a domain in the $(n_B,s)$ plane. This 
domain is the mixed phase. In oder to determine 
thermodynamical quantities in this mixed phase as 
functions of $(n_B,s)$, we should introduce another 
criterion in addition to the phase boundary condition 
of Gibbs. As mentioned above, any point $(T,\mu_b)$ 
on the phase boundary line corresponds to two points 
in the $(n_B,s)$ plane, one for the hadron phase, 
$(n_B^H,s^H)$ and other, $(n_B^Q,s^Q)$ for the QGP 
phase. In the mixed phase, any density of extensive 
quantity, should be a linear function of the qgp/hadron 
concentration ratio. Thus for a given value of the 
baryon number density $n_B$ in the mixed phase, 
$(n_B^H,s^H)$ and $(n_B^Q,s^Q)$ should satisfy 
\begin{equation}
s=\frac{s^Q-s^H}{n_B^Q-n_B^H}(n_B-n_B^H)+s^H. 
\end{equation}
From this equation, we determine the two points on the 
phase boundaries, $(n_B^H,s^H)$ and $( n_B^Q,s^Q)$. 
All the other extensive quantities, say $\varepsilon$, 
can then be obtained as 
\begin{equation}
\varepsilon=\frac{\varepsilon^Q-\varepsilon^H}
{n_B^Q-n_B^H}(n_B-n_B^H)+\varepsilon^H. 
\end{equation}

Finally we can construct the equations of state in the 
whole region of $(n_B,s)$. The parameters of the final 
equations of state are then, 

\begin{itemize}
\item {\it Number of resonances included in the 
hadronic gas}: Here, we take all the mesons with mass 
smaller than 1.5 GeV, and baryons smaller than 2 GeV. 
Resonance widths are not included. 

\item {\it Quark masses}: We may safely take 
$m_u=m_d=0$, but for the strange quark, we take 
$m_s=120$ MeV. 

\item {\it Size of excluded volume}: In the example 
shown in the figures below, $\nu_0=(4\pi r_0^3/3)$ 
with $r_0=0.7$ fm for baryons and $r_0=0$ for mesons. 

\item {\it Bag constant}: We take $B=380$ MeV/fm$^3$.  
\end{itemize}

Fig.~\ref{Tconst} shows the line of constant 
temperature in $(n_B\,,s)$ plane. Fig.~\ref{Adiabat} 
shows the lines of constant entropy per baryon in 
$(\mu_B\,,T)$ plane. 

Note that the present equations of state have, by 
construction, the first order phase transition between 
hadronic gas and quark gluon plasma. Recent lattice 
calculations~\cite{latt} indicate that there exists a 
critical point so\hfilneg\ 

\begin{figure}[!htb] 
\vspace*{-1.6cm}
\begin{center}
\includegraphics*[width=8.cm]{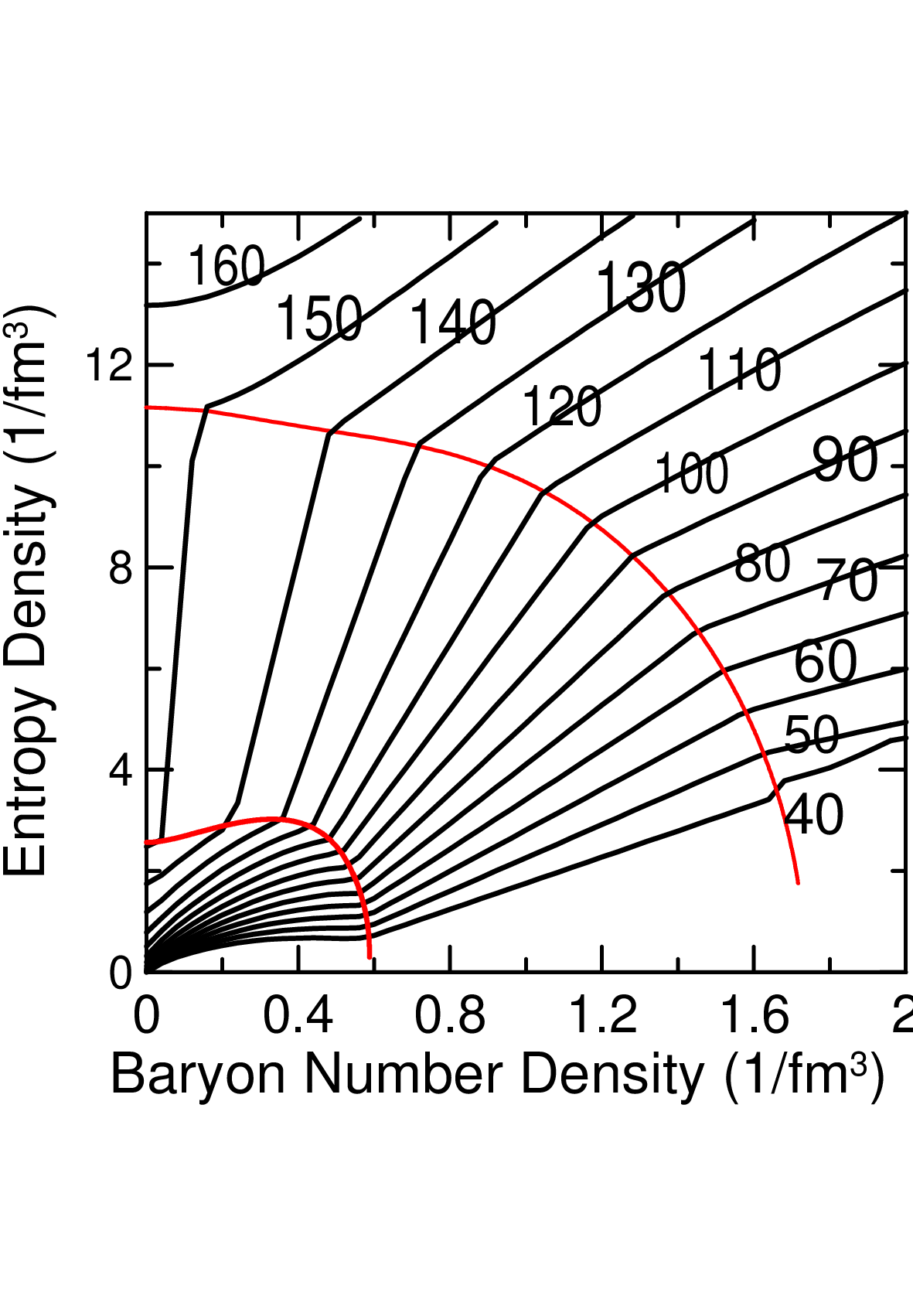} 
\end{center} 
\vspace*{-2.2cm}
\caption{Isotherms in $(n_B,s)$ plane. Values of the 
temperatures are indicated in the figure. Hadron gas, 
mixed phase and quark gluon plasma are clearly seen.} 
\label{Tconst}
\end{figure} 
\newpage

\begin{figure}[!htb] 
\vspace*{-1.8cm}
\begin{center}
\includegraphics*[width=8.cm]{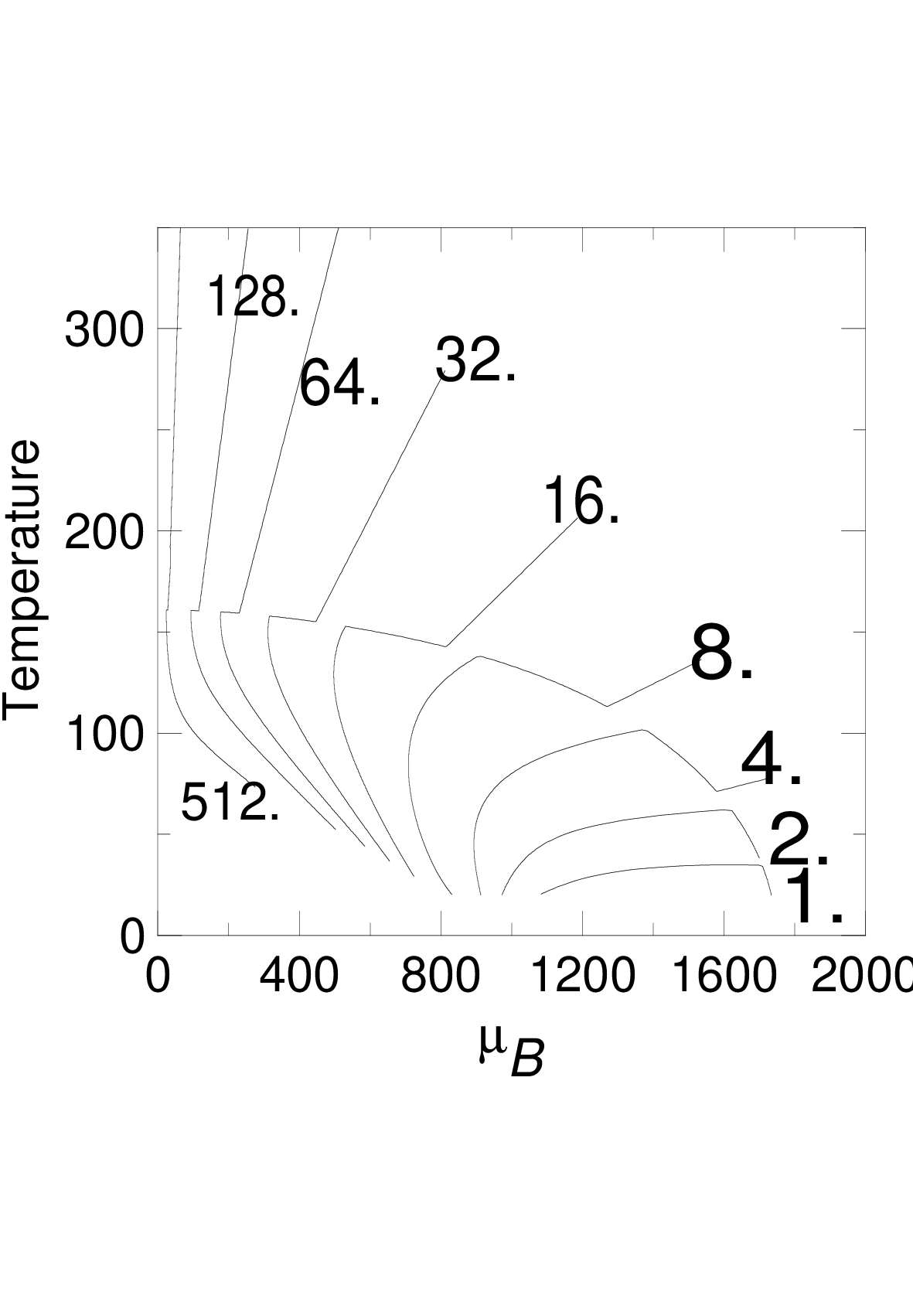} 
\end{center} 
\vspace*{-2.5cm}
\caption{Constant entropy per nucleon curves in the 
$(\mu,T)$ plane. The numbers indicate $(s/n_B)$ 
values.} 
\label{Adiabat}
\end{figure} 

\ni that the phase transition in the 
region starting from that point to zero $\mu_B$ axis is 
not the first order but may be either a higher order 
transition, or a smooth cross over~\cite{Raj}. Effect 
of such possibilities should be investigated within the 
hydrodynamical models. 


\section{Resolution of hydrodynamic equations} 
 \label{hydroequations}

In general, exact analytic resolution of relativistic 
hydrodynamics is a difficult task due to the highly 
non-linear nature of these equations. So, usually one 
resorts to numerical computations. However, since the 
analytical studies are more transparent, it would be 
useful to find analytical soluions, even though they 
correspond to highly ideal cases. Khalatnikov's 
one-dimensional analytical solution~\cite{Khalatnikov} 
to Landau's initial conditions~\cite{Landau}, namely, 
ideal fluid at rest in a Lorentz contracted thin 
spatial region, gave rise to a new approach in 
high-energy physics. The boost-invariant solution 
\cite{scale-invariant} found 20 years later, is 
frequently utilized as the basis for estimations of 
initial energy densities in ultra-relativistic 
nucleus-nucleus collisions~\cite{Bjorken}. Below, we 
shall first describe families of analytical solutions 
we obtained in collaboration with 
T. Cs\"org\H o~\cite{tamas1,tamas2}. 
\smallskip 

Considering that it is not trivial to analytically 
solve the equations of hydrodynamics, it would be 
nice if we could develop a method to obtain an 
approximate but analytical solution of hydrodynamics. 
This has been done with variational 
formulation~\cite{variational}, although we had not 
develop it further by applying it to practical problems 
of high-energy nuclear collisions. However, we did 
apply the variational method to develop a numerical 
code SPheRIO, based on the so-called smoothed-particle 
hydrodynamics (SPH)~\cite{Lucy,Monaghan} tecnique, to 
high-energy nuclear collisions~\cite{sph}, which is 
flexible enough, capable to treat systems with 
configurations without any symmetry and also exploding 
in time. 
\smallskip 

In the following, after presenting some analytical 
solutions, we shall describe in the subsection 
\ref{variational_formalism} the variational 
formulation of hydrodynamics, showing how it could 
used to get approximate solutions. Then, in the 
subsection \ref{smooth}, we shall apply this 
method to adapt the SPH hydrodynamics for relativistic 
heavy-ion collisions.  

\subsection{Analytic solutions}  
            \label{analytic_solutions} 

After Landau's initial proposal~\cite{Landau}, the 
first analytic solution obtained is due to 
Khalatnikov~\cite{Khalatnikov}, considering 
1-dimensional expansion of ideal gas. A more simpler 
boost-invariant solution has been obtained 
later~\cite{scale-invariant} and applied to estimate 
the initial energy densities in ultra-relativistic 
nucleus-nucleus collisions~\cite{Bjorken}. Both of 
these have been frequently used in the study of nuclear 
collisions, showing the usefulness of such simple 
analytical solutions. 
\smallskip 

However, the boost-invariant solution has some 
shortcomings: 
{\it i)} it is scale invariant, having a flat rapidity 
distribution, corresponding to the extreme relativistic 
collisions, which has never been seen; 
{\it ii)} it contains no transverse flow. 
In~\cite{tamas1}, we tried to overcome these shortcomings. 
We started by assuming a simple equations of state, 
corresponding to a gas containing massive conserved 
quanta, namely, 
\bea 
 \varepsilon & = & m n + \kappa p\,, \label{e:eos1}\\
 p        & = & n T\,, \label{e:eos2}
\eea 
having two free parameters, $m$ and $\kappa$. 
Non-relativistic hydrodynamics of ideal gases corresponds 
to the limiting case of $m>>T$, ${\mathbf v}^2 << 1$ 
and $\kappa = 3/2\,$. 
\smallskip 

Then, we looked for {\it similarity flows}, {\it i.e.}, 
\bea  
 \mathbf{v} &=& \left(\frac{\dot X(t)}{X(t)}x, 
                      \frac{\dot Y(t)}{Y(t)}y, 
                      \frac{\dot Z(t)}{Z(t)}z\right), 
 \label{v}\\ 
 f(x^\mu) &=& f_0 \left(\frac{V_0}{V}\right)^a F(s)\,, 
 \label{f}  
\eea 
where $X(t), Y(t), Z(t)$ are the scales of length, in 
three orthogonal directions, and $f(x^\mu)$ is any of 
the thermodynamical quantities, such as $n(x^\mu), 
T(x^\mu), p(x^\mu), \ldots,$ and 
\beq 
s=\frac{x^2}{X^2} +\frac{y^2}{Y^2} + \frac{z^2}{Z^2} 
\label{s} 
\eeq 
is a {\it scaling variable}. 
\smallskip 

Using this parametrization for the one-dimensional 
expansion, it can be easily verified, by direct 
substitution into eqs.(\ref{em_cons},\ref{bn_cons}) 
with eqs.(\ref{Tmn},\ref{e:eos1},\ref{e:eos2},\ref{v} 
and \ref{s}), that a family of solutions can be 
written as 
\bea 
 v &=& z/t\equiv\tanh\eta,\nonumber\\ 
 n &=& n_0 (t_0/\tau){\mathcal{V}}(s),\nonumber\\ 
 p &=& p_0(t_0/\tau)^{1+1/\kappa},\nonumber\\
 T &=& T_0(t_0/\tau)^{1/\kappa}\frac{1}{{\mathcal{V}}(s)}, 
 \label{1d} 
\eea 
with $p_0=n_0 T_0$ and where ${\mathcal{V}}(s)$ is an 
arbitrary non-negative function of 
$s = z^2/(\dot{Z}_0^2 t^2)$. The index $0$ stands for 
the initial values. Thus, this is a family of 
one-dimensional similarity flows, but it is {\bf not} 
scale invariant and $n$ and $T$ are not constant for a 
constant $\tau$. 
\smallskip 

Next, considering cylindrically symmetric flows, with 
boost invariance along $z$ direction (collision axis), 
we could find the following family of solutions, for 
transverse flows:  
\bea
 {\mathbf v} &=& {\mathbf r}/t\,,\quad\mbox{\rm for} 
                 \quad |{\bf r}|\le t,\nonumber \\
 n &=& n_0 (\tau_{z0}/\tau))^3 {\mathcal{V}}(s), 
     \nonumber\\ 
 p &=& p_0 (\tau_{z0}/\tau)^{3 + 3/\kappa},\nonumber\\
 T &=& T_0 (\tau_{z0}/\tau))^{3/\kappa}
       \frac{1}{{\mathcal{V}}(s)}\,. 
 \label{cylindrical}
\eea
Here, $p_0=n_0 T_0$ and ${\mathcal{V}}(s)$ is an 
arbitrary non-negative function of $s$ 
($=r_t^2/(\dot R^2_0\tau_z^2)\,$ in this case), with 
$r_t=\sqrt{x^2+y^2}\,$ and 
$\dot R_0=\sqrt{\dot X_0^2+\dot Y_0^2}\,$ and 
$\tau_z=\sqrt{t^2-z^2}\,$. So, this is a generalization 
of the one-dimensional scale-invariant solution, 
including a class of transverse flows. 
\smallskip 

More recently~\cite{tamas2}, we extended these solutions 
still further, considering less symmetrical flows, but 
still keeping the same EoS (\ref{em_cons},\ref{bn_cons}) 
and similarity flow, with constant velocity, as it 
appears in eqs.(\ref{1d},\ref{cylindrical}). 

\subsection{Variational formulation} 
\label{variational_formalism} 

As shown above, even for very simple equations of state, 
analytic solutions are limited to special cases. For 
realistic situations, even the equations of state 
themselves are available only in the form of numerical 
tables. Therefore, the numerical resources are 
essential for realistic studies of hydrodynamical 
behavior of ultra-relativistic collisional processes. 
However, it is well-known that any numerical method 
for partial differencial equations requires highly 
sophisticated techniques to avoid numerical 
instabilities, and usually it needs a very large 
scale computation, especially when we want to describe 
correctly explosive processes such as relativistic 
heavy-ion collisions. However, as emphasized in the 
Introduction, in the hydrodynamic approach of 
high-energy nuclear collisions, its main ingredients, 
i.e., the equations of state, the initial conditions 
and the freezeout conditions are not quite well known. 
In such a situation, we actually don't need a very 
precise solution of the hydrodynamic equations, 
but a general flow pattern which characterizes the 
final configuration of the system as a response to 
given equations of state, initial conditions and the 
decoupling procedure. So, we prefer a rather simple 
scheme of solving the hydrodynamic equations, not 
unnecessarily too precise but robust enough to deal 
with any kind of geometry. From this point of view, we 
stressed in Ref.~\cite{variational} the advantage of a 
variational approach to relativistic hydrodynamics. 
\smallskip 

Although not commonly found in general textbooks, the 
variational formulation of hydrodynamics has been 
studied by several authors~\cite{Taub+,Yourgrau}. In 
Ref.~\cite{variational}, starting from the action 
\beq 
I=\int d^{4}x\left\{-\varepsilon \right\}, \label{L}
\eeq 
where $\varepsilon$ is the proper energy density, 
the relativistic hydrodynamics was derived from the 
variational principle. 
\smallskip 

Here, we show its derivation, generalizing to include 
the rotational flow. To do this, we introduce the 
two variables, the proper baryon density, $n$, and 
entropy density, $s$, which satisfy the conservation 
laws, 
\bea 
&&\partial _{\mu }(nu^{\mu })=0\;,  \nonumber \\
&&\partial _{\mu }(su^{\mu })=0\;.  \label{conservation}
\eea 
where $u^{\mu }$ is the four velocity of the field, 
satisfying the normalization, 
\begin{equation}
u^{\mu }u_{\mu }=1\,.  \label{norm}
\end{equation} 
The functional form of the energy density, 
\[
\varepsilon =\varepsilon \left( n,s \right) 
\] 
specifies the thermodynamical properties of the fluid. 
The pressure, temperature and chemical potential are 
obtained by the usual thermodynamic relations, 
\begin{eqnarray}
p&=&n\frac{\partial[\varepsilon(n,s)/n]}
          {\partial n}\,,\nonumber\\ 
T&=&\frac{\partial\varepsilon(n,s)}{\partial s} 
          \quad\quad\quad\mbox{and} \nonumber \\  
\mu&=&\frac{\partial\varepsilon(n,s)}{\partial n}\,. 
          \nonumber 
\end{eqnarray}

The hydrodynamical equations of motion for the fluid is 
given by the variational principle with respect to $n$, 
$s$ and $u^\mu$ under constraints, 
Eq.~(\ref{conservation}) and the normalization of the 
four-velocity, Eq.(\ref{norm}). These constraints can be 
incorporated in the variational principle in terms 
of Lagrangian multipliers to write 
\begin{eqnarray} 
\delta\int d^{4}[&-&\varepsilon(n,s)
 +\lambda\partial_{\mu }(nu^{\mu})
 +\zeta\partial_{\mu }(su^{\mu }) \nonumber \\ 
 &-&\frac{1}{2}w\left( u^{\mu }u_{\mu}-1\right)]=0\,, 
 \label{delta I} 
\end{eqnarray} 
where $\lambda,\zeta $ and $w$ are Lagrangian 
multipliers and arbitrary functions of $x$. 
Equivalently, the fluid dynamics is given by the 
effective Lagrangian, 
\begin{eqnarray} 
{\cal L}_{eff}^{(fluid)}(n,s,u^\mu,\lambda,\zeta,w)=\!
   &-&\!\varepsilon(n,s)-nu^{\mu}\partial_{\mu}\lambda 
        \nonumber \\
   &-&\!su^{\mu}\partial_{\mu}\zeta
     -\frac{w}{2}(u^{\mu}u_{\mu}-1)\,, \nonumber \\
   && \label{L_eff}
\end{eqnarray} 
where now all of $n,s,u^{\mu },\lambda ,\zeta ,w$ are 
independent variational variables. 
\smallskip 

The variations with respect to $n,s$ and $u^{\mu }$ 
lead immediately to 
\begin{eqnarray} 
-\mu-u^{\mu}\partial_{\mu}\lambda &=&0\,,\label{Va-1}\\
-T-u^{\mu }\partial _{\mu }\zeta &=&0\,,\label{Va-2}\\
-n\partial_{\mu}\lambda-s\partial_{\mu}\zeta-wu_{\mu} 
&=&0\,.\label{Va-3}
\end{eqnarray} 
Variations with respect to $\lambda ,\zeta $ and $w$ 
give simply the constraints, eqs.(\ref{conservation}) 
and (\ref{norm}). Multiplying the both sides of 
Eq.(\ref{Va-3}) by $u^{\mu}$, and using 
eqs.(\ref{norm},\ref{Va-1},\ref{Va-2}), we get 
\begin{eqnarray}
w &=&n\mu +Ts  \nonumber \\
  &=&\varepsilon +p\,, \label{enthalpy} 
\end{eqnarray} 
where $p$ is the pressure. Eq.(\ref{enthalpy}) shows 
that $w$ is the enthalpy density. Sustiuting back this 
$w$ into Eq.(\ref{Va-3}) and multiplying $u_{\nu }$, we 
have 
\[
wu_{\mu }u_{\nu }=-(nu_{\nu})(\partial_{\mu}\lambda) 
-(su_{\nu})(\partial_{\mu}\zeta)\,. 
\] 
Taking the divergence and using the continuity equations 
$\partial^{\nu}\left( nu_{\nu }\right)=0$ and 
$\partial^{\nu }(su_{\nu})=0$, we get 
\begin{equation}
\partial^{\nu}(wu_{\mu}u_{\nu})=-(nu_{\nu})
(\partial^{\nu}\partial_{\mu}\lambda)-(su_{\nu})
(\partial^{\nu}\partial_{\mu}\zeta)\,.\label{div-1} 
\end{equation} 
But 
\begin{eqnarray}
(nu_{\nu})(\partial^{\nu}\partial_{\mu}\lambda) 
&=&n\partial_{\mu}(u_{\nu}\partial^{\nu}\lambda)
   -n(\partial^{\nu}\lambda)(\partial_{\mu}u_{\nu}) 
   \nonumber \\ 
&=&-n\partial_{\mu}\mu-n(\partial^{\nu}\lambda)
   (\partial_{\mu}u_{\nu}) \nonumber 
\end{eqnarray} 
and analogously 
\[
(su_{\nu})(\partial^{\nu}\partial_{\mu}\zeta) 
=-s\partial_{\mu}T-s(\partial^{\nu}\zeta)
(\partial_{\mu}u_{\nu})\,, 
\] 
so that Eq.(\ref{div-1}) becomes
\begin{eqnarray}
\partial ^{\nu}(wu_{\mu}u_{\nu})  
&=&n\partial_{\mu }\mu
   +s\partial_{\mu}T+(\partial_{\mu}u_{\nu})wu^{\nu} 
   \nonumber \\ 
&=&\partial_{\mu}p+u^{\nu}(\partial_{\mu}u_{\nu})w 
   \nonumber \\ 
&=&\partial_{\mu}p\,. 
\end{eqnarray} 
Here, we have used Eq.(\ref{Va-3}) and the Gibbs-Duhem 
relation, 
\begin{equation}
dp=sdT+nd\mu\,,
\end{equation} 
and the property, 
\[
u^{\nu}(\partial_{\mu}u_{\nu})=0\,. 
\] 
Finally we arrive at the standard form of the 
relativistic hydrodynamic equation (\ref{em_cons}),   
\begin{equation} 
\partial^{\nu}T_{\mu\nu}=0\,, 
\end{equation} 
where  
\begin{eqnarray}
T_{\mu\nu} 
&=&wu_{\mu}u_{\nu}-g_{\mu\nu}p \nonumber \\ 
&=&(p+\varepsilon)u_{\mu}u_{\nu}-g_{\mu\nu}p 
\end{eqnarray}
is the usual energy-momentum tensor of the fluid. 
\smallskip 

It is important to observe that, the effective 
Lagrangian Eq.(\ref{L_eff}) evaluated in the proper 
comoving frame of the fluid motion is 
\begin{equation}
\left.{\cal L}_{eff}^{(fluid)}\right\vert_{comoving} 
=-\varepsilon (n,s)+\mu n+Ts=p\,,  \label{L_eff_prop} 
\end{equation} 
which is nothing but the negative of thermodynamical 
potential of the system. 
\smallskip 

Now, interesting application of this approach appears, 
if we can parametrize possible solutions of continuity 
equations (\ref{conservation}) in terms of a certain 
number of time-dependent parameters 
$\vec{a}(t)=\{a_{i}(t),i=1,\ldots ,N\}$, such that 
\begin{eqnarray*} 
n&=&n\left({\bf r},\vec{a}(t), 
    \frac{d\vec{a}(t)}{dt}\right), \\ 
s&=&s\left({\bf r},\vec{a}(t), 
    \frac{d\vec{a}(t)}{dt}\right), 
\end{eqnarray*} 
together with the velocity field, 
\[
u^{\mu}=u^{\mu}\left( {\bf r},\vec{a}(t), 
        \frac{d\vec{a}(t)}{dt}\right) , 
\] 
then the action, Eq. (\ref{L}), may be written as a 
time integral of an effective Lagrangian 
\begin{equation} 
L_{eff}\left(\vec{a}(t),\frac{d\vec{a}(t)}{dt}\right)
=-\int d{\bf r}\;\varepsilon(n,s)\,.  \label{Leff-1} 
\end{equation} 
The constraint terms vanish for these ansatz. In this 
case, the equations of motion for the variables 
$a_{i}(t)$ are obtained as the Euler-Lagrange equations. 
This method could be applied to relativistic heavy-ion 
collisions, trying to describe them in a simple 
analytic and effective way in terms of few parameters. 
However, the general parametrization which solve 
exactly the continuity equations is not easy. An 
approximate way to solve the continuity equation is 
proposed in a numerical method called {\it Smoothed 
Particle Hydrodynamics}. 

\subsection{Smoothed particle hydrodynamics}
 \label{smooth} 

The SPH algorithm was first introduced for astrophysical 
applications~\cite{Lucy,Monaghan}. In~\cite{sph}, we 
extended this numerical method to heavy-ion collisions 
by the use of the variational approach discussed in the 
preceding subsection. 

\subsubsection{SPH representation of densities} 

The basic idea of the SPH method is to parametrize the 
continous density distribution of any extensive 
physical quantity in terms of sum of base functions 
with finite support. This procedure introduces two 
types of approximations of different nature. To see this, 
let us suppose that $A$ is the physical extensive 
quantity and $a({\bf r},t)$ the corresponding density 
distribution. We start with the identity, 
\[
a({\bf r},t)=\int a({\bf r}^{\prime},t)
\delta({\bf r-r}^{\prime}) d^3{\bf r}^{\prime}. 
\] 
Now, let us introduce the first approximation. 
Substitute the Dirac $\delta $-function by a smooth, 
normalized function $W$ with finite support, say, $h$, 
and transform the density $a$ to $\widetilde{a}$ as  
\beq 
a({\bf r},t)\rightarrow\widetilde{a}({\bf r},t)
 =\int a({\bf r}^{\prime},t)W({\bf r-r}^{\prime};h) 
  d^{3}{\bf r}^{\prime}, 
\label{atil} 
\eeq 
where as mentioned, $W$ is normalized, 
\[
\int W({\bf r-r}^{\prime};h)d^{3}{\bf r}^{\prime}=1\,. 
\] 
and having the property of finite support, 
\[
W({\bf r-r}^{\prime};h)\rightarrow 0,\ {\rm for}\ 
\vert{\bf r-r}^{\prime}\vert >h\,. 
\] 
At this stage, the new density 
$\widetilde{a}({\bf r},t)$ describes the smoothed part 
of the original density $a({\bf r},t)$. 
From the 
Fourier transform, we can see that for this smoothed 
density, the Fourier components with large 
wave numbers, corresponding to 
\[
k>\frac{1}{h}\ ,
\] 
vanish rapidly.
In other words, the kernel function $W$ serves as the 
short wavelength cut-off filter. Physically, it is 
useful to introduce such a filter, since we very often 
want to eliminate very short scale part in order to 
extract the global feature of the dynamics of the 
system. In 60's, similar idea has been used to smooth 
out the spectrum density of the nuclear shell model to 
extract the collective liquid drop potential by 
Strutinski~\cite{Strutinski}. 
\smallskip

Now, we introduce the second approximation. This is 
rather to do with the practical aspect, that is, to 
reduce the degee of freedoms envolved in the 
calculation. We replace the integral Eq.(\ref{atil}) 
by a finite sum over finite discrete set of 
points,$\{{\bf r}_i,\ i=1,..,N\}$.  
\begin{equation} 
\widetilde{a}({\bf r},t)\rightarrow a_{SPH}({\bf r},t) 
 =\sum_i^N A_i W({\bf r-r}_i;h) , 
 \label{aSPH} 
\end{equation} 
where the weight $A_i$ should be chosen appropriately 
to minimize the difference between 
$\widetilde{a}({\bf r},t)$ and $a_{SPH}({\bf r},t)$ 
everywhere. The above expression means that we are 
representing the continous density as sum of finite 
number of unit distributions (kernel) carrying the 
quantity $A_i$. These unit density distributions are 
centered at the position ${\bf r}_i\,$. 
\smallskip

Finally, the correspondence, 
\begin{equation}
a({\bf r},t)\rightarrow a_{SPH}({\bf r},t) 
=\sum_i^{N}A_{i}W({\bf r-r}_i;h) \label{a_SPH} 
\end{equation} 
can be considered as an approximation ansatz for the 
density $a({\bf r},t)$ with finite number of parameters, 
$\{A_i,{\bf r}_i ,i=1..,N\}\,$. Due to the 
normalization of the kernel $W$, we have 
\[
\int a_{SPH}({\bf r},t)d^{3}{\bf r}=\sum_{i}^{N}A_i\,, 
\] 
so that we should choose 
\[
\sum_{i}^{N}A_{i} 
\] 
as the total value of the quantity $A$ of the system. 

\subsubsection{Solution of continuity equations} 

For the application in Hydrodynamics, we can use these 
parameters as the variational variables so that they 
are depending on time. When we deal with one or more 
extensive quantities, we usually choose one conserved 
quantity as the reference density, say $\rho$, and 
represents it by the SPH form, choosing appropriately 
the weights $\{\nu_i\}$ to get 
\begin{equation} 
 \rho_{SHP}({\bf r},t)=\sum_{i}^{N}\nu_{i}
  W({\bf r-r}_{i}(t);h)\,, 
 \label{rho_SPH} 
\end{equation} 
and take ${\bf \nu}_i$ constant in time. Other 
extensive quantity, say $A$, is calculated as in 
Eq.(\ref{a_SPH}) with weights, 
\begin{equation} 
 A_i=\left(\frac{a}{\rho}\right)_{i}\nu_{i}\ .
 \label{Ai} 
\end{equation} 
The quantity $(a/\rho)_i$ then represents the quantity 
$A$ for the unit reference quantity $\rho$ at the 
position ${\bf r=r}_i(t)\,$. Note that the time 
dependence of the density in Eq.(\ref{rho_SPH}) comes 
from those of $\{{\bf r}_i(t)\}\,$. In this sense, 
Eq.(\ref{rho_SPH}) can be understood as if 
$\{{\bf r}_i(t)\}$ are Lagrangian coordinates 
attached to small volumes of the order of $h^3,$ with 
some conserved quantity, such as baryon number or 
entropy in the adiabatic expansion. From now on, we 
refer to these unit density distributions characterized 
by their coordinates $\{{\bf r}_i\}$ as ``SPH 
particles''. From Eq.(\ref{Ai}), the quantity $A_{i}$ 
can be interpreted as the part of $A$ carried by the 
SPH particle $i$. 
\smallskip

The most powerful point of the above scheme of SPH 
representation is that we can solve the continuity 
equation in a very simple manner. Suppose $M$ is a 
conserved quantity. Then the corresponding density 
$\rho$ should satisfy the continuity equation, 
\begin{equation}
 \frac{\partial\rho({\bf r},t)}{\partial t}
  +\nabla\cdot(\rho{\bf v})=0\,,  
 \label{rho_cont} 
\end{equation} 
where {\bf v} is the velocity field. The SPH expression 
for the current ${\bf j}=\rho{\bf v}$ is 
\[
{\bf j}_{SPH}({\bf r},t)=\sum_i{\bf v}_i\,\nu_{i}
W({\bf r-r}_i(t))\, , 
\] 
so that 
\[ 
\nabla \cdot {\bf j}_{SPH}({\bf r},t) =\sum_{i}
{\bf v}_i\,\nu_i\nabla W({\bf r-r}_i)\,. 
\] 
On the other hand, from Eq.(\ref{rho_SPH}), 
\begin{eqnarray*}
\frac{\partial\rho_{SPH}({\bf r},t)}{\partial t} 
&=&\sum_{i}\nu_i\frac{d}{dt}W({\bf r-r}_i(t)) \\ 
&=&\sum_{i}\nu_{i}\frac{d{\bf r}_i(t)}{dt}\cdot 
\nabla W({\bf r-r}_i(t))\, , 
\end{eqnarray*} 
By inspection, if we identify 
\[
{\bf v}_i=\frac{d{\bf r}_i(t)}{dt}\,, 
\] 
then we can see that Eq.(\ref{rho_cont}) is 
automatically satisfied. 
\smallskip

For the application to the relativistic heavy ion 
collisions, we can take the entropy and baryon number as 
the basic conserved quantities. Then, their densities 
(in the space-fixed frame) are parametrized as 
\begin{eqnarray} 
s^{\ast}({\bf r},t)
&=&\sum_{i}^{N}\nu_{i}~W({\bf r}-{\bf r}_{\,i}(t))~,
   \label{entropy} \\
n^{\ast}({\bf r},t) 
&=&\sum_{i}^{N}b_{i}~W({\bf r}-{\bf r}_{\,i}(t))~, 
\label{baryon}
\end{eqnarray} 
where $\nu_{i}$ and $b_{9}$ are the entropy and baryon 
number attached to the $i$-th ``particle''. The total 
entropy and baryon number are then given by 
\begin{eqnarray}
S &=&\int\!d^{3}{\bf r}~s^{\ast}({\bf r},t) 
     =\sum_{i}^{N}\nu_{i}~. \\
B &=&\int\!d^{3}{\bf r}~n^{\ast }({\bf r},t) 
     =\sum_{i}^{N}b_{i}~. 
\end{eqnarray} 
The proper densities of entropy and baryon number are 
related with these space-fixed frame quantities as 
\begin{eqnarray*} 
s &=&\gamma ^{-1}s^{\ast }, \\
n &=&\gamma ^{-1}n^{\ast },
\end{eqnarray*} 
where $\gamma=u^0$ is the Lorentz factor asscociated 
with the fluid velocity. For the hydrodynamical 
description of nuclear and hadronic collisions at 
ultra-relativistic energies, we prefer to use the 
entropy than the baryon number as the reference 
conserved number to write the SPH representation of any 
other extensive quantities. This is because, the 
baryon number may become zero but the entropy density 
never vanishes in the physically interesting region 
within a hydrodynamical description. 

\subsubsection{SPH action and SPH equations} 

In the variational derivation of this method, the set 
of time-dependent variables $\{{\bf r}_i,i=1,...,n\}$ 
are taken as the variational degrees of freedom and 
their equations of motion are determined by minimizing 
the action for the hydrodynamic system. Thus, SPH may 
be considered as an effective description, in which the 
coordinates $\{{\bf r}_i(t)\}$ associated with 
\textquotedblleft particles\textquotedblright\ are the 
optimal dynamical parameters which minimize the model 
action. We observe that $\{\nu_i\}$ or $\{b_i\}$ are 
not dynamical variables and are determined by the 
initital conditions together with the constraints for 
the variational procedure. 
\smallskip 

The effective Lagrangian, Eq.~(\ref{Leff-1}), is 
rewritten in SPH representation as 
\begin{equation} 
L_{SPH}(\{{\bf r}_i,{\bf {\dot{r}}}_i\}) 
=-\sum_{i}\nu_{i}(\varepsilon/s^{\ast})_i=
-\sum_i\left(\frac{E}{\gamma}\right)_i, \label{L_SPH} 
\end{equation} 
where $E_i$ is the ``rest energy'' of the $i$-th 
``particle''. Then, the equations of motion are 
obtained from the usual variational procedure. This 
leads to the following coupled equations 
\begin{eqnarray}
&&\frac{d}{dt}\left(\nu_i\frac{p_i+\varepsilon_i} 
  {s_i}\,\gamma_i\,{\bf v}_i\right)\hspace{4.5cm} 
\nonumber \\ 
&&\hspace{0.5cm}+\sum_j\nu_i\nu_j
\bigg[\frac{p_i}{{s_i^{\ast }}^{2}}+
      \frac{p_j}{{s_j^{\ast}}^2}\bigg]\, 
{\bf \nabla}_i W({\bf r}_{\,i}-{\bf r}_{\,j};h)=0\,. 
\label{SPH_eq}
\end{eqnarray}
\smallskip

\subsubsection{General coordinate system} 

The variational procedure can readily be extended to 
coordinate systems with a non-Cartesian metric. The 
use of generalized coordinate systems is particularly 
important when we consider realistic initial conditions 
for simulations of RHIC processes. As we know, in a 
relativistic heavy-ion collisional process, the initial 
state is a cold, quantum nuclear matter. Just after 
the collision, the hadronic matter stays at a highly 
off-shell state and the materialization occurs only 
after $\sim 1\,$fm/c in the proper time. Therefore, 
the local thermodynamical state would emerge for some 
local proper time and not for the global space-fixed 
time $t$. Thus, it is important to choose a 
convenient coordinate system for the description of 
the relativistic heavy-ion collisions. For example, 
one often uses the hyperbolic time and longitudinal 
coordinates to be described later.  

Let us consider a general coordinate system, 
\begin{equation} 
ds^{2}=g_{\mu \nu }dx^{\mu }dx^{\nu }.
\end{equation} 
However, in order to unambiguously define the conserved 
quantity, we consider only the case that the time-like 
coordinate is orthogonal to the space-like coordinates, 
\begin{equation}
g_{\mu 0}=0\,.
\end{equation} 
The action principle for the relativistic fluid motion 
can be written as~\cite{variational} 
\begin{equation}
\delta I=-\delta\int d^4 x\sqrt{-g}\,\varepsilon=0\,,
\end{equation} 
together with the constraint for the conserved entropy 
current, 
\begin{equation}
(su^{\mu})_{;\mu}=\frac{1}{\sqrt{-g}}\partial_{\mu}
(\sqrt{-g}\,su^{\mu})=0\,, 
\end{equation} 
or 
\begin{equation}
\frac{1}{\sqrt{-g}}
\partial_{\tau}\left(\sqrt{-g}s\gamma\right) 
 +\frac{1}{\sqrt{-g}}\sum_i\partial_i
 \left(\sqrt{-g}s\gamma v^i\right)=0\,, 
\end{equation} 
where 
\begin{equation}
v^{i}=\frac{u^{i}}{u^{0}} 
\end{equation} 
and we use the notation, 
\[
\tau =x^{0},\,\;\gamma =u^{0}.
\] 
The generalized gamma factor $\gamma$ is related to 
the velocity $\vec{v}_a$ through $u_{\mu}u^{\mu}=1$, 
so that 
\beq 
\gamma =\frac{1}
 {\sqrt{g_{00}-\vec{v}^{\,T}\mathbf{g}\,\vec{v}}}\,,  
\label{4}
\eeq 
where $-\mathbf{g}$ is the $3\times 3$ space part of 
the metric tensor. That is 
\begin{equation} 
\left(g_{\mu\nu}\right)=\left(
\begin{array}{cc}
g_{00} & 0 \\ 
0 & -\mathbf{g} 
\end{array} 
\right) .
\end{equation}

Let us now introduce the SPH representation. We may, 
for example, express the entropy density by the ansatz 
\beq 
\sqrt{-g}s\gamma =s^{\ast}\ \rightarrow\ s_{SPH}^{\ast}
 =\sum_i\nu_i W(\vec{r}-\vec{r}_i(\tau))\,,  \label{sph_1}
\eeq 
or by 
\begin{equation}
s\gamma=s^{\ast}\ \rightarrow s_{SPH}^{\ast}
 =\sum_i\nu_i W(\vec{r}-\vec{r}_i(\tau)) 
\label{sph_2} 
\end{equation}
as well. These two possibilities, besides others, are 
simply different ways to parametrize the variational 
ansatz in terms of a linear combination of given 
functions $W(\vec{r}-\vec{r}_a(\tau))$. The most 
important property of an ansatz should be that $W$  
satisfies the normalization condition imposed by the 
basic conserved quantity. Since the total entropy is 
expressed as 
\begin{equation}
S=\int d^3\vec{r}\sqrt{-g}\,s\gamma=\sum_i\nu_i\,, 
\end{equation}
the normalization of $W$ should be taken to be 
\begin{equation}
\int d^3\vec{r}\,W(\vec{r}-\vec{r}\,^{\prime})=1\,, 
\label{norm_1}
\end{equation}
for the parametrization Eq.(\ref{sph_1}) and 
\begin{equation}
\int d^3\vec{r}\sqrt{-g}W(\vec{r}-\vec{r}\,^{\prime}) 
 =1\,,
\label{norm_2}
\end{equation}
for the parametrization Eq.(\ref{sph_2}). In the usual 
SPH calculations, it is not desirable to introduce in 
$W$ the space-time dependence through its normalization 
condition. In this respect, the most natural way to 
introduce the SPH representation is Eq.(\ref{sph_1}). 
With this choice, the SPH action is given by 
\begin{eqnarray}
I_{SPH}&=&-\int d\tau\int d^3\vec{x}\sum_i\nu_i
 \left(\frac{\sqrt{-g}\,\varepsilon}{\sqrt{-g}\,s\gamma} 
 \right)_i W(\vec{r}-\vec{r}_i(\tau)) \nonumber \\ 
&=&-\int d\tau\sum_i\nu_i\left(\frac{\varepsilon} 
{s\gamma}\right)_i.  
\label{SPH_action} 
\end{eqnarray} 
The variational principle leads to the following 
equation of motion, 
\begin{eqnarray} 
\frac{d}{d\tau}\vec{\pi}_i= 
&-&\sum_j\nu_i\nu_j 
   \left[\frac{1}{\sqrt{-g_i}\gamma_i^2} 
   \frac{p_i}{s_i^2}+\frac{1}{\sqrt{-g_j}\gamma_j^2} 
   \frac{p_{j}}{s_{j}^{2}}\right] 
   \nabla_i W_{ij}  \nonumber \\
&+&\frac{\nu_i}{\gamma_i}\frac{p_i}{s_i}
   \left(\frac{1}{\sqrt{-g}}\nabla\sqrt{-g}\right)_i 
   \nonumber \\ 
&+&\frac{\nu_i}{2}\gamma_i
   \left(\frac{p+\varepsilon}{s}\right)_i 
   (\nabla g_{00}
   -\vec{v}_i^{\,T}\nabla\mathbf{g}\vec{v}_i)_i\,,  
\label{EQM}
\end{eqnarray} 
where 
\begin{equation}
\vec{\pi}_i=\gamma_i\,\nu_i
   \left(\frac{p+\varepsilon}{s}\right)_i
   \mathbf{g}\,\vec{v}_i  
\label{q}
\end{equation} 
and the operator $\nabla$ is just the simple 
derivative operator with respect to the coordinate 
variable in use. 

For ultrarelativistic heavy-ion collisions, a useful 
set of variables is 
\bea 
\tau&=&\sqrt{t^{2}-z^{2}},  \label{tau} \\
\eta&=&\frac{1}{2}\ln\frac{t+z}{t-z}\,,\label{eta}\\ 
\vec{r}_T&=&\left( 
\begin{array}{c}
x \\ 
y 
\end{array} 
\right) . 
\eea 
As mentioned above, the initial conditions for RHIC 
processes are specified in terms of the proper time 
rather than of the coordinate time $t$. The variable 
$\tau$ is not exactly the physical proper time of the 
matter, but for the initial times it may approximate  
the proper time. 

The metric tensor for this coordinate system is given 
by 
\begin{eqnarray*} 
g_{00}&=&1, \\ 
\mathbf{g}&=&\left( 
\begin{array}{ccc}
1 & 0 & 0 \\ 
0 & 1 & 0 \\ 
0 & 0 & \tau^2  
\end{array} 
\right) , \\
\sqrt{-g}&=&\tau\,. 
\end{eqnarray*} 
Since the metric is space independent, we can use the 
parametrization 
\[
\tau\gamma_i s_i=s_i^{\ast}=\sum_{j=1}^n\nu_jW(q_{ij}), 
\] 
where 
\[
q_{ij}=\sqrt{(x_i-x_j)^2+(y_i-y_j)^2
             +\tau^2(\eta_i-\eta_j)^2} 
\] 
and $W$ is normalized as 
\[
4\pi\int_0^{\infty}q^2 dq\;W(q)=1\,. 
\]
The SPH equation becomes 
\[
\frac{d}{d\tau}\vec{\pi}_i=-\frac{1}{\tau}\sum_j 
 \nu_i\nu_j\left[\frac{1}{\gamma_i^2}\frac{p_i}{s_i^2}
 +\frac{1}{\gamma_{j}^2}\frac{p_j}{s_j^2}\right]
 \nabla_i W_{ij}\,, 
\] 
where the $\eta$ component of the momentum is related 
to the velocity $d\eta/d\tau$ as 
\[
\pi_{\eta}=\tau^2\nu\gamma
   \left(\frac{p+\varepsilon}{s}\right) 
   \frac{d\eta}{d\tau}, 
\] 
whereas in the transverse direction, we have 
\[
   \vec{\pi}_T=\nu\gamma
   \left(\frac{p+\varepsilon}{s}\right) 
   \frac{d\vec{r}_T}{d\tau}. 
\] 
The Lorentz factor is given by 
\[
\gamma =\frac{1}{\sqrt{1-\vec{v}_T^2-\tau^2 v_{\eta}^2}}. 
\]

\subsubsection{Landau Model}
\label{Landau}

In order to show the efficiency of the method and also 
to show the corrrect choice of the coordinate system, 
let us investigate the Landau model in the SPH scheme, 
using the ordinaly Cartesian coodinates and $\eta-\tau$ 
coodinates. Since the analytical solution is known, we 
can compare the numerical solutions to it. 

\begin{figure}[!hbt] 
\vspace*{-3.cm}
\begin{center}
\includegraphics*[width=9.cm]{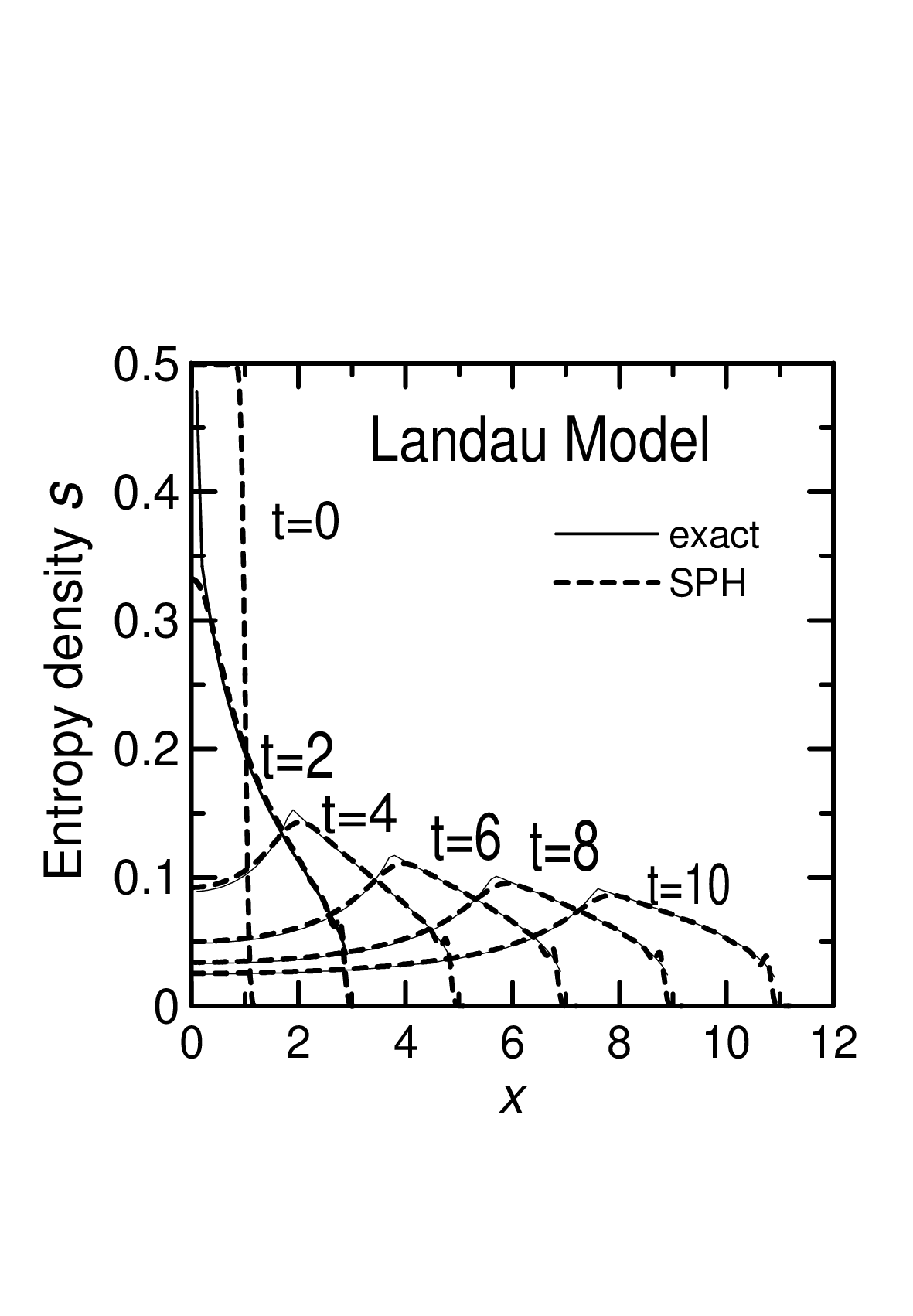} 
\end{center}
\vspace*{-1.9cm}
\begin{center}
\includegraphics*[width=8.cm]{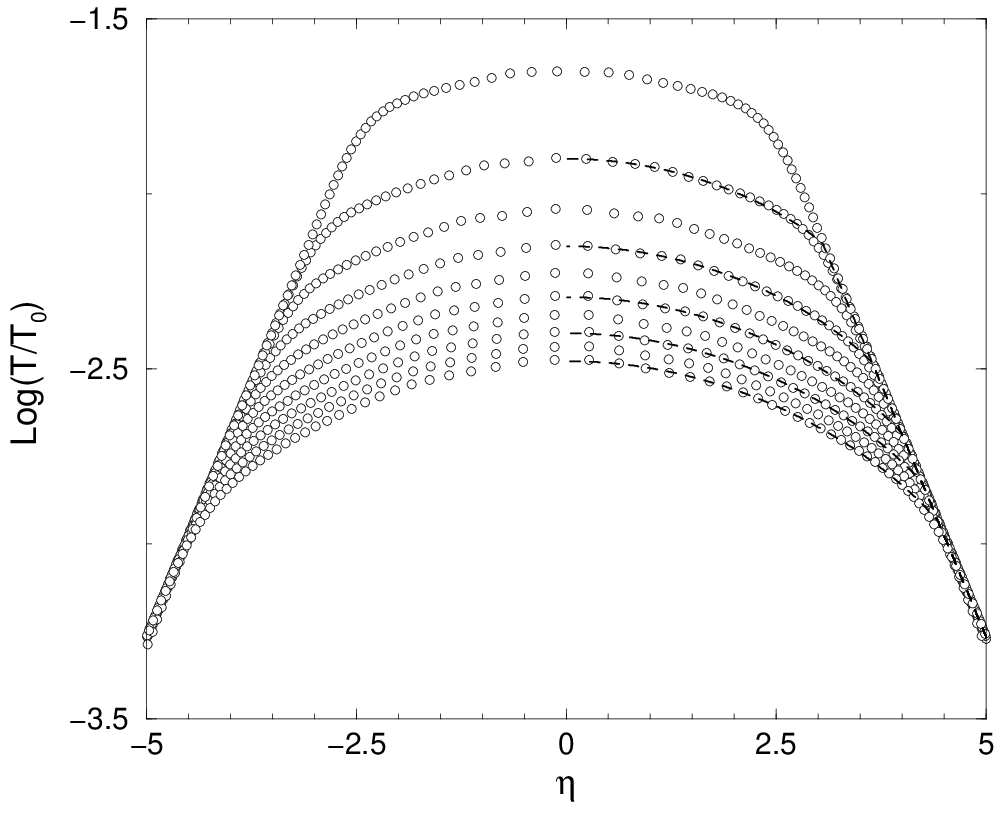} 
\end{center} 
\vspace*{-.2cm}
\caption{$a)$ (above) Entropy profiles of the Landau 
model in Cartesian coordinate for different times. The 
exact results are given by the broken curves. The SPH 
solution is shown by the full curves. $b)$ (below) 
Temperature profiles of the Landau model in the 
hyperbolic coordinate system (see text), for different 
time $\tau$. The SPH calculation is represented by the 
circles, and the exact result by the broken curves.} 
\label{Landau_sol}
\end{figure} 

We thus solve the hydrodynamical evolution of a system 
of one-dimensional relativistic massless baryon-free 
gas initially at rest. The equation of state of a 
relativistic massless boson gas is 

\[
p=\frac{1}{3}\varepsilon =Cs^{4/3}, 
\] 
where 
\[
C=\left( \frac{15}{128\pi ^{2}}\right) ^{1/3}. 
\] 
To apply the SPH method, we introduce the discrete 
one-dimensional space variable ${x_i(t),i=1,..,n}$ 
(and similarly for $\eta-\tau$ coodinates). The 
relation between the momentum and velocity is then 
\begin{equation}
\pi =4C\nu s^{\ast 1/3}\gamma ^{2/3}v,  \label{q(v)} 
\end{equation} 
where, in this case, $v$ can be solved analytically 
with respect to $\pi$. 
\smallskip

In Figs.~\ref{Landau_sol}$\,a$ and $b$, we show the 
results of our SPH calculation together with the exact 
solution~\cite{Landau}. In these examples, we took only 
$100$ particles with equally spaced $x_i$ (or ${\eta_i}$). 
As we see from this example, in spite of rather small 
number of particles, the SPH solution is quite 
satisfactory. In particular, when we use the $\eta-\tau$ 
coordinates with an appropriate distribution of 
$\nu_i^{\prime}s$ (Fig.~\ref{Landau_sol}$\,b$), an 
excellent agreement with the analytical solution can be 
obtained. The computation time needed to get these 
solutions is even less than that needed to numerically  
evaluate the analytical solution. 

\subsubsection{Transverse expansion on longitudinal 
scaling expansion}
\label{trans}

As a further test, closer to a realistic situation than 
that of Figs.~\ref{Landau_sol}, we calculated the 
transverse expansion of a cylindrically symmetric 
homogeneous massless pion gas, undergoing a 
longitudinal scaling expansion, and ini-\hfilneg\ 

\begin{figure}[!hbt] 
\begin{center} 
\includegraphics*[width=8.cm]{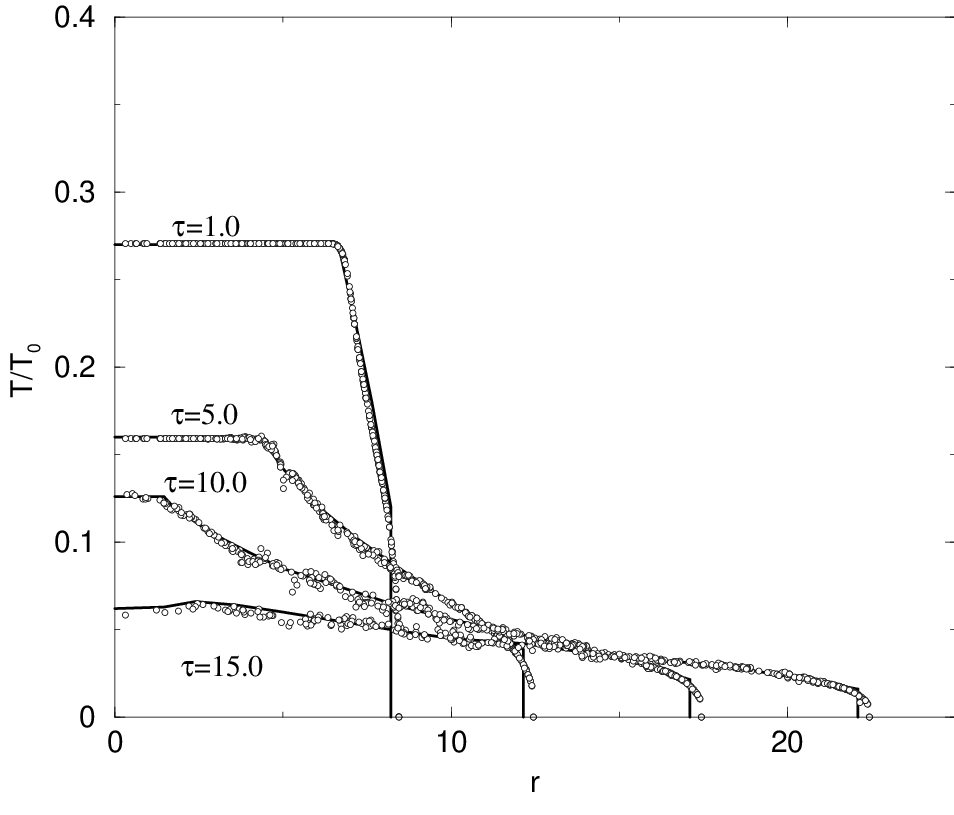} 
\end{center} 
\vspace*{-.3cm} 
\caption{Temperature profile of a cilyndrically 
symmetric flow with longitudinally scaling expansion, 
shown as a function of $r=\sqrt{x^2+y^2}$. The SPH 
results at $\eta=0$ (circles) are compared with the 
numerical solution obtained by a space-fixed grid 
method. The SPH calculation has been perfomed in full 
3D.} 
\label{transverse}
\end{figure} 

\ni tially at rest in transverse directions. Such a 
problem has been discussed by several authors as a 
useful base to understand the transverse expansion. In 
Fig.~\ref{transverse}, we compare our results (a 
full $3D$ calculation without assuming cylindrical 
symmetry) with (2+1) numerical results, obtained by the 
use of the method of characteristics~\cite{Pottag}. 
In this example, we used also $50\times 50\times 50$ 
particles. The result is quite satisfactory. If we 
decrease the accuracy by $10\%$, we can reduce the 
particle number almost by one order of magnitude. 

\subsubsection{Shock formation and Neumann-Richtmeyer 
pseudo viscosity}

As seen in the previous examples, our entropy-based 
relativistic SPH method works quite well for the 
adiabatic dynamics of the massless pion gas. However, 
for the application to realistic problems, it is 
fundamental to see how this scheme works for 
non-adiabatic cases, too. This is because, whenever a 
piece of fluid matter flows into another region of the 
fluid with a speed exceeding the sound velocity of the 
fluid, there appears a shock wave, and this is 
essencially a nonadiabatic process. Thus, except for a 
really quasi-static dynamics, there should be an 
entropy production mechanism. This becomes especially 
important in a domain close to the phase transition 
region, because there the velocity of sound tends to 
zero. In the following, we study some examples of one 
dimensional shock problems in the scheme of the SPH 
methods. 

The shock front manifests as a discontinuity in 
thermodynamical quantities in a hydrodynamic solution. 
Mathematically speaking, the shock front should be 
treated as a boundary connecting two distinct 
hydrodynamic solutions. The smoothed particle ansatz 
excludes such a possibility from the beginning. Since 
short-wavelength excitation modes do not exist in the 
SPH ansatz, the energy and momentum conservation 
required by the hydrodynamics results in very rapidly 
oscillating motion of each SPH particle. Such a 
situation occurs, for example, when a very high-energy 
density gas is released into a low density region. This 
kind of shock, for the case of a baryon gas, is 
discussed in~\cite{Maruhn} and also, in the SPH 
context, in~\cite{Siegler}. Here, we appliy our 
entropy-based SPH approach to the massless pion gas. 

Fig.~\ref{shock} gives the typical behavior of SPH 
solution for such a situation, if entropy production 
is not taken into account. As discussed above, there 
appear in fact rapid oscillations in thermodynamical 
quantities just behind the shock front. Actually, such 
oscillations always appear in any numerical approach 
if entropy production is not included. 
\begin{figure}[!htb] 
\vspace*{-1.cm}
\begin{center}
\includegraphics*[width=8.cm]{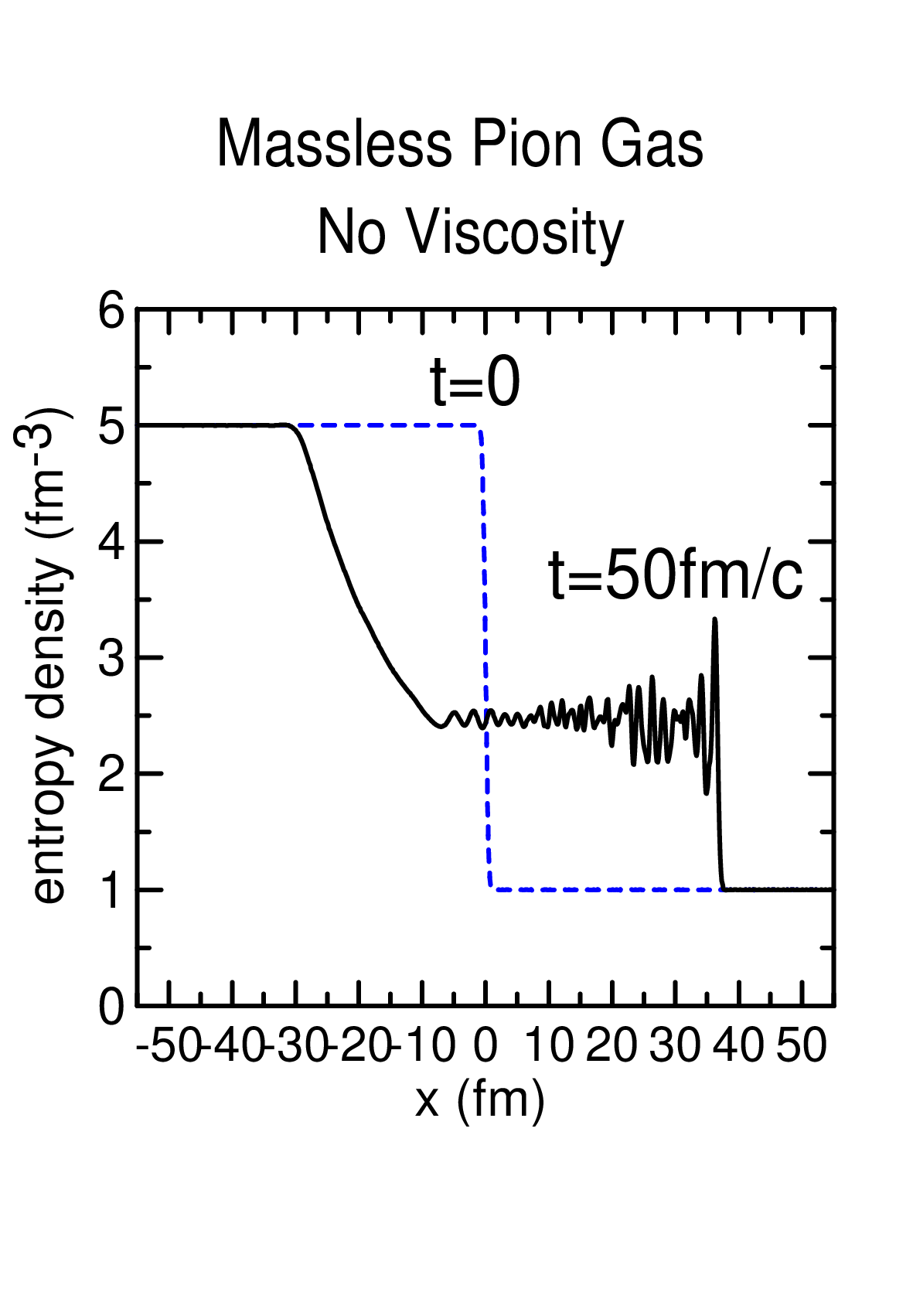} 
\end{center} 
\vspace*{-2.cm}
\caption{Shock wave formation in one-dimensional pion 
gas, calculated with SPH. No viscosity is used.} 
\label{shock}
\end{figure} 

In order to avoid these unphysical oscillations, von 
Neuman and Richtmeyer~\cite{Neuman} introduced the 
concept of pseudoviscosity. The idea is to set the 
dissipative pressure where the shock wave discontinuity 
is present. To do this, Neuman and Richtmeyer proposed 
to replace the pressure by 
\[
p\rightarrow p+Q,
\] 
where $Q$ is the pseudoviscosity and they took the 
following ansatz, 
\[
Q=\left\{ 
\begin{array}{ccc}
(\alpha\Delta x)^2\rho\;(\dot{\rho}/\rho)^2, 
& \;\;\; & \dot{\rho}>0 \\ 
0\,,\;\;\;\;\;\;\;\;\;\;\;\;&\;\;\;&\;\dot{\rho}<0 
\end{array} 
\right. .
\]
The above formula is for nonrelativistic 
one-dimensional hydrodynamics. Here, $\rho $ is the 
mass density, $\Delta x$ is the space grid size and 
$\alpha$ is a constant of the order of unity. In order 
to generalize the above pseudoviscosity for 
relativistic SPH case, we replace the quantity 
$\dot{\rho}/\rho\,$ by $-\theta=-\partial_{\mu}u^{\mu}$ 
and $\Delta x$ by $h$, where $h$ is as before the width 
of the smoothing kernel $W$. More precisely, we take 
the following form which is a slightly modified 
expression suggested by Ref.~\cite{Siegler}, 
\begin{equation} 
Q=\left\{ 
\begin{array}{ccc}
p\left[-\alpha h\theta+\beta(h\theta)^2\right]\,, & 
\;\;\; & \theta <0\,, \\ 
0\,,\;\;\;\;\;\;\;\;\;\;\;\; & \;\;\; & \theta \geq 0\,.
\end{array} 
\right.   
\label{Q}
\end{equation} 
where 
\begin{align}
\theta & =\frac{1}{V}\frac{dV}{dt}  \nonumber \\
& =\partial _{\mu }u^{\mu }.
\end{align} 

Actually, $Q$ is equivalent to the bulk viscosity and 
therefore there is no heat flow associated with it. 
What this artificial viscosity does is to convert the 
collective flow energy into the microscopic thermal 
energy. As a consequence, the total energy, that is, 
the sum of the collective flow energy and the internal 
thermal energy is still conserved. In order to 
incorporate the internal energy conservation in the SPH 
scheme, we substitute all the pressure $p_{i}$ by 
$p_{i}+Q_{i},$ and we add the following equation for 
the entropy production,  
\begin{equation}
\frac{1}{\nu_i}\frac{d\nu_i}{dt}
=-\frac{Q_i\gamma_i}{Ts_i^{\ast}}\theta_i. 
\label{dnudt}
\end{equation}

\begin{figure}[!b] 
\vspace*{-1.cm} 
\begin{center}
\includegraphics*[width=8.cm]{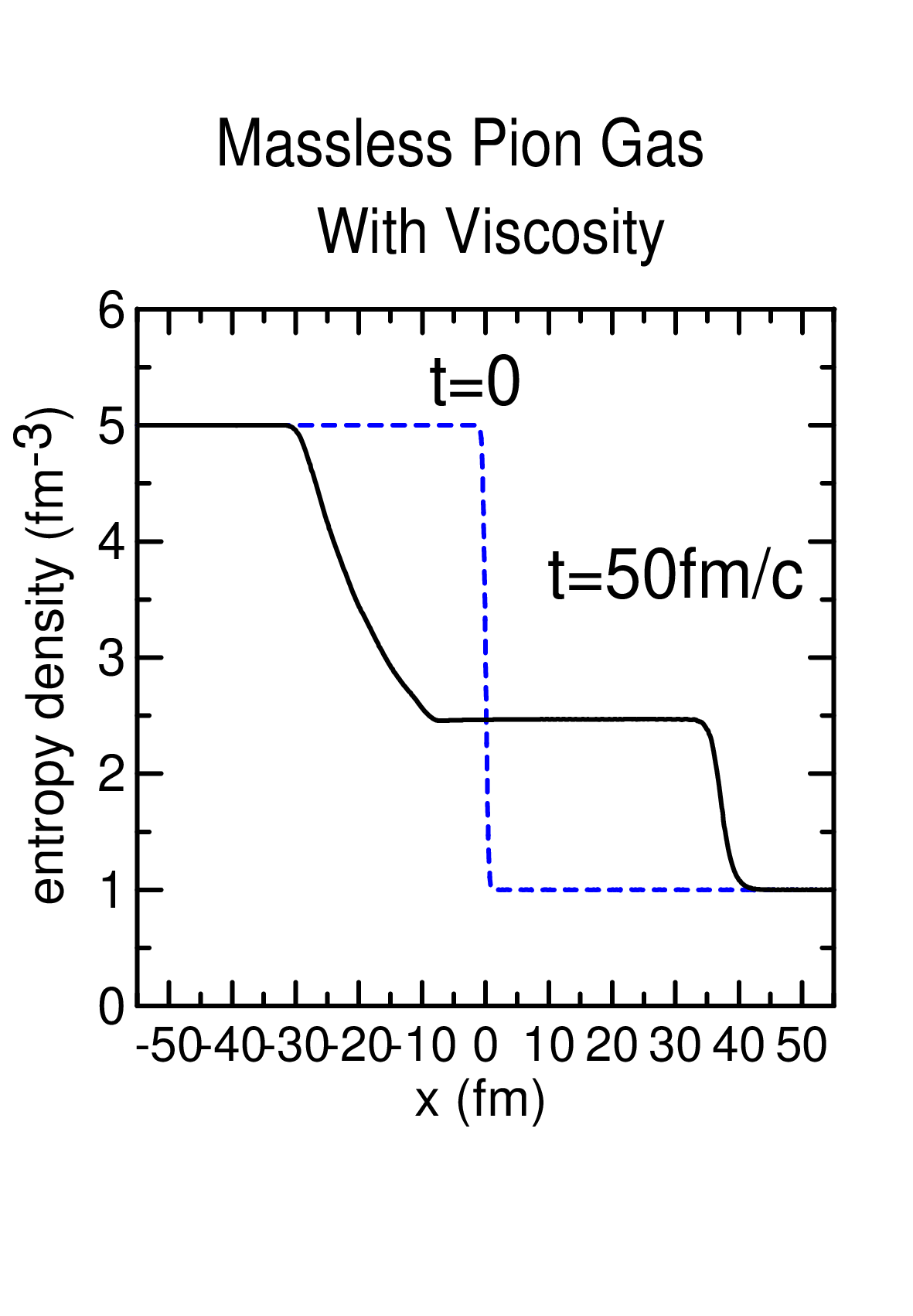} 
\end{center} 
\vspace*{-2.cm} 
\caption{After the introduction of the $Q$ term in the 
SPH calculation.} 
\label{shockQ}
\end{figure} 

Fig.~\ref{shockQ} is the solution of the same problem 
as in Fig.~\ref{shock}, but with the entropy production 
taken into account. In this calculation, the parameters 
have been chosen as 
\[
\alpha =2,\;\beta =4
\] 
and $h=0.5fm$ for 1000 SPH particles. As we see, the 
rapid oscillations have been smoothed out (and in turn, 
the numerical calculation became much more efficient). 

It is known that the energy- and momentum-flux 
conservations 
through a sock front
relate the ratio $s_2/s_1$ of entropy 
densities after and before the shock to the velocity 
$v_s$ of the shock front as (Hugoniot-Rankine relation) 
\beq 
\frac{s_2}{s_1}=\frac{2}{3^{3/4}}v_s
 \frac{(9v_s^2-1)^{1/4}}{(1-v_s^2)^{5/4}}\,. 
\label{H-R}
\eeq 
In Fig.~\ref{HR}, we show the velocity of the shock 
front obtained in our SPH calculations as function of 
the entropy ratio(dots). Each point corresponds to the 
different initial condition. They are compared with 
the Hugoniot-Rankine relation Eq.~(\ref{H-R}) (curve). 
The accordance shows that our SPH calculation 
reproduces faithfully the conservation of kinetic energy and 
momentum of the flow through the shock front.

\begin{figure}[!htb] 
\vspace*{-1.8cm} 
\begin{center}
\includegraphics*[width=8.cm]{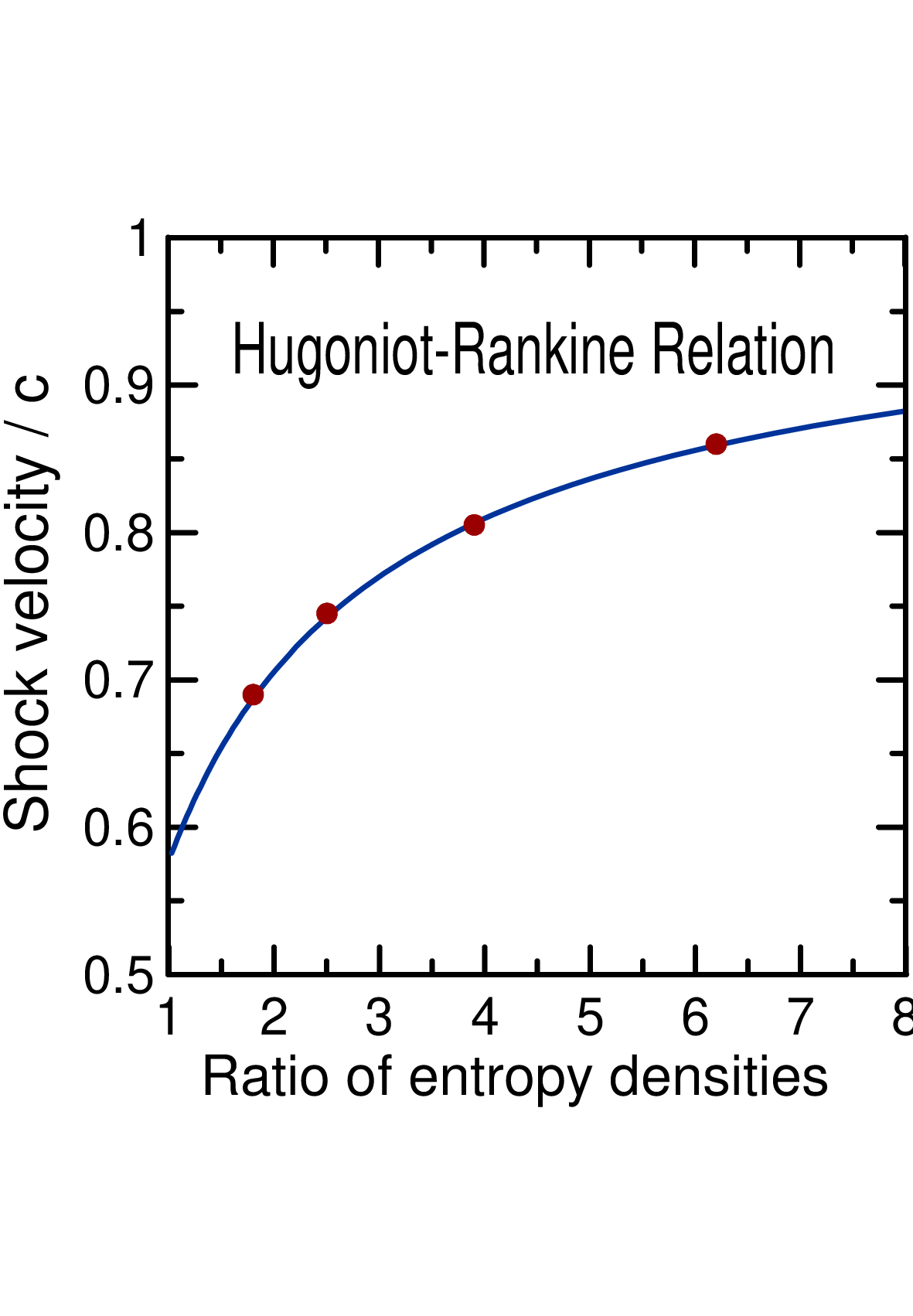} 
\end{center} 
\vspace*{-2.cm} 
\caption{Test of Hugoniot-Rankine relation. The 
circles are results of SPH calculations with different 
initial conditions, and the full curve is 
Eq.~(\ref{H-R}).} 
\label{HR}
\end{figure}

In the usual hydrodynamic computations using space 
grids, the symmetry of the problem is often a crucial 
factor to perform a calculation of reasonable size. 
The SPH method cures this aspect and furnishes a robust 
algorithm particularly appropriate to the description 
of processes where rapid expansions of the fluid should 
be treated. The equations of motion are derived by a 
variational procedure from the SPH model action with 
respect to the Lagrangian comoving coordinates. This 
guarantees that the method furnishes the maximal 
efficiency for a given number of degrees of freedom, 
keeping strictly the energy and momentum conservation. 
For this reason, solutions can be obtained with a very 
reasonable precision, with a relatively small number of 
SPH particles. This is the basic advantage of the 
present method, when we analyze the event-by-event 
dynamics of the relativistic heavy-ion collisions.

On the other hand, the precision of this method 
increases rather slowly with the number of SPH 
particles. Therefore, a relatively large number of 
particles is required if one wants a very precise 
numerical solution. However, for the application to the 
RHIC physics, we may need rather crude precision 
especially if we consider the dubious validity of the 
rigorous hydrodynamics. For a calculation with 
typically $10\%$ errors, the SPH algorithm presented 
here furnishes a very efficient tool to study the flow 
phenomena in the RHIC physics. 

A fundamental difficulty of the relativistic 
hydrodynamics for viscous fluid~\cite{Israel,Hiscock} 
is that the dissipation term causes an intrinsic 
instability to the system. This instability basically 
comes from the fact that the dissipation term contains 
$\theta=\partial^{\mu}u_{\mu}$ (see 
Eqs.~(\ref{Q},\ref{dnudt})), so that it necessarily 
introduces the third time-derivative into the equation. 
This means that we have to specify, at least, a part of 
the acceleration as the initial condition. Even we 
specify the initial acceleration, the requirement of 
the internal self-consistency among the equations above 
leads to intrinsically unstable solutions. Israel 
proposed~\cite{Israel,Hiscock} to cure these 
difficulties by introducing higher-order thermodynamics 
with respect to deviations from the equilibrium. 
Recently this ``second order thermodynamics'' formalism 
was discussed in the context of Bjorken type 
solution~\cite{Muronga}. In the examples presented in 
the present paper, we did not address this question and 
simply estimated the quantity $\theta$ from the 
quantities one time step before. In practice, this will 
cause no numerical instability and the behavior of the 
solution is quite satisfactory. 

In spite of the above conceptual difficulties when 
nonadiabatic process is involved, the SPH approach has 
a nice feature as its flexibility, allowing the 
treatment of problems with initial conditions without 
any symmetry, as happens in small systems as those 
resulting in relativistic nuclear collisions. It should 
be stressed that, due to the use of Lagrangian 
coordinates, the method is most suitable for 
explosive processes like the relativistic heavy ion 
collisions. (Lagrangian coordinates have been used 
for treatment of relativistic nuclear collisions also 
by Nonaka, Honda and Muroya~\cite{nonaka}). 
Furthermore, the variatioal approach guarantees that 
the SPH equations (\ref{SPH_eq}) give the optimal 
description of motions for a given total number of 
``particles'' $\{{\bf r}_i(t)\}$, which are our 
parameters. In this approach, no numerical 
instabilities will occur, since the whole system is a 
Lagrangian system. A numerical code, called SPheRIO 
has been developped by us on the basis of this 
algorithm. We shall discuss, in Sec.\ref{application}, 
some results obtained using this code. 


\section{Decoupling criteria} 
\label{decoupling} 

\subsection{Cooper-Frye prescription} 
 \label{CFp} 

As mentioned in the Introduction, the decoupling 
process is customarily described using the Cooper-Frye 
prescription~\cite{C-F}, which gives the invariant 
momentum distribution as 
\beq 
 E\frac{d^3N}{dp^3}
 =\int_{\sigma}d\sigma_{\mu}p^{\mu}f(x,p)\,. 
 \label{CFf} 
\eeq 
This description of decoupling introduces a sharp 
freezeout hypersurface $\sigma\,$, usually characterized  
by a constant temperature $T_{f.o.}\,$. Before crossing 
it, particles have a hydrodynamical behavior and, 
when they cross it {\bf suddenly} decouple, 
free-streaming toward the detectors, keeping memory 
of the conditions (flow, temperature) of where and when 
they crossed the three dimensional surface. 
\smallskip

In SPH representation, we write 
\bea 
 E\frac{d^3N}{dp^3}
  =\sum_j\frac{\nu_j n_{j\mu}p^\mu}
          {s_j |n_{j\mu}u_j^\mu|}\;f(u_{j\mu}p^\mu)\,, 
  \label{CF-SPH} 
\eea 
where the summation is over all the SPH particles, 
which should be taken where they cross the 
hyper-surface $T=T_{f.o.}\,$ and $n_{j\mu}$ is the 
normal to this hyper-surface. 

Another often used procedure is to take such a 
freezeout temperature not only constant for a given 
energy but also energy-independent. 
\smallskip 

Though operationally simple, and actually useful for 
obtaining a nice comprehension of several aspects of 
the phenomena, such a concept of sharp freezeout 
hypersurface and also of a constant freezeout 
temperature are clearly highly idealized when applied 
to finite-volume and finite-lifetime systems as those 
formed in high-energy heavy-ion collisions. 

\subsection{Finite-size effect} 
 \label{finite_size}

Before going further, let us for a moment assume 
that such a freezeout temperature is meaningful. 
At least, as an average temperature, it should exist. 
Then, how can we estimate it based on the properties 
of the system? A simple and natural criterion has 
already been given by Landau~\cite{Landau}, by which a 
particle decouples when its mean free-path $\ell$ in 
the medium becomes larger than the system size $L$, 
\beq 
\ell>L\,. \label{Land_cond} 
\eeq 
This means that $T_{f.o.}\,$ is not an intrinsic 
thermodynamic property of the fluid, but it depends 
also on the size of the system. In~\cite{fernando}, 
we applied this idea to estimate $T_{f.o.}\,$ as 
function of the incident energy both for $pp$ ($\bar 
pp$) and nucleus-nucleus collisions, obtaining 
approximately 
\beq 
T_{f.o.}\sim(\sqrt{s})^{-1/12}. 
\eeq 

Here, the energy dependence appears as a consequence 
of the increase in the initial energy density, which 
implies longer expansion time, so larger size $L$ of 
the system (both longitudinally and tranversally) at 
the moment of decoupling, requiring lower density, so 
lower decoupling temperature, too. 
Figure \ref{fernando} shows the comparison made in 
\cite{fernando} of an estimate of the incident-energy 
dependence of $T_{f.o.}\,$, using Landau's criterion 
mentioned above, with $T_{f.o.}$ obtained in a data 
analysis of $\pi$ and $K$ transverse-momentum spectra 
in $pp$ (and $\bar pp$) collisions, in terms 
of a hydrodynamic parametrization of transverse 
velocity distribution and temperature. The data were 
taken from \cite{ISR,collider,Fermilab}. 

\begin{figure}[!htb]
\begin{center}
\includegraphics*[angle=1.7,width=8.cm]{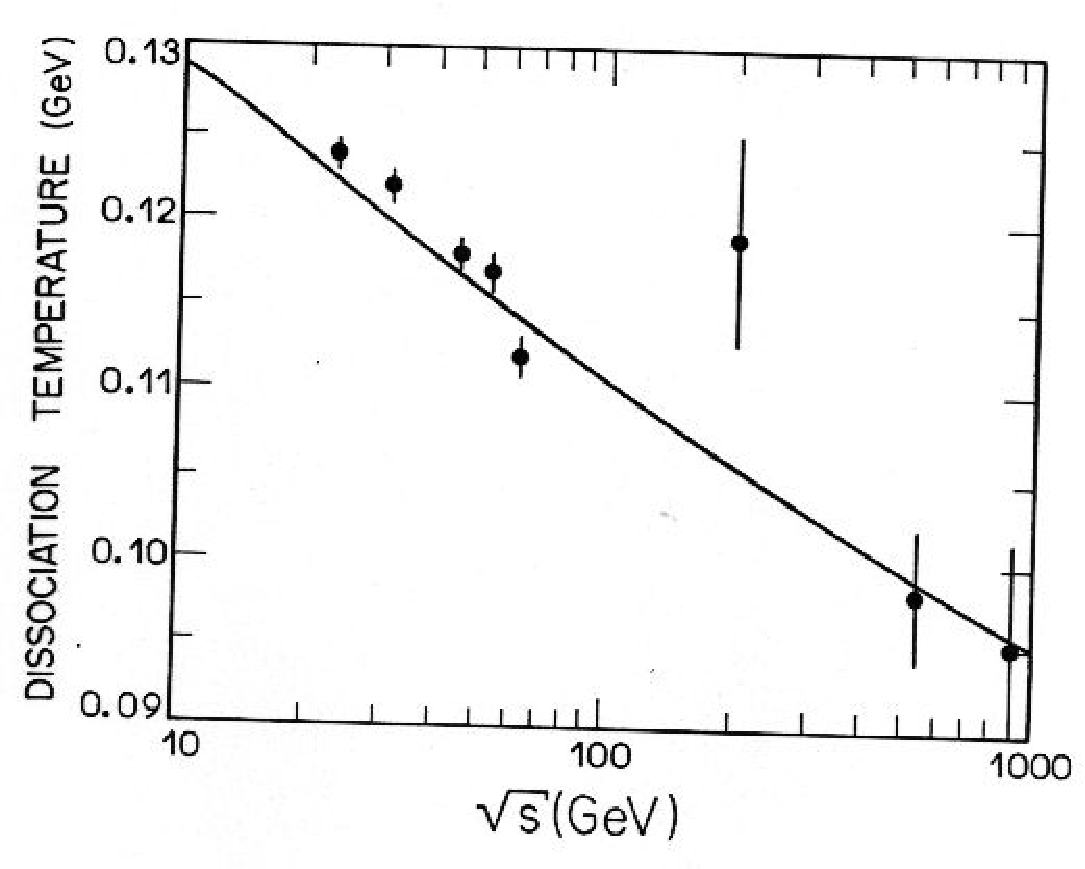} 
\end{center}
\caption{Energy dependence of $T_{f.o.}$ in $pp$ 
 (and $\bar pp$) collisions. The solid line is the 
 estimate as explained in the text. The points were 
 obtained from a data analysis~\cite{fernando}.} 
\label{fernando} 
\end{figure} 

An independent estimate of $T_{f.o.}(s)$ for $pp$ 
was made by Navarra {\it et al.}~\cite{udo}, based on 
somewhat different but related freeze-out criterion 
\[
\tau_{hydro}\sim\tau_{col}\,,
\] 
where $\tau_{hydro}$ and $\tau_{col}$ are, 
respectively, characteristic time for hydrodynamic 
expansion and particle collision, obtaining a similar 
result. 
\smallskip 

When Au + Au collisions data began to appear, Xu and 
Kaneta \cite{Nu} reported that $T_{f.o.}$ seems to 
decrease with $\sqrt{s}$ also in nuclear collisions at 
high energies, and we could verify that the reported 
results are consistent with 
$T_{f.o.}\sim(\sqrt{s})^{-1/12}$. It is interesting 
that, more recently~\cite{eff-t}, in trying to 
understand the experimentally\hfilneg\ 

\begin{figure}[!htb]
\begin{center}
\includegraphics*[width=8.cm]{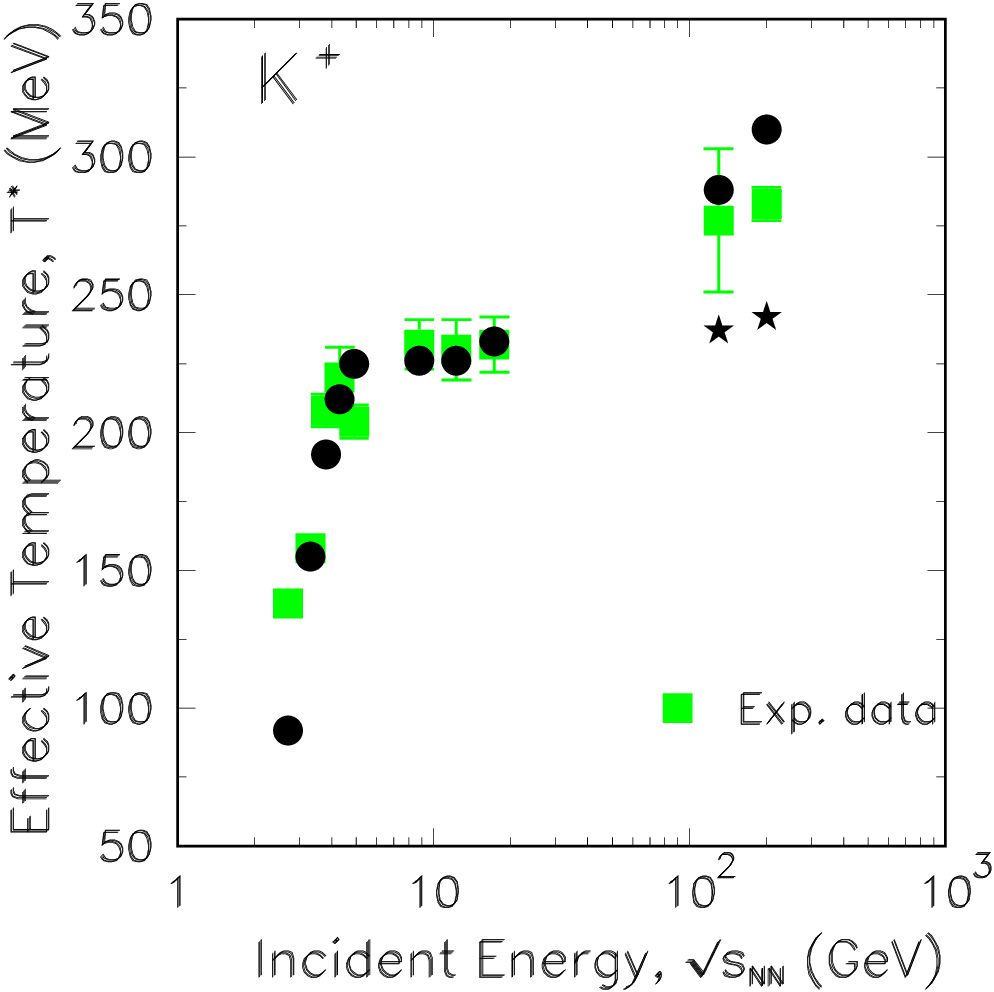} 
\end{center}
\medskip 
\vspace*{-.5cm}
\caption{Energy dependence of the inverse slope 
parameter for $K^+$ in central Pb + Pb (Au + Au) 
collisions. Here, averaged IC were used. Stars 
correspond to $T_{fo}$=155~MeV (see Table~\ref{table1}). 
Data are from~\cite{k_slope}}. 
\label{teff} 
\vspace*{-.5cm}
\end{figure} 

\begin{table}[!hbt] 
\caption{\label{table1}Freezeout temperature 
$T_{fo}\,$, averaged transverse velocity $\bar{v}_T$ in 
$-0.5\le y\le 0.5$ and the inverse slope parameter 
$T^*$ for $K^+$ production in central Pb + Pb (Au + Au) 
collisions, obtained using SPheRIO with averaged NeXuS 
IC. $T_0$ and $\varepsilon_0$ are the initial values at 
the midpoint of the fluid.} 
\medskip
\begin{tabular}{|c|c|c|c|c|c|}
\hline
$\sqrt{s}$&$T_0$&$\varepsilon_0$&$T_{fo}$&$\bar{v}_T$&
$T^*$\\ 
(A$\cdot$GeV)&(MeV)&(GeV/fm$^3$)&(MeV)&  & (MeV) \\
\hline\hline
2.7   &   98  &   0.75  &   85  &  0.067 &   92  \\
\hline
3.3   &  128  &   0.66  &   94  &  0.28  &  155  \\
\hline
3.8   &  131  &   1.01  &   97  &  0.41  &  192  \\
\hline
4.3   &  135  &   1.38  &  115  &  0.37  &  212  \\
\hline
4.9   &  140  &   1.55  &  120  &  0.39  &  225  \\ 
\hline
8.8   &  198  &   4.06  &  144  &  0.31  &  226  \\ 
\hline
12.3  &  248  &   9.04  &  147  &  0.32  &  226  \\ 
\hline
17.3  &  265  &  11.37  &  148  &  0.33  &  233  \\ 
\hline
130	&  281  &  13.22	&  128  &  0.54  &  288  \\ 
\cline{4-6} 
      &       &         &  155  &  0.35  &  237  \\ 
\hline 
200	&  288  &  14.54	&  125  &  0.57  &  310  \\ 
\cline{4-6}
      &       &         &  155  &  0.37  &  242  \\ 
\hline
\end{tabular}
\end{table}

\ni observed anomalous 
behavior of the inverse slope parameter $T^*$ of kaon 
transverse-momentum spectra in central Pb + Pb (Au + Au) collisions~\cite{mg2}, we could 
succeed to obtain the increase in $T^*$ when going 
from SPS to RHIC domain only with decreasing 
freezeout temperature $T_{f.o.}(s)$ with increasing 
$\sqrt{s}$ as shown in Fig. \ref{teff} and Table 
\ref{table1}. Here, SPheRIO code has been used with 
averaged IC. This is because, as will be shown in 
Sec.~\ref{application}, Figs.~\ref{mt_sps}, 
\ref{mt1_sps}, \ref{mt2_sps}, \ref{rhic_pt}, the slope 
parameter is not sensitive to IC fluctuations. 

\subsection{Continuous emission} 
 \label{cem} 

The decoupling temperature discussed in the preceding 
paragraphs should actually be interpreted as an 
average value of such a temperature. When applied 
to finite-volume and finite-lifetime systems as those 
formed in high-energy heavy-ion collisions, a sharply 
defined freeze-out hypersurface is clearly too 
idealized and for a more precise analysis of data, 
one needs a more elaborate description of the 
decoupling process. In~\cite{ce}, we introduced a 
method that we call {\it Continuous Emission} (CE) 
and, as compared to the usual Cooper-Frye one, we 
believe closer to what happens in the actual 
collisions. The essential point is the introduction 
of momentum-dependent escape probability 
\begin{equation} 
 {\cal P}(x,p)
 =\exp\left[-\int_\tau^\infty\rho\sigma v d\tau'\right] 
 \label{P} 
\end{equation} 
of a particle from a space-time point $x$ without 
collision in the medium, so that the emission may occur 
from any point of the fluid and at any time, according 
to this probability. This means that we are 
interpreting probabilistically the Landau condition, 
(\ref{Land_cond}), as should be and also giving the 
system size $L$ a more precise meaning, namely, the 
quantity of matter the escaping particle encounters in 
his trajectory. The integral above is evaluated in the 
proper frame of the particle. Then, the distribution 
function $f(x,k)$ of the expanding system has two 
components, one representing the portion of the fluid 
already free and another corresponding to the part 
still interacting, {\it i.e.}, 
\begin{equation}
f(x,p)=f_{free}(x,p)+f_{int}(x,p)\,. 
\end{equation} 
We may write the free portion as  
\begin{equation}
f_{free}(x,p)={\cal P}f(x,p)\;. 
\label{ffree} 
\end{equation} 
The inclusive one-particle distribution, for instance, 
is then written as 
\begin{eqnarray} 
 E\frac{d^3N}{dp^3} 
 &=&\int_{\sigma_0}d\sigma_{\mu}p^{\mu}f_{free}(x_0,p) 
 \nonumber \\ 
 &+&\int d^4x\,\partial_\mu[p^{\mu}f_{free}(x,p)], 
 \label{idce} 
\end{eqnarray} 
where the surface term corresponds to particles already 
free at the initial time. 
\smallskip

As particles can be emitted in different stages of the 
fluid expansion, it is natural that the appearance of 
the observables in CE becomes different from that of 
the usual Cooper-Frye prescription at a constant 
temperature~\cite{C-F}. In general, in CE, the 
large-$p_T$ particles are mainly emitted at early times 
when the fluid is hot and mostly from its surface, 
whereas the small-$p_T$ components are emitted later 
when the fluid is cooler and from a larger spatial 
domain. We shall discuss, in Sec.~\ref{application}, 
the manifestation of this effect in two-pion 
interferometry. 
\smallskip

More detailed description of this method and several 
of its predictions are accounted for by Grassi in 
a separate paper of this issue \cite{grassi}. 

\section{Applications}\label{application} 

In the following, we shall present some applications 
of NeXuS+SPheRIO code~\cite{ebe1,ebe2,fic-hbt}, which 
has been constructed by coupling the NeXuS event 
generator to SPheRIO code, described in 
Subsection~\ref{smooth}.
\smallskip

As mentioned in the Introduction, the main ingredients 
of hydrodynamic calculations are IC, EoS and the 
decoupling procedure. None of these are well known at 
the present moment. 
\smallskip

Since we are mainly concerned with the effects of IC 
fluctuations and consequences of different decoupling 
descriptions, here we just take the IC created by NeXuS 
event generator and use the equations of state 
described in Section~\ref{practicalEOS}. The 
strangeness conservation has not been taken into 
account. As for the decoupling procedure, we adopted 
both the conventional sharp freeze-out prescription and 
the continuous emission description, because one of our 
purpose is to see the differences resulting from these 
two descriptions. 
\smallskip

We do not expect that these options will reproduce all 
the experimental data. In fact, we found that the use 
of the same version of NeXuS code for high- and 
low-energy nuclear collisions caused some discrepancies 
in reproducing the rapidity distribution of charged 
particle. Therefore, we have introduced  additional 
parameters to ajust at least the overall rapidity 
distributions.   

\subsection {Effects of fluctuating initial conditions}  

In Sec.~\ref{IC}, we stressed that, due to the finite 
size of\hfilneg\  

\begin{figure}[!htb]
\hspace*{-.cm} 
\begin{center}
\includegraphics*[width=8.cm]{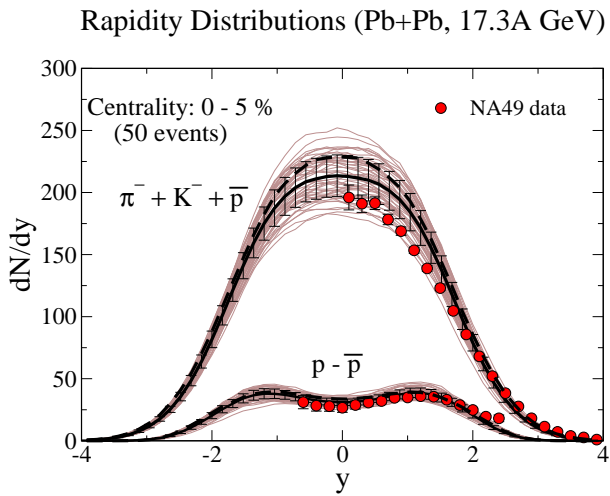} 
\end{center}
\caption{Rapidity distributions as indicated, with 
 $T_{fo}=140\,$MeV. The solid lines represent the 
 averages over the fluctuating distributions (with 
 dispersions), whereas the dashed lines are results 
 with the averaged initial conditions. The data are 
 from NA49$\,$\cite{na49}.} 
\label{y_sps}
\end{figure} 

\begin{figure}[!htb]
\center{\includegraphics*[width=8.cm]{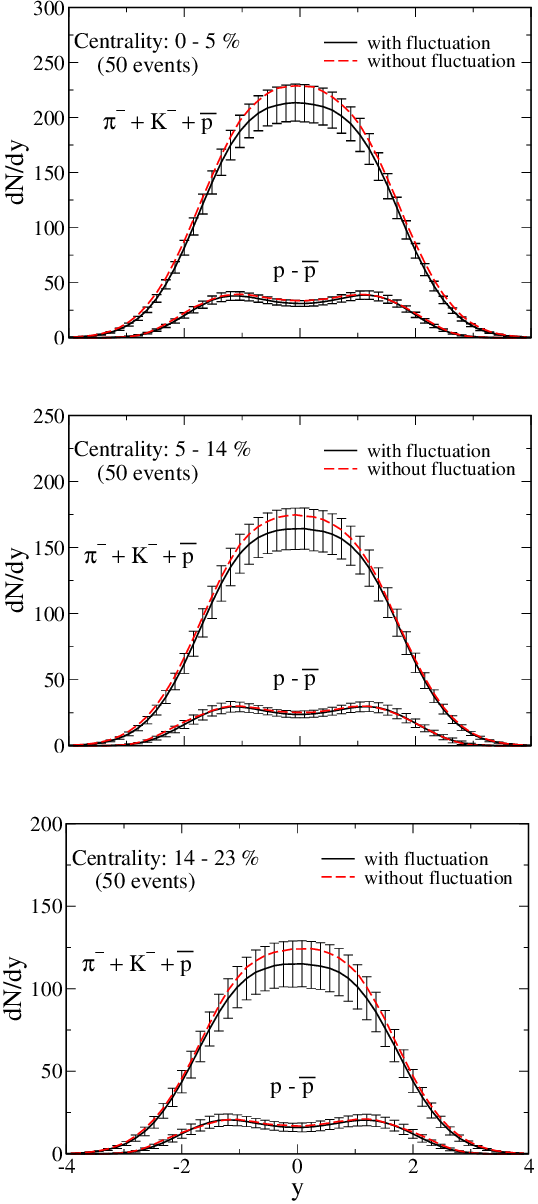}} 
\caption{Centrality dependence of rapidity 
 distributions, for Pb~+~Pb collisions 
 at $\sqrt{s}=17.3\,A\,$GeV (SPS). Here, ``with 
 fluctuation'' (the solid lines) means results with 
 NeXuS fluctuating IC and ``without fluctuation'' 
 (dashed lines) those with the averaged IC.} 
\label{y1_sps} 
\end{figure} 

\ni the systems, large fluctuations are expected in 
the initial stage of actual nuclear collisions, even 
for a fixed impact parameter, and that IC generated by 
realistic event generators do show such effects 
\cite{gyulassy,nexus}. Let us see in this Section what 
are the effects of fluctuating IC on some of the 
observables. 

\begin{figure}[!htb]
\center{\includegraphics*[width=8.cm]{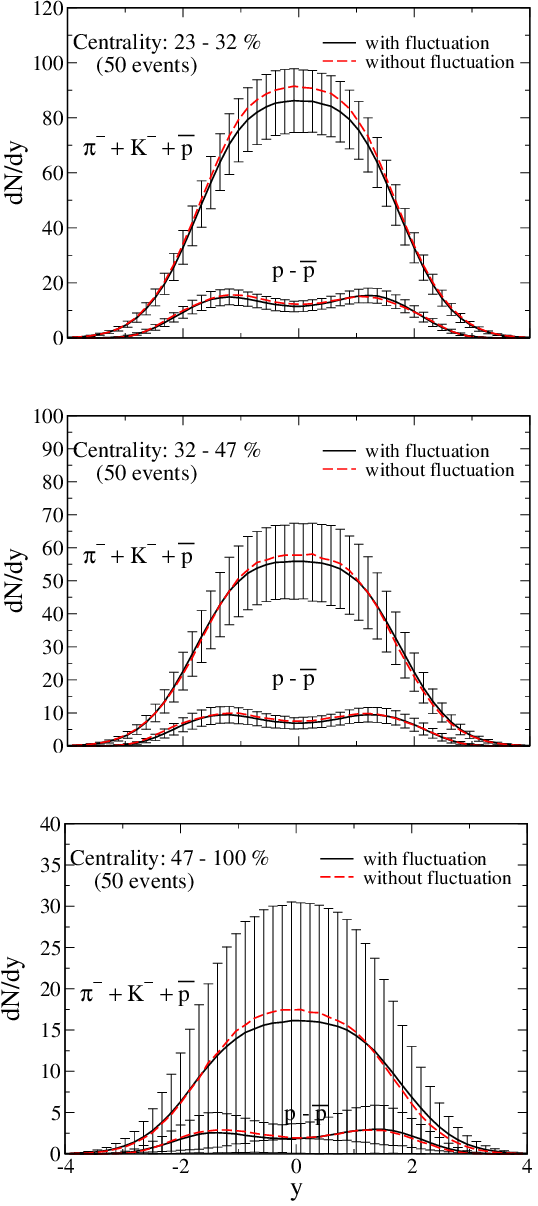}} 
\caption{The same as the previous Figure, for less 
 central events.} 
\label{y2_sps}
\end{figure} 

In \cite{gyulassy}, by solving the hydrodynamic 
equations with longitudinal boost-invariance, it has 
been shown that the bumpy IC {\it i}) develop 
azimuthally asymmetric flows, even for central 
collisions, because there is no symmetry in each 
fluctuating event; and also {\it ii}) enhance 
high-$p_T$ direct photon yields due to the high 
temperature in the blobs. 
\smallskip 

Since our interest here is just to show the effects 
of fluctuating IC, we chose the simplest decoupling 
criterion, namely the usual Cooper-Frye sudden 
decoupling with freezeout temperature $T_{fo}$ in this 
Subsection.

\subsubsection{Rapidity and $m_T$ distributions} 
 \label{mt}

Let us first consider Pb~+~Pb collisions at SPS. 
In Fig.~\ref{y_sps}, we show the rapidity distributions 
for negative particles and $p-\bar{p}\,$, respectively, 
for the most central Pb~+~Pb collisions at 
$\sqrt{s}=17.3\,A\,$GeV. Each event, computed from 
randomly generated IC like the one shown in 
Figs.~\ref{fic} and \ref{fic2} (left), is represented 
by a thin curve (for each type of particle, either 
negatives or $p-\bar{p}$). The thick solid lines 
represent the averages over 50 events, with the 
corresponding {\it dispersions}. We compare the average 
distributions with the distributions computed starting 
from the averaged IC like the one shown in 
Figs.~\ref{fic} and \ref{fic2} (right), represented by 
dashed lines here. Here, we took $T_{fo}=140\,$MeV. 
The data points are shown for comparison \cite{na49}. 
\smallskip 

As is seen in this Figure, the rapidity distributions 
show large fluctuations from event to event and, 
although similar, there is a non-negligible difference 
between the average distributions and the ones obtained 
from the average IC, especially in the case of negative 
particles, constituted mostly of pions. For the same 
average initial energy, the multiplicity decreases 
about $6\sim7\%$ if fluctuations exist, confirming what 
we showed in Sec. \ref{IC}. Finally, the data points 
in Fig.~\ref{y_sps} are closer to the averaged 
distribution over fluctuating IC. 

\begin{figure}[!b]
\center{\includegraphics*[width=8.cm] {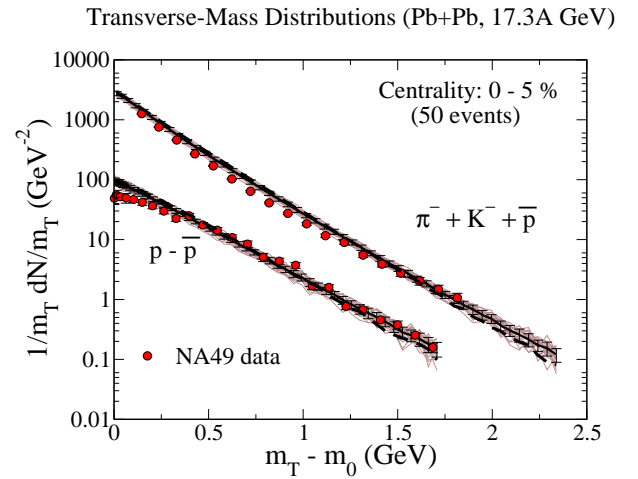}} 
\medskip 
\caption{Transverse mass distributions as indicated, 
 computed at $T_{fo}=140\,$MeV. The solid lines 
 represent the averages over the fluctuating 
 distributions (with dispersions), whereas the dashed 
 lines are results with the averaged initial 
 conditions.  The data are from NA49$\,$\cite{na49}.} 
\label{mt_sps} 
\end{figure} 

What happens with other centralities is similar as 
seen in Figs.~\ref{y1_sps} and \ref{y2_sps}. Some 
differences are: {\it i)} as the collisions become 
more peripheral, naturally the dispersions increase; 
{\it ii)} the differences between the average 
distributions and the ones produced by average IC 
seems to decrease. 

\begin{figure}[!t]
\center{\includegraphics*[width=8.cm]{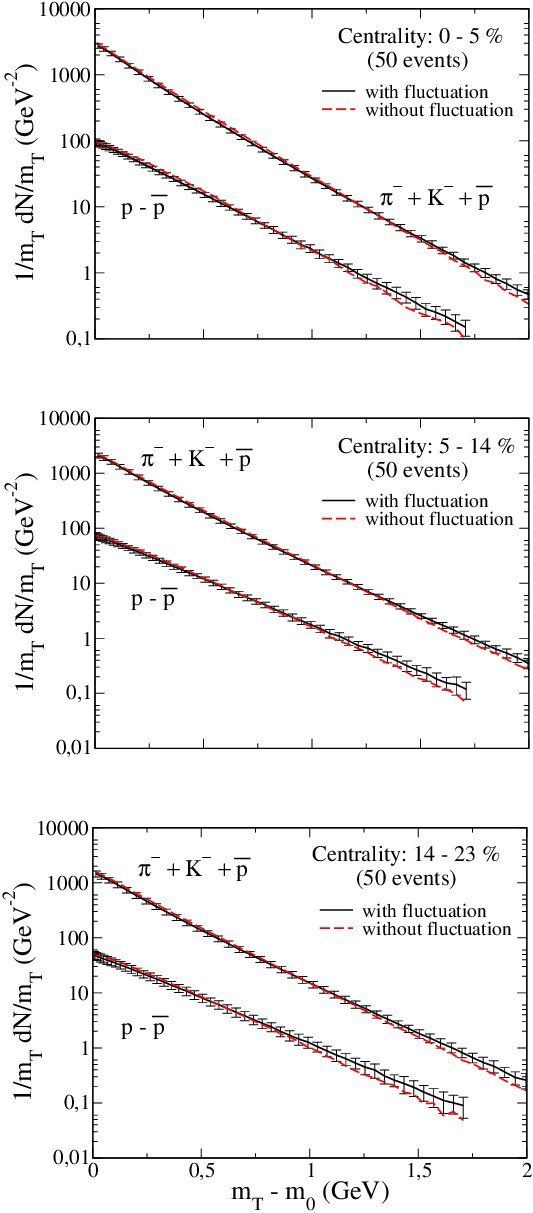}} 
\caption{Centrality dependence of transverse mass 
 distributions, for Pb~+~Pb collisions at 
 $\sqrt{s}=17.3\,A\,$GeV (SPS). Here, ``with 
 fluctuation'' (the solid lines) means results with 
 NeXuS fluctuating IC and ``without fluctuation'' 
 (dashed lines) those with the averaged IC.} 
\label{mt1_sps}
\end{figure} 

\begin{figure}[!t]
\center{\includegraphics*[width=8.cm]{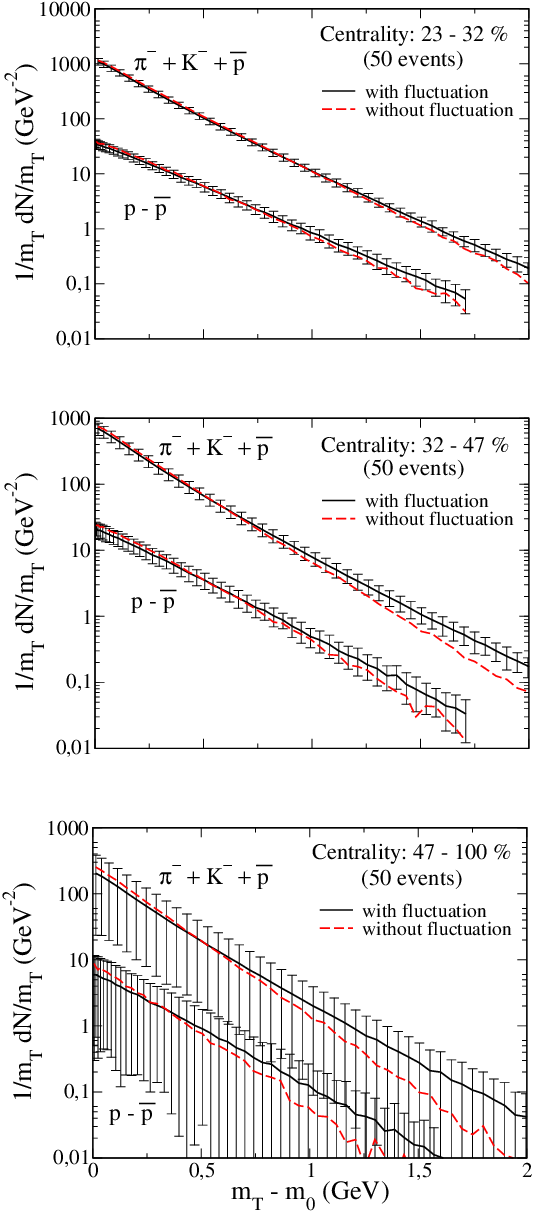}} 
\caption{The same as the previous Figure, for less 
 central events.} 
\label{mt2_sps}
\end{figure} 

\smallskip 
In Fig.~\ref{mt_sps}, we show the corresponding $m_T$ 
distributions, for the most central Pb~+~Pb collisions 
at $\sqrt{s}=17.3\,A\,$GeV. One can see that apparently 
the fluctuation effects on the transverse-momentum 
spectra are very small and the averaged spectra are in 
good agreement with those computed with averaged IC, 
and also with data. This smallness of the fluctuation 
effects on the transverse-momentum spectra is not in 
contradiction with the large fluctuations we saw above 
of rapidity distributions. In part this is due to the 
logarithmic scale used here. One can conclude, however, 
that the IC fluctuations affect very little the 
{\it slope} of the $m_T$ spectra. This conclusion is 
valid for other centralities, as seen in 
Figs.~\ref{mt1_sps} and \ref{mt2_sps}, except for the 
most peripheral case, where the dispersions are very 
large. If we look more carefully, we can perceive a 
small difference between the averaged spectra 
and those computed with the averaged IC, which appears 
in all the centralities, that is the tail of the 
averaged spectra is more concave and the distributions 
become higher, probably because the expansion is more 
violent in this case, due to the high density spots in 
the IC. 

\begin{figure}[!b]
\vspace*{-.6cm} 
\begin{center}
\includegraphics*[angle=-90, width=9cm]{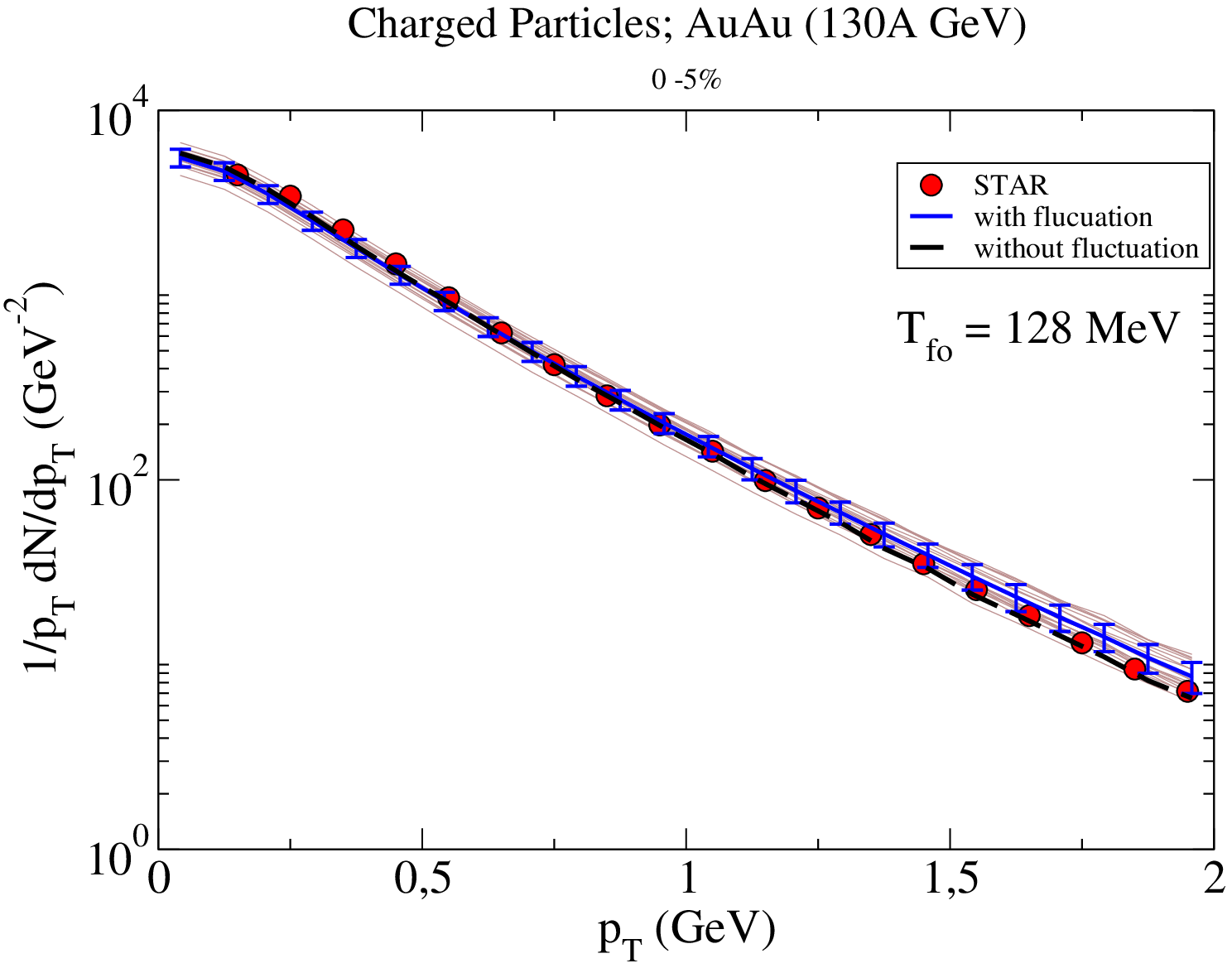}
\end{center}
\caption{$p_T$ distributions for the most central Au+Au 
 at 130A GeV. The data are from STAR \cite{star1}.} 
\label{rhic_pt}
\end{figure} 

At RHIC energies the results are similar. In 
Fig.~\ref{rhic_pt}, we show the $p_T$ distributions 
of charged particles in the most central Au~+~Au 
collisions at 130A GeV, calculated both with NeXuS 
fluctuating IC and with averaged IC. Like in the case 
of Pb~+~Pb collisions at SPS, we used sudden freezeout, 
but with lower freezeout temperature, 
$T_{fo}=128\,$MeV, the same value used in \cite{eff-t} 
to fit the inverse slope parameter of kaons, in the 
same collisions, as shown in Fig.~\ref{teff} and 
Table~\ref{table1} of Section~\ref{decoupling}. As is 
seen, the fluctuation effects are small in $p_T$ 
distribution, both curves agree each other and with 
data, and the average over fluctuating events gives a 
slightly more concave shape, just like in Pb~+~Pb 
collisions at SPS. 
\smallskip

We show in Fig.~\ref{rhic_eta}, the pseudorapidity 
distribution of charged particles in the most central 
Au~+~Au collisions at 130A GeV, calculated with 
NeXuS fluctuating IC and the corresponding averaged IC, 
and compared with data~\cite{phobos}. 
Qualitatively, the behavior is similar to the case of 
Pb~+~Pb collisions at SPS, namely the average $\eta$ 
distribution is close to the distribution with the 
averaged IC, being the latter slightly higher than the 
former. The difference here is smaller than at the 
lower energy case. Probably this is due to the decrease 
of $\Delta E_i/\!<E>$ in Eq.~(\ref{entrop}) as the incident energy increases.
\begin{figure}[!htb]
\vspace*{-1.cm} 
\begin{center}
\includegraphics*[width=8.3cm]{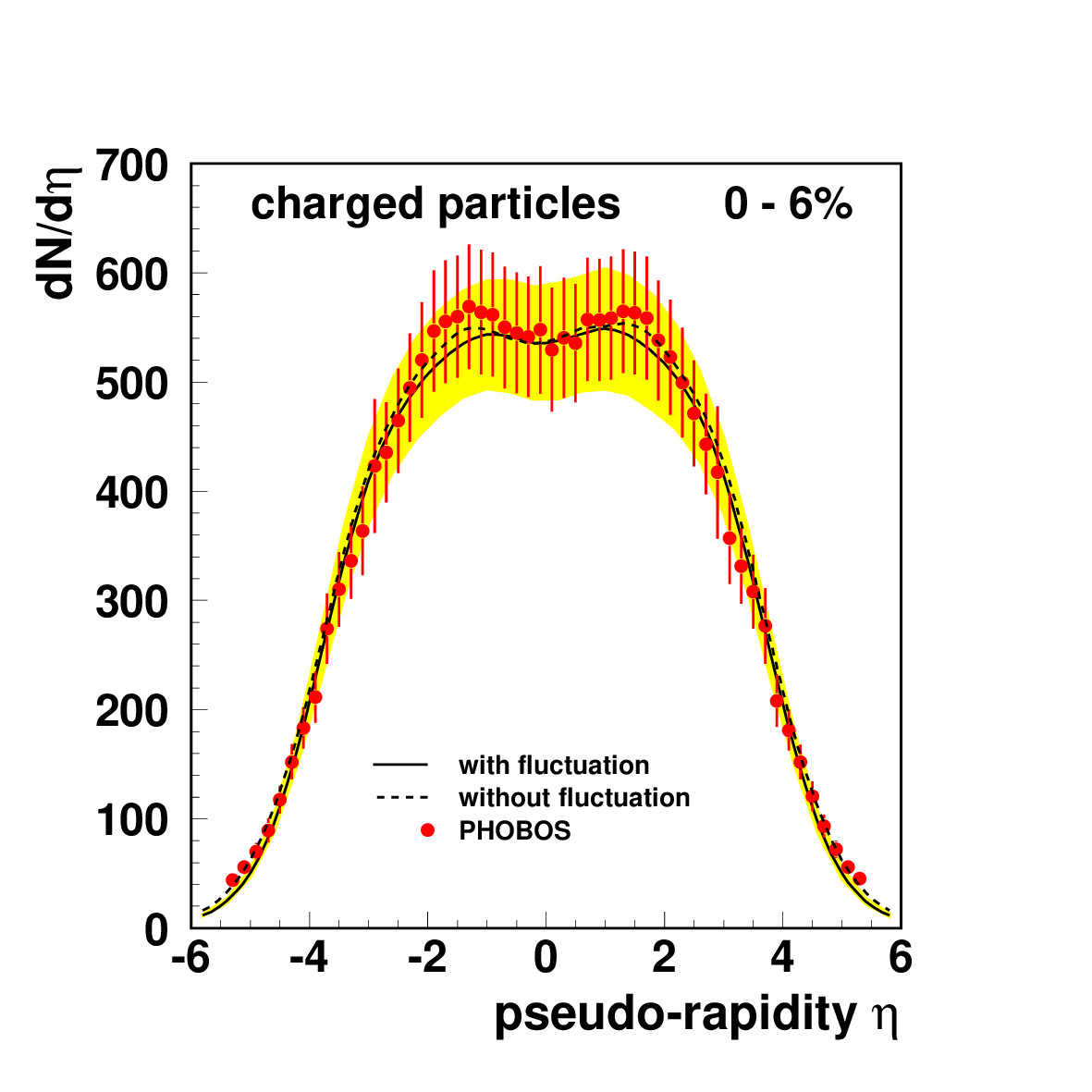}
\end{center}
\vspace*{-.5cm}
\caption{Charged-particle pseudo-rapidity 
 distributions for the most central Au+Au at 130A GeV. 
 The data are from PHOBOS~\cite{phobos}.} 
\label{rhic_eta}
\end{figure} 



\subsubsection{Elliptic-flow parameter $v_2$}
 \label{v2} 

Let us turn to the elliptic-flow parameter $v_2\,$, 
defined as the second Fourier coefficient of the 
azimuthal distribution~\cite{voloshin}  
\beq 
\frac{dN}{d\phi}\propto
\left(1+2\sum_n v_n \cos\,[n(\phi-\psi)]\right)\,.  
\eeq 
Thus, 
\beq 
v_2=<\cos\,[2(\phi-\psi)]>\,, 
\eeq 
where the bracket denotes the average value and $\psi$ 
gives the event-dependent collision plane. 
\smallskip

Usually, $v_2$ parameter is interpreted as indicating 
a flow asymmetry caused by the initial-condition 
asymmetry associated with the non-zero impact 
parameter. In this case, as the produced matter in 
the collision is likely to be flattened in the 
impact-parameter direction, the pressure gradient would 
be larger in this direction and so would be the flow.  
However, as mentioned at the top of this section, it was 
shown in Ref.~\cite{gyulassy} that, even for central 
collisions, fluctuating IC develop azimuthally 
asymmetric flows, because in this case there is no 
symmetry in each event and, experimentally, the impact 
parameter cannot be determined unambiguously. So, one 
expects larger $v_2$ for fluctuating IC as compared to 
averaged IC case. Remark that the fluctuation we are 
talking about is not the often 
discussed~\cite{voloshin,ollitrault} finite-multiplicity  
fluctuation at the end of the process. 
In Fig.~\ref{v2}, results of NeXuS+SPheRIO for 
$v_2$ of pions produced in Pb+Pb collisions at 
17.3A GeV are shown, both with fluctuating IC and 
averaged IC and compared with data~\cite{na49}. Large 
discrepancy is seen between the two ways of calculating  
this parameter, and the data are closer to $<v_2>$, 
calculated with fluctuating IC. 

\begin{figure}[!t] 
\begin{center}
\includegraphics*[angle=-90, width=8cm]{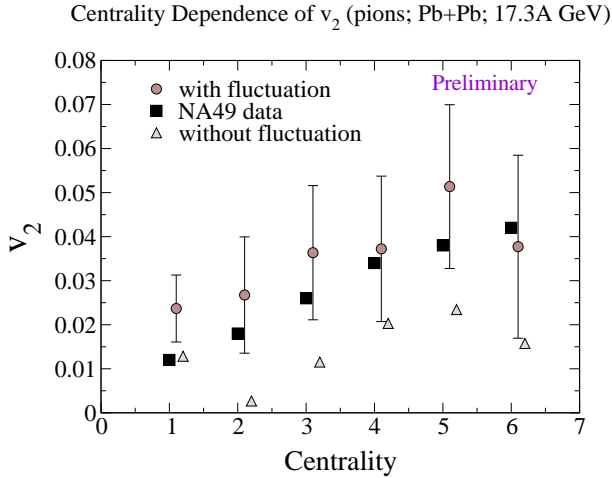}
\end{center}
\caption{Centrality dependence ov $v_2$ for pions for 
 Pb+Pb collisions at SPS. The data are from NA49  \cite{na49}.} 
\label{v2}
\end{figure} 

\subsubsection{Two-pion interferometry}\label{hbt} 

As is well known, the identical-particle correlation, 
also known as Hanbury-Brown-Twiss effect (HBT 
effect)~\cite{hbt1} is a powerful tool for probing 
geometrical sizes of the space-time region from which 
they were emitted. If the source is static like a star, 
it is directly related to the spatial dimensions of the 
particle emission source. When applied to a dynamical 
source, however, several non-trivial effects 
appear~\cite{hama,pratt}, reflecting its time evolution 
as happens in high-energy heavy-ion collisions. Being 
so, the inclusion of IC fluctuations may affect 
considerably the so-called HBT radii, because, as 
discussed in Sec.~\ref{IC}, the IC in the 
event-by-event base often show small high-density spots 
in the energy distribution, and our expectation is 
that such spots manifest themselves at the end when 
particles are emitted, giving smaller HBT radii. 
\smallskip

We shall discuss here only the recent application of 
NeXuS+SPheRIO code for HBT effect with IC 
fluctuations, for Au+Au collisions at 130A$\,$GeV  
\cite{fic-hbt}. A more detailed account of two-particle 
correlations is given by Padula~\cite{sandra} in a 
separate paper in this issue. 
\smallskip

For studying the fluctuation effects of IC on HBT 
correlation, first we assume Cooper-Frye sudden 
freezeout (FO) at $T_{f.o.}=128\,$MeV. This freezeout 
temperature is the same one previously found by 
studying the energy dependence of kaon slope parameter 
$T^*$~\cite{eff-t}, discussed in 
Subsection~\ref{finite_size} and appears in 
Table~\ref{table1}. We showed in Subsection~\ref{mt} 
that indeed this choice of $T_{f.o.}$ reproduces the 
$p_T$-distribution data at 130A$\,$GeV, both with 
averaged and fluctuating IC. We also neglect 
the resonance decays. It is argued~\cite{csorgo} that, 
since resonance decays contribute to the correlations 
with very small $q$ values 
($q\lower.1cm\hbox{$\,\buildrel<\over\sim\,$}q_{min}\,$, 
where $q_{min}$ is the minimum measureable $q$), the 
experimentally determined HBT radii are essentially due 
to the direct pions. Then the two-particle correlation 
function is expressed in terms of the distribution 
function $f(x,p)$ as 
\begin{equation} 
C_2(q,P)=1+\frac{|I(q,P)|^2}{I(0,p_1)I(0,p_2)} \label{c_hbt}
\end{equation} 
where $P=(p_1+p_2)/2$ and $q=(p_1-p_2)$ and $p_i$ is 
the momentum of the $i$th pion. Usually 
\begin{equation} 
 I(q,P)\equiv\langle a_{p_1}^+a_{p_2}\rangle
 =\int_{T_{f.o.}}d\sigma_\mu P^\mu f(x,P)e^{iqx}. 
\label{I} 
\end{equation} 
In SPH representation, we write $I(q,P)$ as 
\begin{equation} 
I(q,P)=\sum_j 
\frac{\nu_j n_{j\mu} P^\mu}{s_j |n_{j\mu} u_j^\mu|}\;
{\mathrm{e}}^{iq_\mu x_j^\mu}f(u_{j\mu} P^\mu)\,,
\label{corr_sph} 
\end{equation} 
where the sumation is over all the SPH particles. In the 
Cooper-Frye freezeout, these particles should be taken 
where they cross the hyper-surface $T=T_{f.o.}$ and 
$n_{j\mu}$ is the normal to this hyper-surface. 
Notice that, if we put $p_1=p_2=P$, so that $q=0$, 
Eqs.~\ref{I} and \ref{corr_sph} are reduced, 
respectively, to Eqs.~\ref{CFf} and \ref{CF-SPH}, that 
is, to the inclusive one-particle distribution. 
\smallskip 

\begin{figure}[!b] 
\vspace*{8.5cm}
\includegraphics{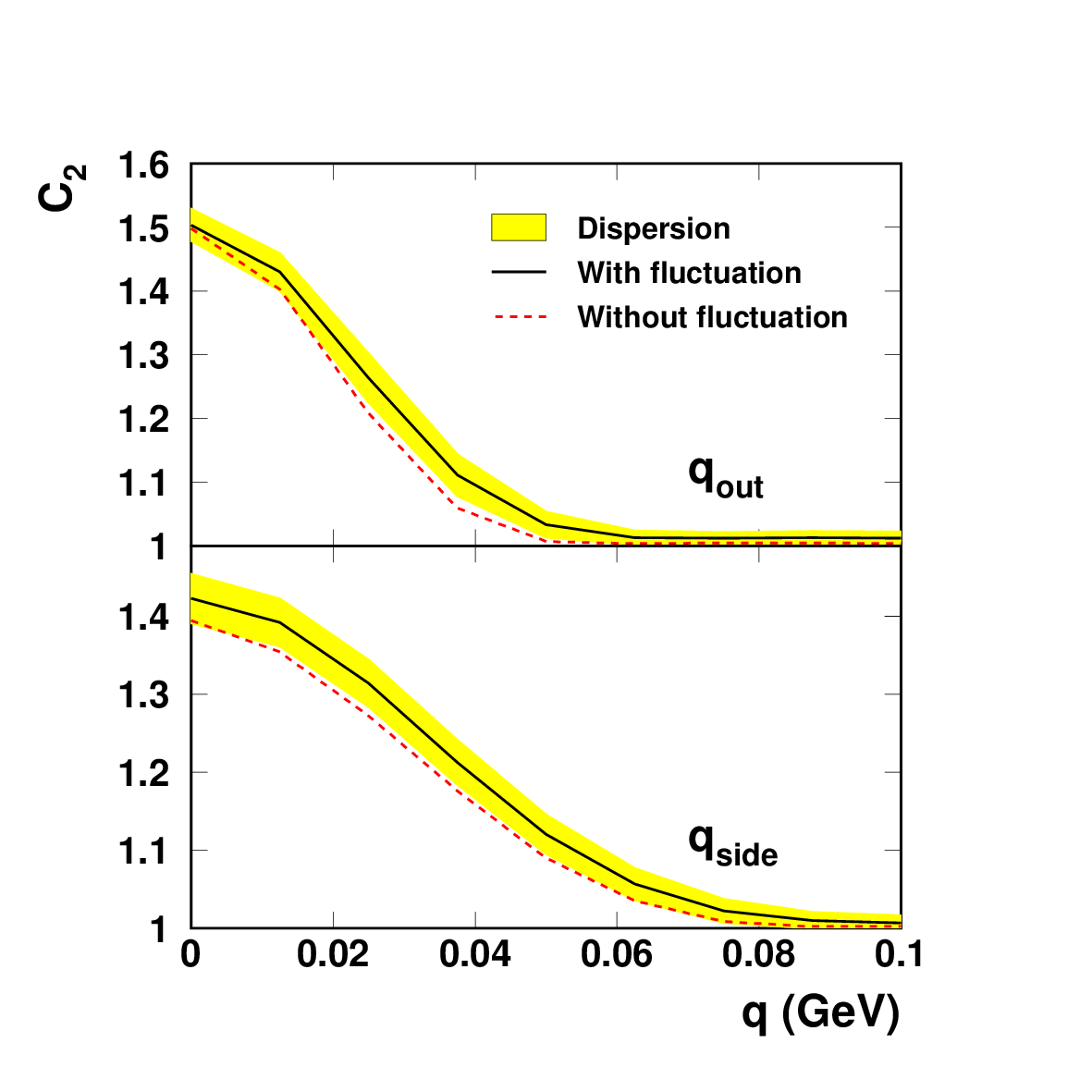} 
\caption{\small Correlation functions from 
fluctuating IC and averaged IC. Sudden freeze-out is 
used here. The rapidity range is $-0.5\leq Y\leq 0.5$ 
and $q_{o,s,l}$ which do not appear in the horizontal 
axis are integrated over $0\leq q_{o,s,l}\leq 35$MeV.} 
\label{hbt-fic}
\end{figure} 

In Fig.~\ref{hbt-fic}, we compare the correlation 
function $C_2$ averaged over 15 fluctuating events with 
those computed starting from the averaged IC (so, 
without fluctuations). One can see that the IC 
fluctuations are reflected in large fluctuations also 
in the HBT correlations. When averaged, the resulting 
correlation $<C_2>$ are broader than those computed 
with averaged IC, so giving smaller radii as expected. 
Also the shape of the correlation functions changes. 
\smallskip

We plot the $m_T$ dependence of HBT radii, with 
Gaussian fit of $C_2\,$, in Fig.~\ref{radii}, together 
with RHIC data~\cite{star2,phenix} and results with 
CE, which will be discussed in Sec. \ref{hbt-ce}. It is 
seen that the smooth IC with sudden FO makes the $m_T$ 
dependence of $R_o\,$ flat or even increasing, which is 
in agreement with other hydro calculations~\cite{morita} 
but in conflict with the data. The fluctuating IC make 
the radii smaller, especially in the case of $R_o\,$, 
however without changing the $m_T$-dependence. 

\begin{figure}[!htb]
\vspace*{13.8cm}
\begin{center}
\includegraphics{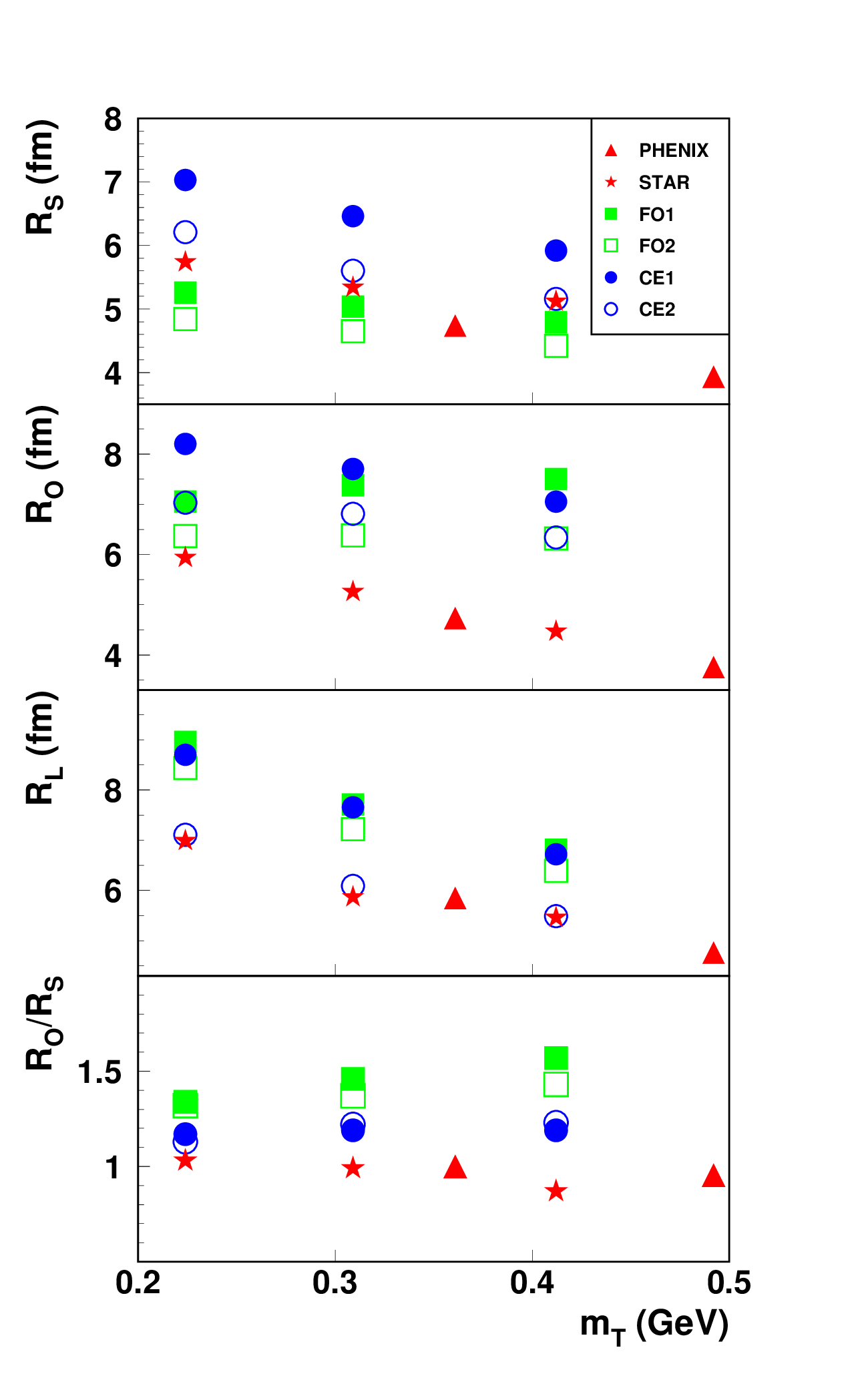} 
\end{center}
\vspace{-.4cm} 
\caption{\small HBT radii and the ratio $R_o/R_s$ 
for sudden freeze-out (FO) and CE. 1 stands for averaged 
IC and 2 fluctuating IC. Data are 
from~\cite{star2,phenix}.} 
\label{radii} 
\end{figure} 

\subsection{Effects of continuous emission} 
 \label{ce}  

As discussed in Sec.~\ref{decoupling}, it is more 
likely that the decoupling occurs not suddenly, but 
continuously from any point of the fluid and at any 
instant of time, according to some escaping probability 
${\cal P}(x,p)$ given by Eq.~(\ref{P}). There are 
several nice predictions of the {\it Continuous 
Emission Model} (CEM) as discussed in~\cite{grassi}. 
However, although more realistic, this description is 
not handy because, ${\cal P}$ depends on the momentum 
of the escaping particle and, moreover, on the future 
of the fluid as seen in Eq.~(\ref{P}). In order to make 
the computation practicable, in~\cite{fic-hbt} we first 
took ${\cal P}$ on the average, {\it i.e.}, 
\[
{\cal P}(x,p)\Rightarrow<{\cal P}(x,p)>
\equiv{\cal P}(x)\,.
\] 
Then, approximated linearly the density 
\[
\rho(x^\prime)=\alpha s(x^\prime) 
\] 
(where, for example in the ideal massless pion gas 
case, $\alpha=(45\,\zeta(3))/2\pi^4=.278$) in the 
integral of Eq.~(\ref{P}). Thus, 
\begin{equation} 
{\cal P}(x,p) \Rightarrow {\cal P}(x) 
=\exp\left(-\kappa\frac{s^2}{|ds/d\tau|}\right)\,,
\label{prob} 
\end{equation} 
where $<\sigma v>$ has been included in 
$\kappa=0.5\,\alpha<\sigma v>$. 
Although approximately, now we can compute ${\cal P}$ 
in each space-time point. 
\smallskip

In Fig.\ref{probability}, we show the time evolution 
of the escaping probability ${\cal P}(x)$ given by 
Eq.(\ref{prob}) in the mid-rapidity plane for the 
most central Au+Au collisions at 130$\,$A GeV. The 
IC have been computed at $\tau=1\,$fm and averaged 
over 30 NeXuS events. Here and in the computations 
of observables below, the parameter $\kappa$ has 
been estimated to be .3, corresponding to 
$<\sigma v>=2\,$fm$^2$ in the zero temperature limit 
and some 20\% larger at $T=m_\pi\,$. As seen, the 
probability remains $0.1<{\cal P}(x)<0.8$ in a 
quite large domain, especially for $\tau>10\,$fm, 
indicating that both the emission zone and duration 
are expected to be large, in opposition to the 
standard sudden freezeout case. For comparison, we 
show also the temperature distribution in 
Fig.\ref{probability}, with some isotherms. 
\smallskip 

To calculating the spectra, now Eq.~(\ref{idce}) 
is translated into SPH language, which is given by 
Eq.~(\ref{CF-SPH}), but where the sum is computed in 
the present case not over $T=T_{f.o.}$ hypersurface 
but picking out SPH particles according to the 
probability ${\cal P}$, given by Eq.~(\ref{prob}) 
with the normal $n_{j\mu}$ pointing to the 4-gradient 
of ${\cal P}$. Since our procedure favors emission 
from fast outgoing SPH ``particles'', because $\rho$ 
decreases faster there and so does $s$ in this case 
making ${\cal P}$ larger, we believe that the main 
feature of CEM is preserved in our approximation. 

\begin{figure}[!htb] 
\vspace*{-3.cm}
\begin{center}
\includegraphics*[width=8.cm]{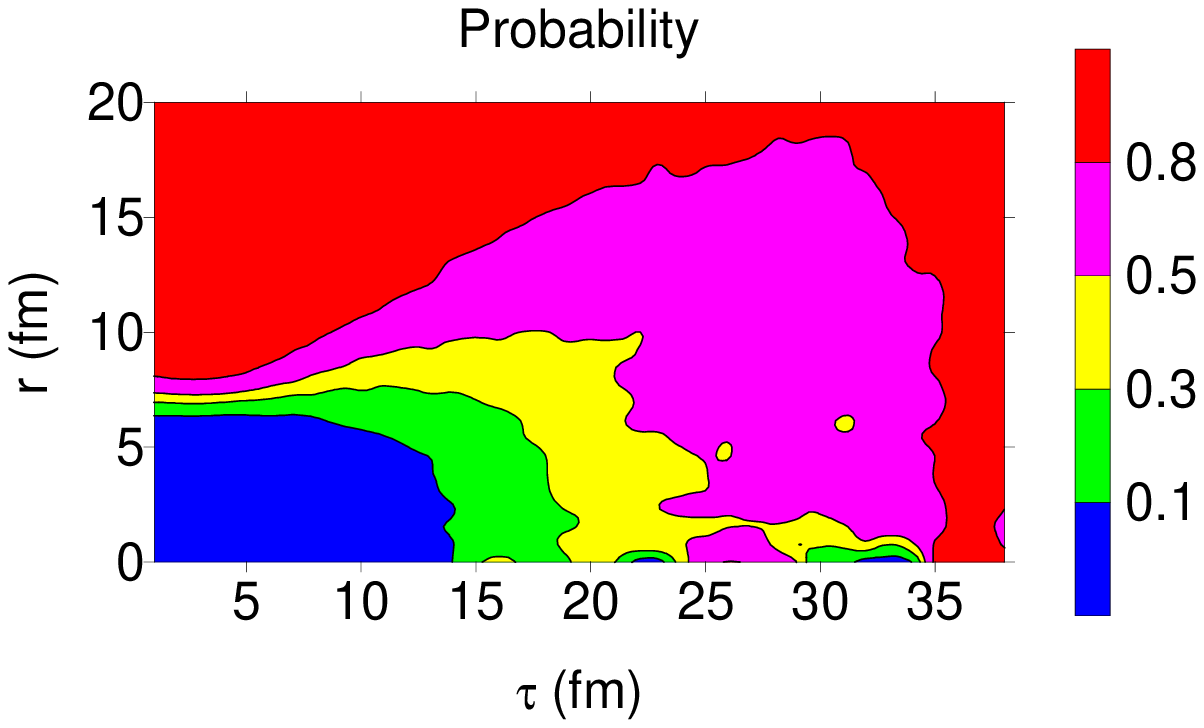} 
\end{center}
\vspace*{-7.cm}
\begin{center}
\includegraphics*[width=8.cm]{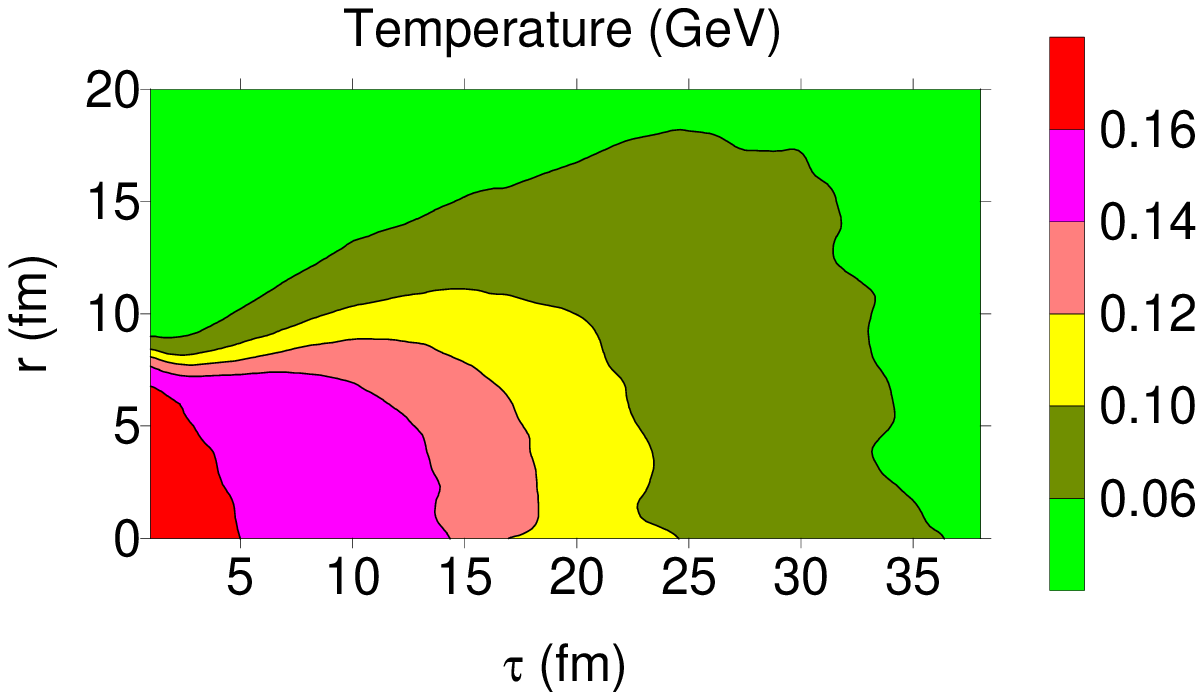} 
\end{center} 
\vspace*{-3.8cm}
\caption{\small Upper panel: Probability distribution 
 as given by eq.(\ref{prob}) for the most central 
 Au+Au collisions at 130$\,$A GeV, in the mid-rapidity 
 plane for averaged IC. Lower panel: Corresponding 
 temperature distribution.} 
\label{probability} 
\end{figure} 

\smallskip

\subsubsection{$m_T$ distributions} 
 \label{mt-ce}

\begin{figure}[!b] 
\vspace*{6.5cm}
\includegraphics{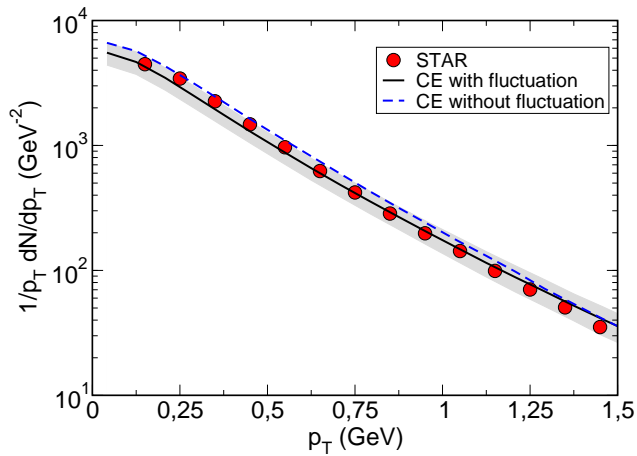}  
\vspace*{-.2cm}
\caption{Charged-particle $p_T$ distributions in CEM 
 for the most central Au~+~Au at 130A GeV. The data 
 are from STAR~\cite{star1}.} 
\label{rhic_pt_ce}
\end{figure}

In Fig.~\ref{rhic_pt_ce}, we show the charged $m_T$ 
distribution at mid-rapidity in the most central 
Au~+~Au collisions at 130A GeV, computed by using 
CEM, with and without fluctuations. Compare this 
figure with Fig.~\ref{rhic_pt}, where the same data 
are compared with calculations using sudden 
freezeout. One sees that, although the emission 
mechanisms are quite different, the results are 
similar, for the choice of the parameters, $T_{fo}$ in 
one case and $\kappa$ in the other. However, the 
origins of the spectrum shape are different. Whereas, 
in the freezeout case, all the particles are emitted 
at the same temperature $T_{fo}$ and the concave shape 
of the spectra is due to different transverse 
velocities of the fluid at different instants of 
time, in the continuous emission case, the large-$p_T$ 
particles are in general emitted earlier and at higher 
temperature, the small-$p_T$ particles are emitted 
later, when the fluid is cooler. 

\subsubsection{Two-pion interferometry} 
 \label{hbt-ce}  

Let us now consider the effect of continuous emission 
on HBT effect. Since HBT is sensitive to the 
space-time geometry of the fluid, and CEM produces 
important modification of the emission zone, we expect 
considerable changes in the so-called HBT radii. 
As mentioned above, according to this picture, the 
large-$p_T$ particles are mainly emitted at early times 
when the fluid is hot and mostly from its surface, 
whereas the small-$p_T$ components are emitted later 
when the fluid is cooler and from larger spatial 
domain. 
\smallskip

In \cite{ce-hbt}, we considered this effect, using 
the boost-invariant solution~\cite{scale-invariant}, 
and showed that whereas the so-called {\it side} radius 
is independent of the average $p_T\,$, the {\it out} 
radius decreases with $<p_T>$, because of the reason 
mentioned above. This behavior is expected to 
essentially remain in the general 3-dimensional 
expansion, described by SPheRIO code. 
\smallskip

To compute the correlation function $C_2(q,P)$ in CEM, 
we first rewrite the integral~(\ref{I}) as 
\bea 
 I(q,P)
 &=&\int_{\sigma_0}d\sigma_{\mu}P^{\mu}f_{free}(x_0,P) 
 e^{{iqx}} \nonumber \\ 
 &+&\int d^4x\,\partial_\mu[P^{\mu}f_{free}(x,P)] 
 e^{{iqx}}, 
 \label{Ice}  
\eea 
which is similar to Eq.~(\ref{idce}). Then, translate 
it into SPH language by using Eq.~(\ref{corr_sph}), but 
with the sum evaluated by picking out SPH particles 
according to the probability ${\cal P}$, given by 
Eq.~(\ref{prob}) with the normal $n_{j\mu}$ pointing to 
the 4-gradient of ${\cal P}$, exactly in the same way as 
done in calculating the inclusive spectrum. 
\smallskip

We plot in Fig.~\ref{radii}, some results for the $m_T$ 
dependence of HBT radii computed in this way, both with 
fluctuating IC and averaged IC, for Au~+~Au collisions 
at 130A$\,$GeV. For comparison, we show also the results 
with sudden freezeout and data~\cite{star2,phenix}. 
Comparing the averaged IC case (with CEM (CE1)), with 
the corresponding freezeout (FO1), one sees that, while 
$R_L$ remains essentially the same, $R_s$ decreases 
faster and as for $R_o\,$, it decreases now inverting 
its $m_T$ behavior. 
\smallskip

In Fig.~\ref{radii}, one can also see the combined 
effect of fluctuating initial conditions together with 
continuous emission (curve CE2). All the radii are 
smaller than CE1, with averaged IC (but using CEM), as 
happend in the sudden freezeout case. The agreement with 
data is excelent both for $R_L$ and $R_s\,$, and 
improved considerably the results for $R_o$ with 
respect to the usual hydrodynamic description. 


\section{Conclusions and Outlook} 
 \label{conclusion} 

In this survey, we discussed on several aspects of 
applications of hydrodynamic model to nucleus-nucleus 
collisons, giving especial emphases on i) method of 
solving the hydrodynamic equations for arbitrary 
configurations; ii) accounting for the probable 
event-by-event fluctuations of the initial conditions; 
and iii) the decoupling criteria for obtaining the 
observables. 
\smallskip

The Smoothed-Particle Hydrodynamic approach, using a 
special hyperbolic coordinate system, is particularly 
interesting tool for describing rapidly expandig matter 
as that in high-energy nucleus-nucleus collisons, 
because of its efficiency and flexibility, as 
demonstrated through examples considered in 
Sec.~\ref{Landau}, \ref{trans} and through 
applications, shown in Sec.~\ref{application}, where 
the systems, in general, do not present any symmetry. 
\smallskip

The initial conditions for hydrodynamic expansion of 
matter formed in relativistic heavy-ion collisions are 
likely neither smooth nor symmetrical, because 
of small size of these systems. This property has also 
been shown by some event generators, which take the 
microscopic dynamics into account. The fluctuations in 
the initial conditions may produce large discrepancies 
in the computed results for some observables, in 
comparison with those obtained with the usual averaged, 
smooth and symmetrical initial conditions. So, it is 
our opinion that, to extracting correct information 
from certain experimental data, the inclusion of the 
fluctuations mentioned above is mandatory.  
\smallskip

We find that, although Cooper-Frye formula can 
give good results in many cases, a more realistic 
decoupling description is needed for obtaining many 
other observables. A description we proposed is the 
Continuous Emission Model. One typical example we 
showed here, in which this model improve considerably 
the description of data is the $m_T$ dependence of HBT 
radii, observed at RHIC. 
\smallskip

In addition to the points summarized above, 
there are several basic questions to be addressed in 
the hydrodynamical scenary. First of all, although for 
the bulk properties of observables so far studied the 
ideal hydrodynamical description works fairly well, 
we still do not know how the non-equilibrium processes 
affect the results of these studies. This question 
refers both to the initial conditions and to 
the final particle decoupling process. For example, 
the present study of continuous emission showed that 
the real particle emission mechanism seems to be very 
far from the conventional Cooper-Frye scenary of sharp 
freeze-out surface. In Fig.~\ref{probability}, we 
showed that the escape probability ${\cal P}$ varies 
rather slowly both in space and time, and there is no 
indication of a sharp freezeout hypersurface. Another 
way of showing the same is given in Fig.~\ref{SxT}, 
where we plotted the entropy vs. temperature of the 
SPH particles at the emission point. The temperature 
values corresponding to the particle emission spread 
widely. This shows that, even for a large system 
(Au+Au) there exists a substantial part of the system 
which remains out of equilibrium for a long time. Up 
to now, no dynamical reaction of these non-equilibrium 
components on the hydrodynamical evolution is studied. 
In this sense, it is very important to develop 
transport theoretical investigations~\cite{yuri,lazslo} 
related to hydro description. 

\begin{figure}[!t]
\begin{center}
\includegraphics[width=8.8cm]{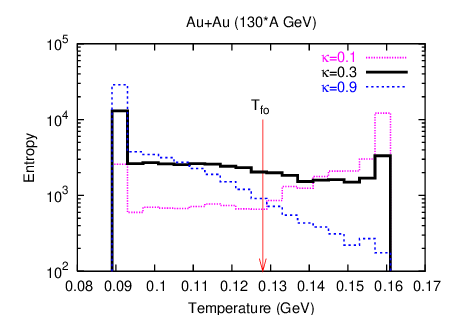}
\end{center}
\caption{The entropy liberated vs. temperature of SPH particles 
at the decoupling point. The freeze-out temperature
$T_{fo}$ used in the studies above is also shown.}
\label{SxT}
\end{figure}

Another interesting type of phenomena related to the 
hydrodynamical model is the ocurrence of instabilities 
associated to the phase transition and surface tension 
(finite-size effects), such as spinodal instabilities \cite{fraga04}. 
Some studies in this direction is in progress. 

\bigskip

\ni {\bf Acknowledgments} 
\medskip

The authors are grateful to all the SPheRIO team, 
namely, F. Grassi. C.E. Aguiar, T. Osada, B.M. Tavares, 
L.L.S. Portugal, G. Grunfeld, R. Andrade, for their 
collaboration and stimulating discussions. Especially, 
the most of the hard task in the development of the 
code is due to T. Osada. The main part of the 
consideration  on hadronic resonance is due to 
C.E. Aguiar. B.M. Tavares, L.L.S. Portugal, 
A.G. Grunfeld helped the construction of equations 
of state. We are indebted to K. Werner for valuable 
discussions on NeXuS event generator and for 
providing us with the correspondent code. Analytical 
solutions have been obtained in a fruitful 
collaboration with T. Cs\"org\H o. We are thankful to 
Y.M. Sinyukov and L.P. Csernai for profitable 
discussions on transport theoretical description of 
continuous emission. We also appreciated the 
discussions and the encouragements received from 
F. Pottag, S.S. Padula, F.S. Navarra, S.B. Paiva, 
M. Makler, M. Gorenstein, M. Ga\'zdzicki, J. Rafelski, 
H.T. Elze and E. Fraga. 

We acknowledge financial support by FAPESP 
(1990/4074-5, 1993/2463-2, 1995/4635-0, 1998/2249-4, 
2000/04422-7 and 2001/09861-1), CAPES/PROBRAL, CNPq, 
FAPERJ and PRONEX.

\end{document}